\documentclass[a4paper,fleqn,usenatbib]{mnras}

\usepackage{newtxtext,newtxmath}
\usepackage[T1]{fontenc}
\usepackage{ae,aecompl}

\usepackage{graphicx}	% Including figure files
\usepackage{amsmath}	% Advanced maths commands
\usepackage{amssymb}	% Extra maths symbols
\usepackage{arydshln}
 
\usepackage{verbatim}
\usepackage{breqn}

\newcommand{\lya}{Ly$\alpha$}
\newcommand{\lyb}{Ly$\beta$}

\newcommand{\zem}{$z_{em}$}
\newcommand{\zabs}{$z_{abs}$}
\newcommand{\kms}{$km s^{-1}$}

\newcommand{\HI}{\mbox{H\,{\sc i}}}

\newcommand{\OVI}{\mbox{O\,{\sc vi}}}

\newcommand{\CII}{\mbox{C\,{\sc ii}}}

\newcommand{\CIV}{\mbox{C\,{\sc iv}}}

\newcommand{\SiII}{\mbox{Si\,{\sc ii}}}
\newcommand{\SiIII}{\mbox{Si\,{\sc iii}}}
\newcommand{\SiIV}{\mbox{Si\,{\sc iv}}}

\newcommand{\MgII}{\mbox{Mg\,{\sc ii}}}

\newcommand{\ang}{\textup{\AA}}
\title[Spatial correlations]{Three- and two-point spatial correlations of intergalactic medium at $z\sim2$ using projected quasar triplets }

\author[Maitra et al.]{Soumak Maitra,$^{1}$\thanks{E-mail: soumak@iucaa.in}
	Raghunathan Srianand,$^{1}$
	Patrick Petitjean$^{2}$,
	Hadi Rahmani$^{3}$,
	\newauthor{Prakash Gaikwad$^{4,5}$, 
	Tirthankar Roy Choudhury$^{6}$ \& Christophe Pichon$^{2,7}$  }
	\\
	$^{1}$ IUCAA, Postbag 4, Ganeshkhind, Pune, 411007, India\\
	$^{2}$ Institut d'Astrophysique de Paris, CNRS-SU, UMR 7095, 98bis bd Arago, 75014, Paris, France\\
	$^{3}$  GEPI, Observatoire de Paris, PSL Universite, CNRS, 5 Place Jules Janssen, 92190 Meudon, France\\
	$^{4}$ Institute of Astronomy, University of Cambridge, Madingley Road, Cambridge, CB3 0HA, UK\\
	$^{5}$ Kavli Institute for Cosmology, University of Cambridge, Madingley Road, Cambridge, CB3 0HA, UK\\
	$^{6}$ National Centre for Radio Astrophysics, Tata Institute of Fundamental Research, Pune 411007, India \\
	$^{7}$ School of Physics, Korea Institute for Advanced Study (KIAS), 85 Hoegiro, Dongdaemun-gu, Seoul, 02455, Republic of Korea \\
}

\date{Accepted XXX. Received YYY; in original form ZZZ}

\pubyear{0000}

\begin{document}
	\setlength{\parskip}{0pt}
	\label{firstpage}
	\pagerange{\pageref{firstpage}--\pageref{lastpage}}
	\maketitle
	
	\begin{abstract}
	We present analysis of two- and three-point correlation functions of \lya\ forest (at $2\le z\le 2.5$) using X-Shooter spectra of three background quasar triplets probing transverse separations of 0.5-1.6 pMpc. We present statistics based on transmitted flux and clouds identified using Voigt profile fitting. We show that the observed two-, three-point correlation functions and reduced three-point correlation (i.e Q) are well reproduced by our simulations. We assign probabilities for realising all the  observed correlation properties simultaneously using our simulations.
	Our simulations suggest an increase in correlation amplitudes and Q with increasing $N_{\rm HI}$. {We roughly see this trend in the observations too.} We identify a concurrent gap of 17\AA\ (i.e 14.2 $h^{-1}$cMpc, one of the longest reported) wide in one of the triplets. Such  gap is realised only in 14.2\% of our simulated sightlines and most of the time belongs to a void in the matter distribution. 
	In the second triplet, we detect DLAs along all three sightlines (with spatial separations 0.64 to 1.6 pMpc) within a narrow redshift interval (i.e $\Delta z = 0.088$). Detection of a foreground quasar ($\sim$ 1 pMpc from the triplet sightlines) and excess partial Lyman Limit systems around these DLAs suggest that we may be probing a large over-dense region. We also report positive \CIV-\CIV\ correlations up to $\sim 500$~\kms only in the longitudinal direction. Additionally, we conclude a positive \CIV-\lya\ correlations for higher $N_{\rm HI}$ thresholds up to a scale of $\sim 1000$~\kms both in transverse and longitudinal directions.
	\end{abstract}
	

	%%%%%%%%%%%%%%%%%%%%%%%%%%%%%%%%%%%%%%%%%%%%%%%%%%
	
	%%%%%%%%%%%%%%%%% BODY OF PAPER %%%%%%%%%%%%%%%%%%
	
	\section{Introduction}

 The \lya\ forest absorption seen in the spectra of distant quasars directly probe structures in the inter-galactic medium (IGM) and are therefore used to constrain (i) primordial density fluctuations \citep[][]{bi1992,mcdonald2003}, (ii) cosmic reionization \citep[][]{fan2006, worseck2018}, (iii) thermal history of the universe \citep[][]{hui1997,schaye2000} and (iv) the impact of various
  feedbacks processes (such as SNe and AGN driven outflows) on the IGM that operate during the formation and evolution of galaxies over cosmic time \citep{aguirre2001,openheimer2006}. Numerical simulations
  and analytical modelling of a warm photo-ioinized IGM in
  the framework of $\Lambda$CDM models successfully reproduce many observational properties of the \lya\ forest absorption: the column density ($N_{\rm HI}$)  distribution, the
  Doppler b-parameter distribution, the flux probability distribution function, power-spectrum of transmitted flux and the redshift evolution of absorption lines above a certain $N_{\rm HI}$ threshold \citep[see ][]{cen1994, Petitjean1995, springel2005, smith2011, bolton2012, rudie2012,gaikwad2017a,gaikwad2017b}. Through these models we can constrain the  H~{\sc i} photoionization rate, mean  IGM temperature and effective equation of state
  over a large redshift range \citep[see for example,][]{schaye2000,becker2013}.

 \begin{figure*}
     \centering
    
\textbf{Field 1\qquad\qquad\qquad\qquad\qquad\qquad\qquad\qquad\quad Field 2 \qquad\qquad\qquad\qquad\qquad\qquad\qquad\qquad Field 3}
     \includegraphics[width=5.5cm]{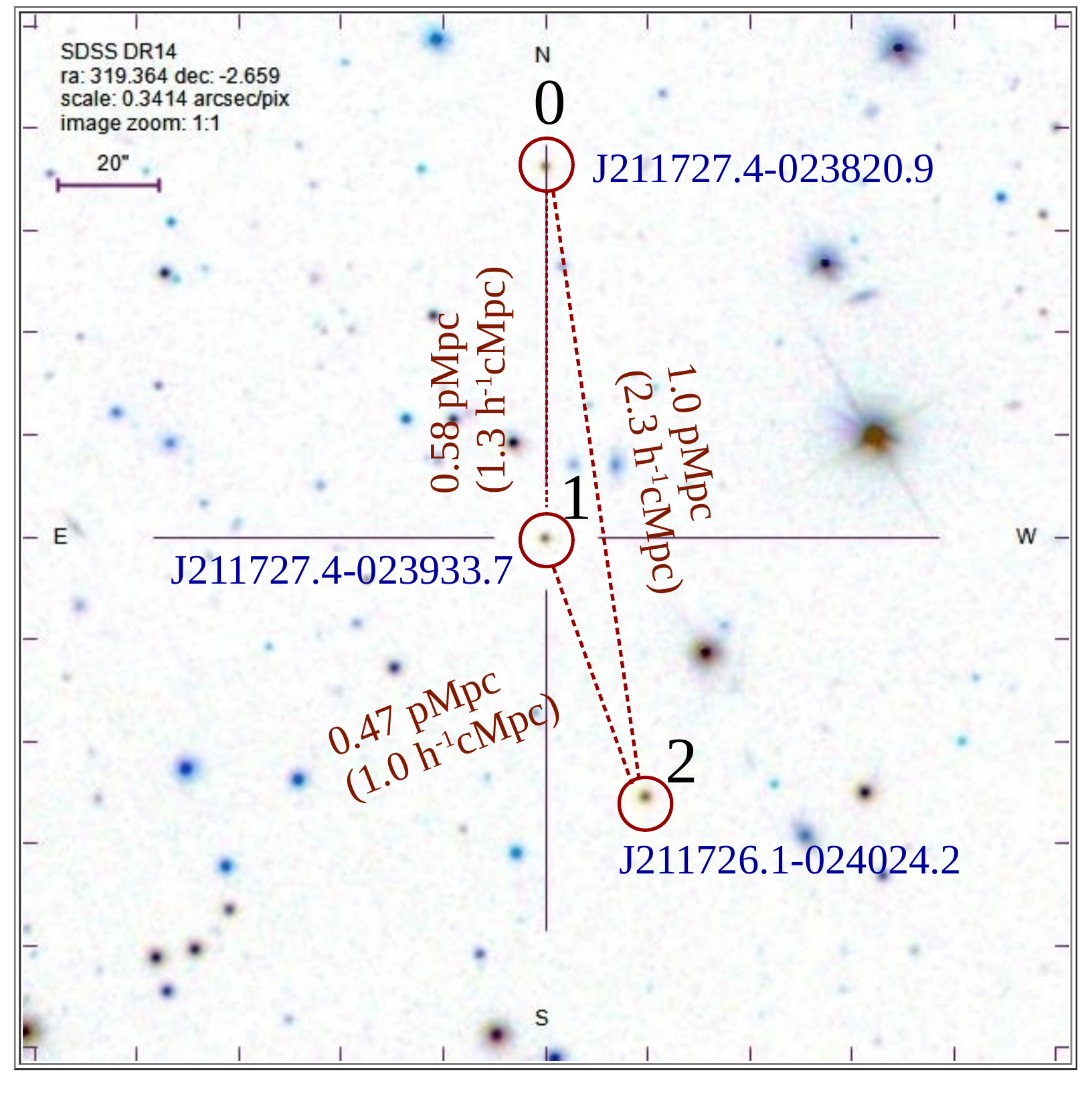}%
     \includegraphics[width=5.5cm]{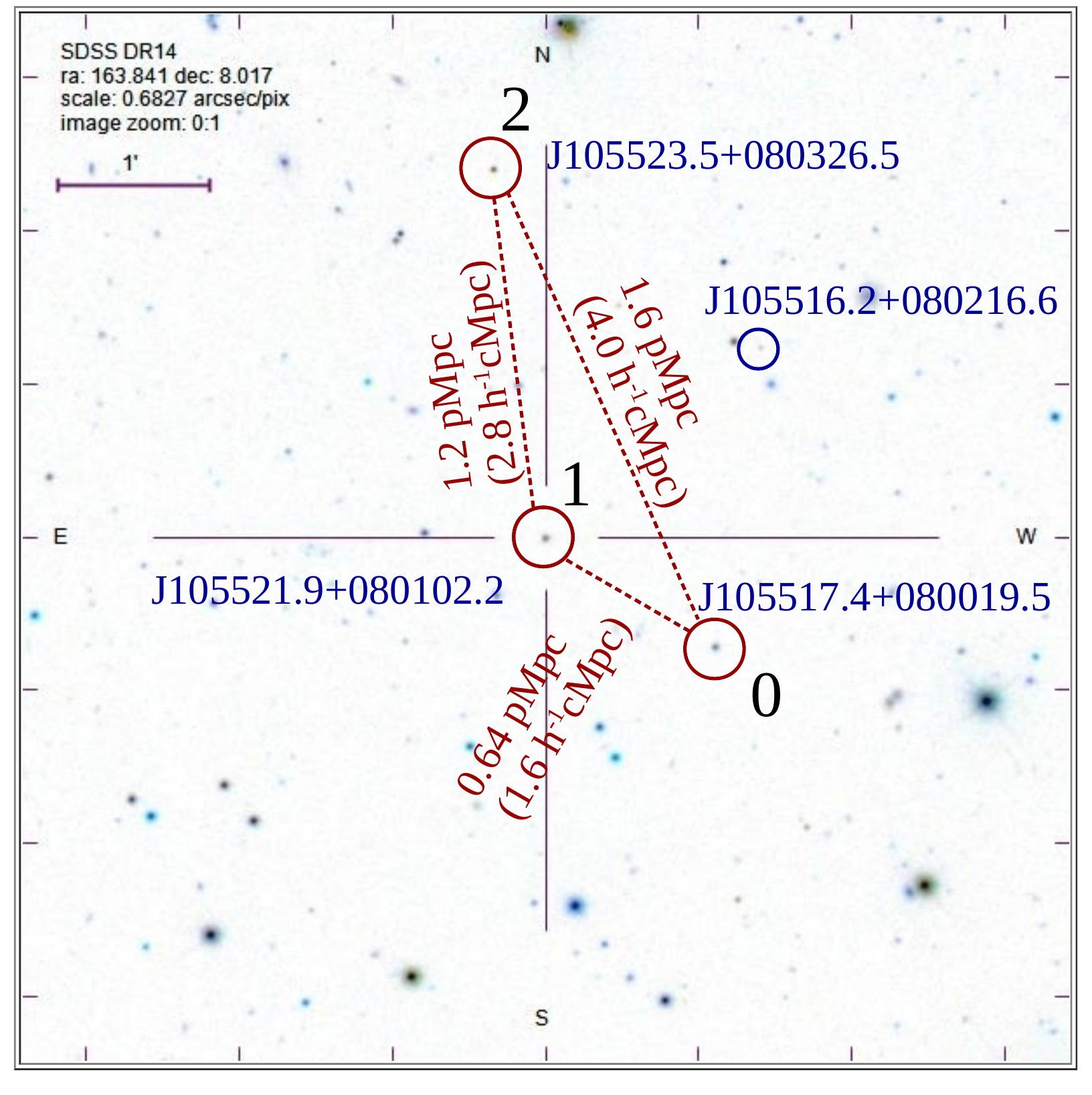}%
     \includegraphics[width=5.5cm]{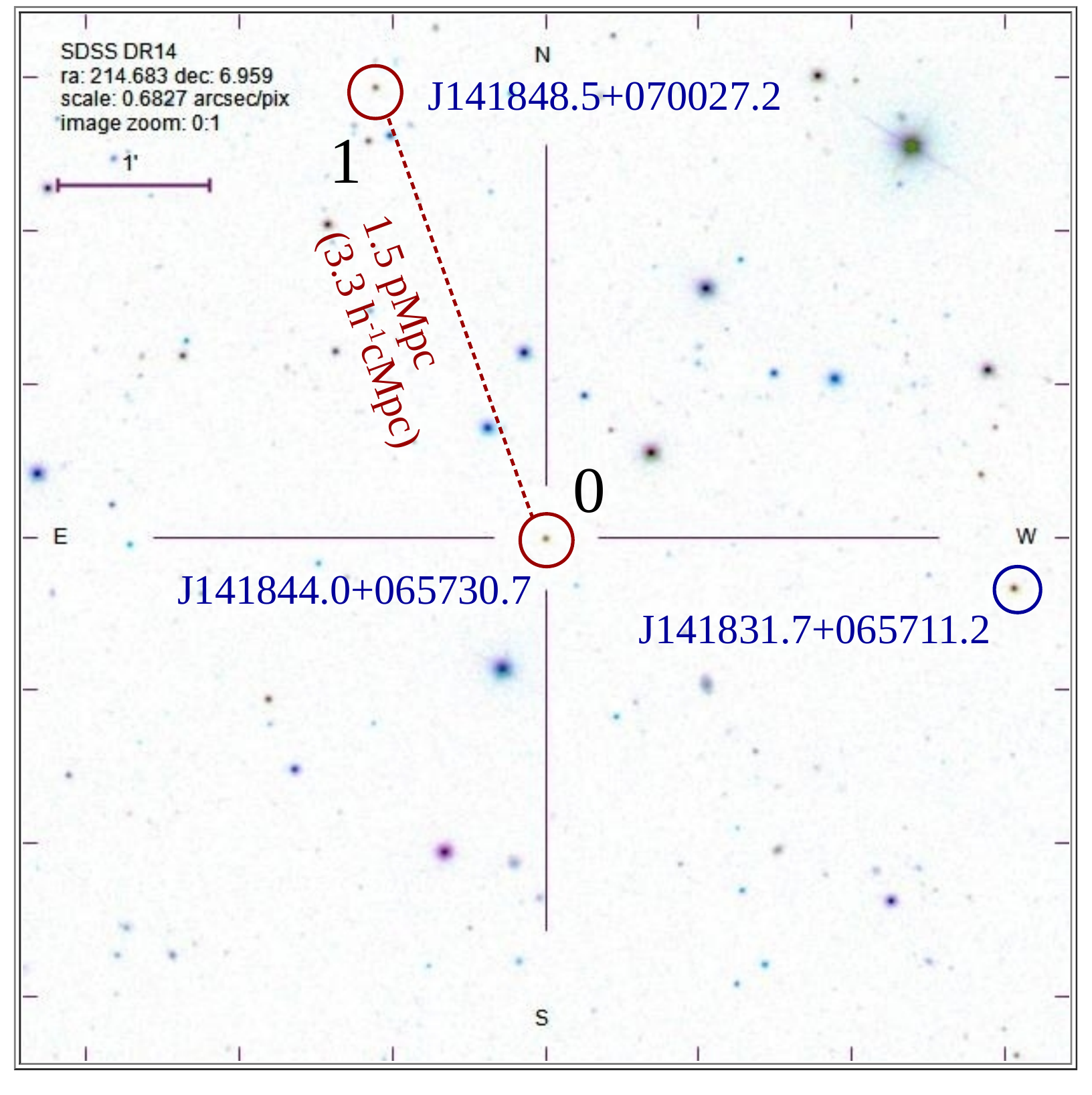}%
     \caption{Fields of quasars used in this study. Red circles mark the position of quasars with X-Shooter spectra. Blue circles mark quasars for which we have only the SDSS spectra. The physical (as well as comoving) distances quoted in the figure are computed at the lowest emission redshift among the triplets.}
     \label{config}
 \end{figure*}

In these simulations, the H~{\sc i} density fluctuations responsible for the \lya\ absorption closely trace those of the underlying dark matter density field on scales
larger than the pressure smoothing scales \citep[$>$ few 100 ckpc; see for example,][]{miralda1996,schaye2001}. Most of the baryons producing  absorption having $N_{\rm HI}$ $\sim$ 10$^{14}-10^{15}$ cm$^{-2}$ are found to be  located in mildly non-linear regimes probed by filaments and sheets at $z\sim2$ \citep{Petitjean1995}.  Most of the volume is, however, occupied by under-dense regions that produce unsaturated \lya\ absorption lines if any. 
 Therefore a rich insight into
  the morphological properties of the cosmic web (filaments, sheets, voids and connection between them),
  but also the large scale ionization and chemical inhomogeneities and their redshift evolution
  can be gained by
  simultaneous analysis of \lya\ absorption detected along closely spaced sightlines.
  Virtual experiments performed on simulated data suggest that it is possible to reconstruct a 3D map
  of the full density field using a dense
  enough grid of spatially close lines of sight \citep[see,][]{Pichon2001, mcdonald2003, caucci2008, Lee2018}.
  
  Using a group of background quasars, we can probe different physical
  processes at different scales:  (i) The \HI\ density and velocity fields at the scales of few 100 cKpc may have thermal memory of cosmic reionization in the form of pressure broadening\citep{peeples2010,rorai2018}; (ii) at the scale of $\sim$ 1 pMpc we can probe
  matter clustering around massive galaxies (quasar hosts and intervening metal systems) and various feedback processes
  connecting gas flows between galaxies and the IGM. At present these scales are best probed using quasar pairs \citep[see for example,][]{Prochaska2013};
  (iii) at the scales of one to few Mpc, one is probing the cosmic structure of filaments and
  voids and the effect of radiative feed back from bright persistent objects like quasars \citep[e.g][]{finley2014,Lee2018}; and  (iv) the Baryon Acoustic Oscillations (BAO) 
at $\sim$100 Mpc probes primordial density fluctuations at very large scales \citep[][]{ata2018}.

  Our understanding of small scale \lya\ clustering
  is mainly dominated by observations of quasar pairs and gravitationally lensed images of quasars \citep[][]{Smette1995, rauch1995, petitjean1998,aracil2002,coppolani2006,dodorico2006}.
  \citet{cappetta2010} have studied the 3D distribution of \lya\ forest at $z\sim2$ using high resolution spectroscopy of the quasar pairs, one triplet and a sextet. However, such high signal to noise ratio (SNR) and high spectral resolution studies of triplets or multiple quasar sightlines are rare due to lack of bright targets. Recently, \citet{Krolewski2018} presented 3D \HI\ density field reconstruction at $z\sim 2$ using wiener filtering technique (with an effective smoothing scales of 2.5 h$^{-1}$ cMpc) applied to \lya\ absorption detected in moderate resolution spectra of high-$z$ star-forming galaxies.  While we wait for the arrival of
  extremely large telescopes to probe IGM tomography over a large range of scales with better sensitivity \citep[][]{skidmore2015,evans2016}, SDSS quasar catalog offers rare possibilities to make some important progress. Recently, \cite{tie2019} has made a theoretical predication of three-point correlation in \lya\ forest at large scales (10-30$h^{-1}$ Mpc).

  In this paper, we present detailed analysis spatial correlations of \lya\ forest and metal absorption lines using our X-Shooter spectra along the line of sight towards two quasar triplets and a quasar pair (see Fig.~\ref{config}). Details of the quasars studied here are summarised in Table~\ref{Tab_obs}. 
The "Field 1" consists of quasars J211727.4$-$023820.9 (\zem = 2.323), J211727.4$-$023933.7 (\zem = 2.309) and J211726.1$-$024024.2 (\zem = 2.309). We will subsequently refer to these as "Triplet 1". The spatial separation between the sightlines ranges from 0.47 - 1.00 pMpc.
  The "Field 2" consists of quasars J105517.4+080029.5 (\zem = 2.897), J105521.9+080102.2 (\zem = 2.709) and J105523.2+080326.5 (\zem = 2.627) with the spatial separations probed in the range 0.63 to 1.6 pMpc which we will subsequently refer as "Triplet 2". In the SDSS-DR12 database we find a 4th quasar J105516.23+080216.6 (\zem = 2.320) with a typical separation of 1 pMpc from the other quasars (see Fig.~\ref{config}). We also study the distribution of \HI\ gas around this quasar along the line of sight to the three background quasars.
 The "Field 3" consists of quasars J141848.5+070027.2 (\zem = 2.2305), J141844.03+065730.7 (\zem = 2.403) and J141831.72+065711.2 (\zem = 2.389). In this case, we got X-Shooter spectra only for the first two quasars which we will refer to as the "Doublet".
    Therefore, our analysis of this triplet is restricted to two-point correlation function of the IGM and gas distribution around the lowest redshift quasar.
    
 		\begin{table*}
		\caption{Details of quasars studied in this work.} 
		\begin{tabular}{lcccccc}
			\hline
			\textbf{QSO} & $z_{e}$     & $\lambda$ range & \multicolumn{1}{c}{SNR} & \multicolumn{1}{c}{FWHM} & \multicolumn{1}{c}{$L_{912\ang}$} & \multicolumn{1}{c}{$r_{eq }$} \\ 
			 & & (\AA) & & (\kms) &$(10^{49}erg s^{-1} \ang^{-1}$) &(pMpc)\\
			\hline
		    \hline
			\multicolumn{7}{c}{\underline{\textbf{Field 1} ($r_{01}$ = 0.579 pMpc , $r_{12}$ = 0.472 pMpc , $\theta=160^{\circ}$    )}}\\
%			\\
			J211727.4-023820.9 (X-Shooter)   & 2.3230 &  3460-3940          & 17.3                     & 50           & 17.1           & 1.60                                   \\
			J211727.4-023933.7 (X-Shooter)   & 2.3090 &   3460-3940         & 19.7                     & 47           &9.5          & 1.20                                            \\
			J211726.1-024024.2 (X-Shooter)   & 2.3090 &    3460-3940        & 18.0                     & 51  		&33.6          & 2.26                               \\
%\\
			\multicolumn{7}{c}{\underline{\textbf{Field 2} ($r_{01}$ = 0.635 pMpc , $r_{12}$ = 1.155 pMpc , $\theta=130^{\circ}$    )}}\\
%			\\
			J105517.4+080019.5 (X-Shooter)   & 2.897 &    4055-4333        & 25.3                     & 68           & 43.9           & 2.77                                   \\
			J105521.9+080102.2 (X-Shooter)   & 2.709 &   4055-4333         & 13.1                     & 73           &43.4          & 2.72                                             \\
			J105523.5+080326.5 (X-Shooter)   & 2.627 &   4055-4333         & 28.8                     & 73  		&58.6          & 3.12
			\\
%			\hdashline
			J105516.2+080216.6 (SDSS)   & 2.320 &     -      &   -                   &   -	&     3.2     & 0.69
			\\
%			\hdashline
%			\\
			\multicolumn{7}{c}{\underline{\textbf{Field 3} ($r_{01}$ = 1.520 pMpc    )}}\\
 %           \\
%            \hdashline
            J141831.7+065711.2 (SDSS)	& 2.389	&	-	& -    	&	-	& 106.4		& 4.05	
			\\
%			\hdashline
			J141844.0+065730.7 (X-Shooter)	& 2.4030	&	3490-3856	&     16.7	&	73	&	41.0	&	2.53
			\\
			J141848.5+070027.2 (X-Shooter)	& 2.2305	&	3490-3856	&     27.7	&	63	&	33.8	&	2.23\\
			\hline         
		\end{tabular}
		\label{Tab_obs}
	\end{table*}

   This paper is organized as  follows. In section 2, we provide the details of observations, quality of spectra achieved and properties of the quasars in our sample. In section 3, we provide details of our simulations and generation of spectra. In section 4, we validate our simulations by reproducing some observational results.
  In section 5, we present the observed transverse and longitudinal two-point and three-point correlations of \lya\  forest measured based on transmitted flux as well as using Voigt profile decomposed "clouds". We also quantify the probability of obtaining sighlines in our simulations similar to what we find along the triplets discussed here. In addition to this, we also study the distribution of coherent gaps. In section 6, we present QSO-\lya, DLA-\lya\ and \CIV\ transverse correlations. Our main results are summarised in section 7.
  
 \section{Details of observations}
 
 Spectra of quasars were obtained 
 with X-Shooter \citep{vernet2011} at the European Southern Observatory (ESO) Very Large
Telescope (VLT) in service mode under the programme ID: 096.A-0193 (PI: Petitjean).  The X-Shooter spectrograph covers a wavelength range of 0.3-2.3 $\mu$m at medium resolution in a simultaneous use of three arms in UVB, VIS and NIR. To have a robust sky subtraction, the nodding mode was used following an ABBA scheme. Slit width of 1.2 arcsec was used for all arms of X-Shooter in all our observations. This choice of slit widths results in formal spectral resolutions of 4000, 6700 and 3900 for the UVB, VIS and NIR, respectively. However, under good seeing conditions where the QSO point spread function (PSF) is less than slit width, the spectral FWHM is better than the predicted ones. Hence, for such cases we use the method described in \citet{krogager2017} to obtain the spectral resolution.

We have used the X-Shooter Common Pipeline Library \citep[][]{goldoni2006} 
release 6.5.1
%\footnote{http://www.eso.org/sci/facilities/paranal/instruments/xshooter/doc/} 
for reducing the science raw images and produce the final 2D spectra of the QSOs.  We first compute an initial guess for the wavelength solution and position of the center and edges of the orders. Then we trace the accurate position of the center of each order and follow this step by generating the master flat frame out of five individual lamp flat exposures. Next we find a 2D wavelength solution and modify it by applying a flexure correction to correct for the shifts that can be of the order of the size of a pixel. 
Finally, having generated the required calibration tables we reduce each pair of science frames to obtain the flat-fielded, wavelength calibrated and sky subtracted 2D spectrum. To extract the 1D flux of the QSO we follow a spectral point spread function (SPSF) subtraction as described in \citet{rahmani2016}. 
In summary we model the QSO's PSF using a Moffat function which is a smooth function defined by the centroid wavelength and FWHM. We then integrate the light profile at each wavelength pixel to obtain the flux of the QSO.

We fit the continuum to the 1D extracted spectra using lower order polynomial smoothly connecting the identified  absorption line free regions. 
In Table~\ref{Tab_obs}, we summarise various details of quasars used in our study.  For each triplet we provide $r_{01}$ (the projected separation between the 	first and the second quasar of the triple as listed in column 1),  $r_{12}$ (the projected separation between quasar 2 and 3) and $\theta$ (the angle between the two pairs). 
The median SNR obtained and typical FWHM of our spectral PSF (obtained as discussed above) are given in columns 4 and 5 respectively.

 \subsection{Quasar redshifts using narrow emission lines}
 
 \begin{figure}
 	\begin{center}
	\includegraphics[viewport=50 20 600 800, height=0.35\textheight,angle=270,clip=true]{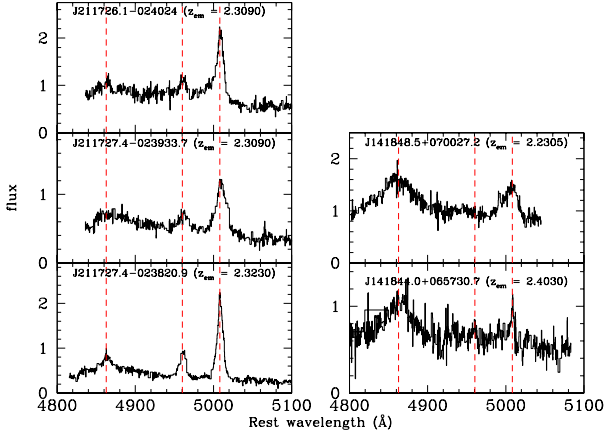}
		\end{center}
		\caption{H$\beta$ and [O~{\sc iii}] emission lies
detected in the spectra of 5 quasars in our sample. Vertical dashed lines mark the locations of emission lines. The measured systemic redshift is quoted in each panel.
} 
\label{fig_nlr}
	\end{figure}

The systemic redshift of the quasars is better determined using the low ionization broad and narrow emission lines \citep{gaskell1982}. 
In our X-Shooter NIR spectra [O~{\sc iii}], H$\beta$ and H$\alpha$ emission lines are clearly detected for all the quasars in  "Triplet 1" (See left panels in Fig.~\ref{fig_nlr}). This together with the detection of [C~{\sc iii}] and Mg~{\sc ii} emission lines in the VIS spectrum allow us to measure the systemic redshift of the quasars accurately. The second column in Table~\ref{Tab_obs} gives our measurement of the systemic redshift of the quasar. 
Based on the NIR spectrum both J2117-0240 and J2117-0239 are at the same redshift with a projected separation of 472 pKpc. Thus these quasars could be physically associated with each other. All three quasars show associated absorption (see Fig.~\ref{fig_CIV_a} in the Appendix). In the case of J2117-0238 and J2117-0240 narrow \CIV\ associated absorption are seen with both inflowing and outflowing signatures. In the case of J2117-0239 the \CIV\ absorption is consistent with broad absorption lines (BALs) with some of the narrow components  at very large ejection velocities (i.e $v\sim$ 4500 \kms) showing signatures of partial coverage. 

In the case of "Triplet 2", we do not clearly detect [O~{\sc iii}], H$\beta$ or H$\alpha$ emission lines in the NIR spectrum. Therefore, we base our systemic redshift determination mainly on the \MgII\ and [C~{\sc iii}] broad emission lines. Unlike quasars in "Triplet 1", in this case the redshifts of the quasar are very different. The quasar J1055+0800 shows associated broad \CIV\ absorption with clear signatures of high velocity outflowing components. In the case of J1055+0801 and J1055+0803 narrow associated \CIV\ absorption is clearly detected.
In the case of two quasars in "Doublet", in our NIR spectra we clearly detect [O~{\sc iii}], H$\beta$ and H$\alpha$ emission lines (see Fig.~\ref{fig_nlr}). These lines were used to determine the systemic redshift of the quasars.

We consider the common redshift range between \lya\ and \lyb\ emission lines of the quasars avoiding the proximity regions (i.e within 5000 \kms to the quasar redshift) for our IGM correlation studies. These wavelength ranges are provided in column 3 of Table~\ref{Tab_obs}.  We also avoid \lya\ of the associated absorption systems.

We identified \CIV\ and \MgII\ doublets  and DLAs in all our spectra. The redshifts of these absorbers are summarized in Table~\ref{tab_metal} in the Appendix.
We mask the \lya\ range that may be contaminated by metal absorption associated with these redshifts during our correlation analysis.

%\vspace {-1.cm}
\subsection{Ionization sphere of influence of quasars}
Assuming an isotropic continuum emission and for a given quasar Spectral Energy Distribution (SED), we can compute the radius of influence of quasar ionization for a given metagalactic UV ionizing background \citep[here we assume the one computed by ][]{khaire2019}. The \HI\ photoionization rate at distance $r$ from the quasar is given as
\begin{equation}
\Gamma(\HI,r) =
   \int_{100 \ang}^{912 \ang} \frac{L_{\lambda}/4\pi r^2}{hc/\lambda}\sigma_{\lambda}(\HI)d\lambda \ ,
\label{eq:gamma}
\end{equation}
where $L_{\lambda}$ is the specific \HI\ ionizing luminosity ($erg s^{-1} \ang^{-1}$) of the quasar, %$\Gamma(\HI, r)$ is the $\HI$ photoionization rate at separation $r$ from the quasar 
and $\sigma_{\lambda}(\HI)$ is the wavelength dependent ionization cross- section for \HI\ by photons with energy above 13.6 eV. 
%The luminosity is related to the flux as $L_{\lambda}=4\pi D_L^2 f_{\lambda}$ where $D_L$ is the luminosity distance of the quasar from the observer. 
We have assumed %the wavelength dependence for the rest UV spectrum 
the UV SED of our quasars as adopted by \citep{khaire2019} for computing their UV background.
We have taken the far UV spectral index $\alpha=-1.8$ for the flux calculation. 
The Lyman continuum luminosity inferred for each quasar using the observed flux at rest frame $\lambda\sim1450$\AA\  is given in 6th column of Table~\ref{Tab_obs}.
The ionization radius is then defined to be the radius ($r_{eq}$) at which
$\Gamma(r,\HI)$ from the quasar is equal to the background photoionization rate.
The computed $r_{eq}$ values for all the quasars are given in the last column of the Table~\ref{Tab_obs}. It is clear from the table that all the quasars in the "Triplet 1" and "Triplet 2" will influence the ionization state of the IGM along the other two sightlines if the continuum emission is isotropic.  However this may not be the case for J1055+0802 found close to the sightlines along "Triplet 2" as inferred
$r_{eq}$ is less than the separation between this quasar and the nearest quasar sightline.

	\section{Simulation}
	We use the smoothed particle hydrodynamical code {\sc gadget-3} \citep[a modified version of
the publicly available
{\sc gadget-2}\footnote{\url{http://wwwmpa.mpa-garching.mpg.de/gadget/}} code , see][]{springel2005} for generating $100 h^{-1} $cMpc simulation box with $2\times 1024^3$ particles. We use standard flat $\Lambda$CDM cosmology with parameters ($\Omega_{\Lambda}$, $\Omega_{m}$, $\Omega_{b}$, $h$, $n_s$, $\sigma_8$, $Y$ ) $\equiv$
	(0.69, 0.31, 0.0486, 0.674, 0.96, 0.83, 0.24). The initial conditions are generated at $z=99$ using the publicly available {\sc 2lpt}\footnote{\url{https://cosmo.nyu.edu/roman/2LPT/}} \citep{scoccimarro2012} code. The gravitational softening length has been taken as $1/30^{th}$ of the mean inter-particle separation. The {\sc gadget-3} simulation incorporates radiative heating and cooling of SPH particles internally for a given UV background assuming ionization and thermal equilibrium  but solves time-dependent temperature evolution equation.  In our case, we have used the ionization and heating rates as given by \citet{khaire2019} for the assumed far-UV spectral index of $\alpha=-1.8$. In order to run the simulation faster, we also use the {\sc quick\_lyalpha} flag in the simulation which converts gas particles with $\Delta>10^3$ and $T<10^5K$ to stars \citep[see][]{viel2004a}. The simulation does not include AGN or stellar feedback or galactic outflows. We have stored the simulation outputs between $z=6$ and $z=1.8$ with a redshift interval of 0.1. Considering the median redshift intervals, we use simulation box at $z=2$ and $z=2.5$ for "Triplet 1" and "Triplet 2" respectively. In these redshifts our assumed box size provide a line of sight wavelength coverage of $\sim$122\ang\ and 151.4\ang\ respectively. The resolution of the final simulation spectrum we obtain is sufficient to resolve the features in X-Shooter spectrum.

	\subsection{Transmitted flux and Voigt profile fitting}
	To generate triplet (or a doublet) sight lines  having configurations similar to the observed one, we place the triplet (or doublet) source configuration along one face of the box and shoot lines of sight (LOS) parallel to the other faces. For the generated LOSs, neutral hydrogen density $n_{\rm HI}$, temperature  $T$ and the peculiar velocity $v$ are assigned along the LOS using SPH smoothing of the nearby particle values. We typically sample each line of sight with 2048 equally sampled grids in comoving length of the box. 
	According to the SPH formulation \citep{monaghan1992,springel2005}, the value of a quantity $f_i$ at the $i^{th}$ grid point is expressed as
	\begin{equation}
	f_i=\sum_{j} f_j \frac{m_j}{\rho_j}W_{ij} \ ,
	\end{equation}
	where the summation is done over all the particles. $f_j$, $m_j$ and $\rho_j$ are the values of the quantity, mass and density of the $j^{th}$ particle, respectively.  $W_{ij}$ is the SPH kernel which is a window function that depends on the distance between the $i^{th}$ grid and $j^{th}$ particle ($r_{ij}$) and the smoothing length $h_{j}$. 
	We use the SPH kernel of \citet{springel2005}:

	\begin{equation}
	W(r,h)\equiv \frac{8}{\pi h^3}
	\begin{cases}
	1-6\left(\frac{r}{h}\right)^2+6\left(\frac{r}{h}\right)^3, & \text{if}\  0\leq\frac{r}{h}\leq\frac{1}{2} \\
	2\left(1-\frac{r}{h}\right)^2, & \text{if}\  \frac{1}{2}\leq\frac{r}{h}\leq1 \\
	0 & \text{if}\  \frac{r}{h}>1
	\end{cases} 
	\end{equation}

	Next, using the $n_{\rm HI}$, temperature and velocity fields, we obtain the \lya\ optical depth $\tau$ along the sightlines \citep[see Eq.30 of][]{choudhury2001}. 
	The \lya\ transmitted flux $F$ is obtained as the negative exponential of the optical depth, i.e, $F=e^{-\tau}$.
	We add the effects of instrumental resolution and noise to the simulated \lya\ transmitted flux skewers. The transmitted flux is convolved with the instrumental LSF (line spread function) which we assume as a gaussian with FWHM $\sim 50$ \kms for the "Triplet 1" and FWHM $\sim 70$ \kms for "Triplet 2". The data is then rebinned to $\sim 15$ \kms pixels to match the pixel sampling in our X-Shooter spectra. 
	Next, to incorporate the effects of noise, we add a simple Gaussian noise to the skewers corresponding to the SNR values mentioned in columns 4 of Table~\ref{Tab_obs}. This simulated spectra are then used for all the statistics that are based on the transmitted flux. 

    An alternative approach to the flux based statistics is that instead of treating the IGM as a continuous fluctuating density field, 
    is to decompose the \lya\ forest into Voigt profile components, i.e, distinct absorbers  parameterised by $z$, $N_{\rm HI}$ and b.
    For simplicity we denote individual Voigt profile components as "clouds". 
  We use the automated parallel Voigt profile fitting code {\sc viper} to identify the \lya\ absorption lines and obtain the column density and line width of the absorbers \citep[see][for details regarding VIPER]{gaikwad2017b}. The code assigns a rigorous significance level \citep[RSL, as described in][]{keeney2012} to these fitted absorption features. We consider only Voigt profile components for which the RSL$>3$ to avoid false identifications. Voigt profile fits to all the observed quasar sightlines used in our study are shown in Fig.~\ref{fig_J2117_spectra}, ~\ref{fig_J1055_spectra} and ~\ref{fig_J1418_spectra} in the Appendix. One of us (HR) fitted the \lya\ forest using {\sc vpfit} \citep{carswell2014} and we found a good agreement between the decompositions obtained using {\sc viper} and {\sc vpfit}. 
    %{\color{red} You can mentioned that we (Hadi) also fitted observed spectra by VPFIT and the estimated parameters are consistent. }
   
\section{Validation of our simulations}	
In this section we try to reproduce some of the well known properties of the high-$z$ \lya\ forest to validate our simulations before applying them to understand spatial correlations in the \lya\ forest.

	\subsection{Flux Probability distribution function}
	
	\begin{figure*}
		\begin{minipage}{0.24\textwidth}
			\includegraphics[width=4.5cm]{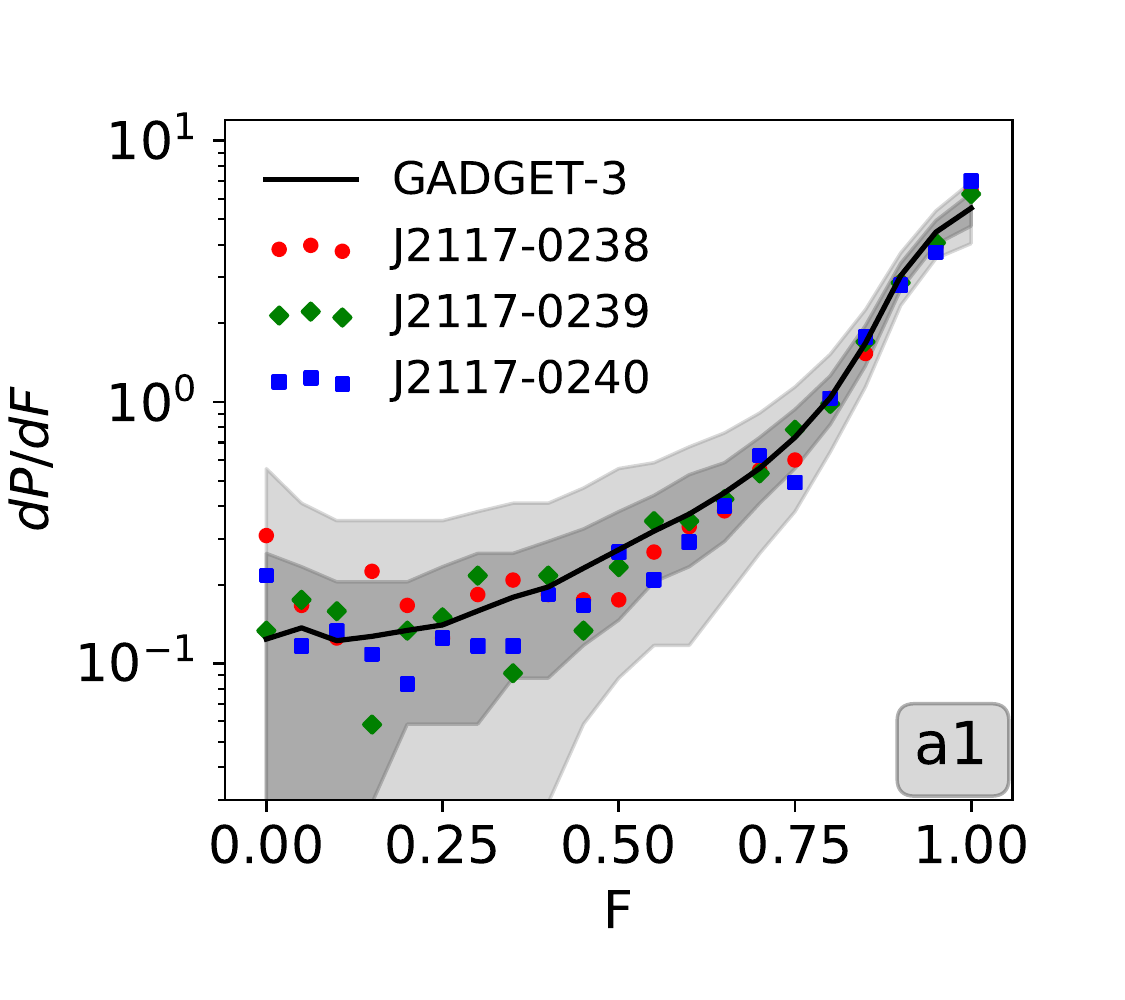}
		\end{minipage}%
		\begin{minipage}{0.24\textwidth}
		    \includegraphics[width=4.5cm]{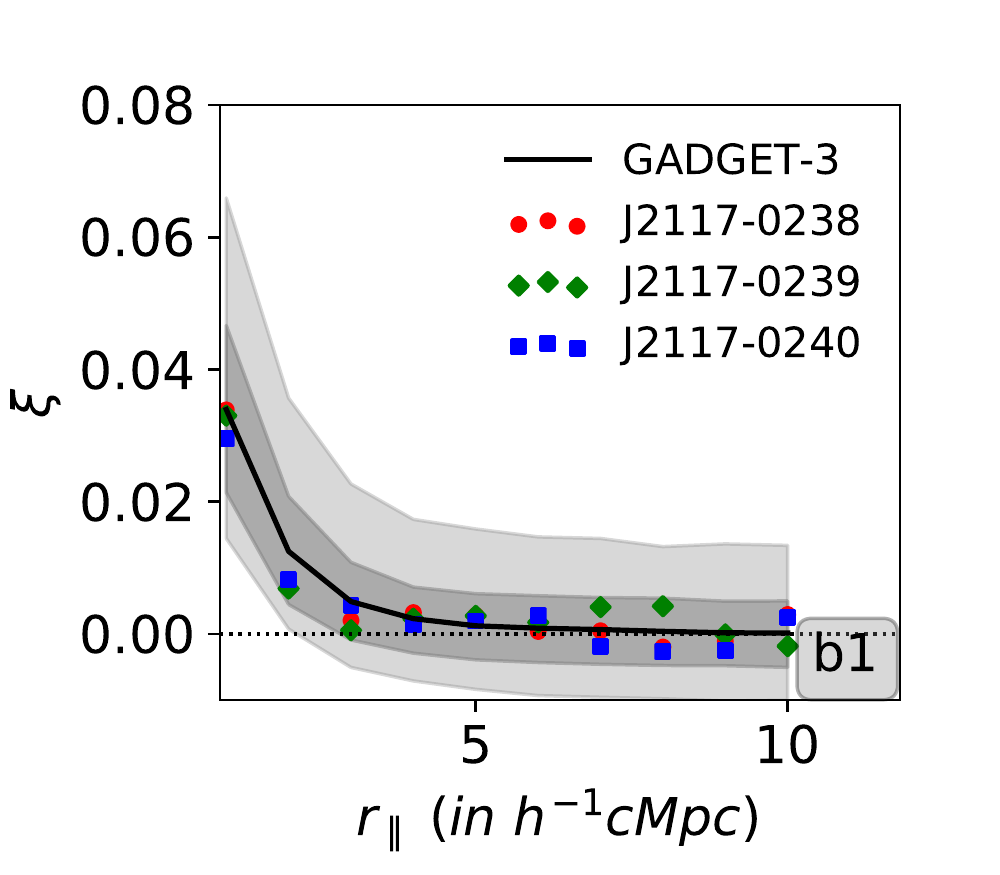}
		
		\end{minipage}
		\begin{minipage}{0.24\textwidth}
		    \includegraphics[width=4.5cm]{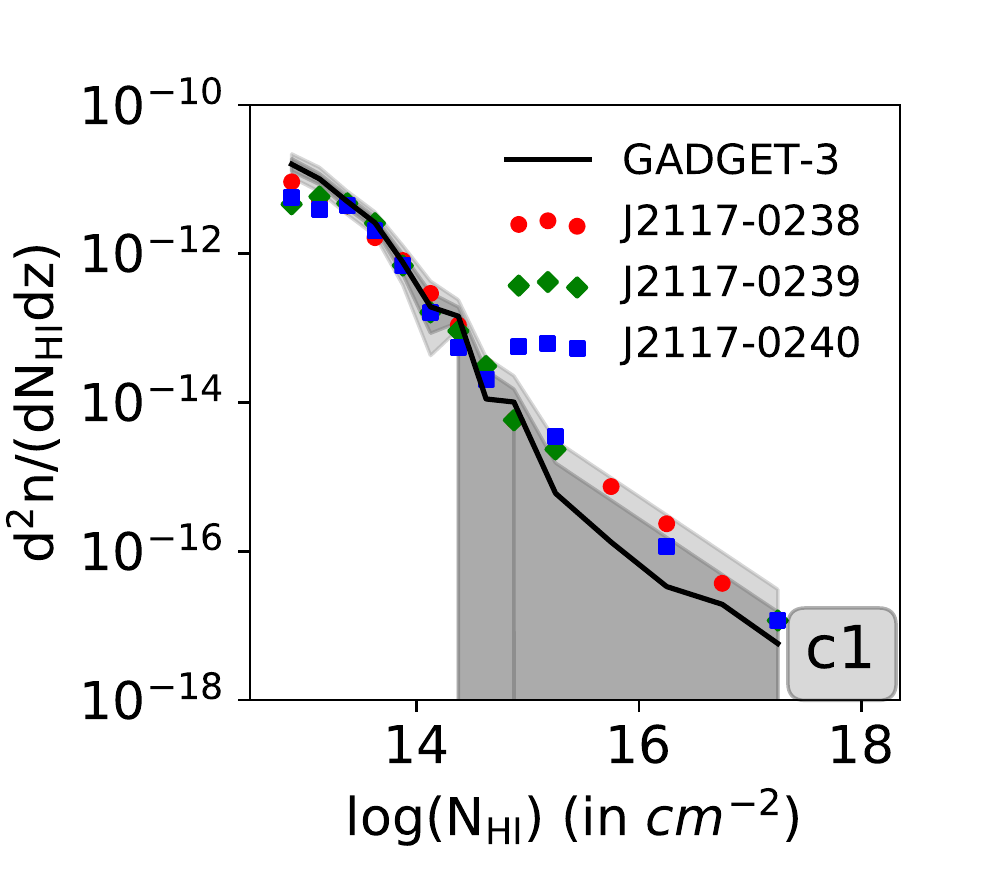}
		
		\end{minipage}%
		\begin{minipage}{0.24\textwidth}
		    \includegraphics[width=4.5cm]{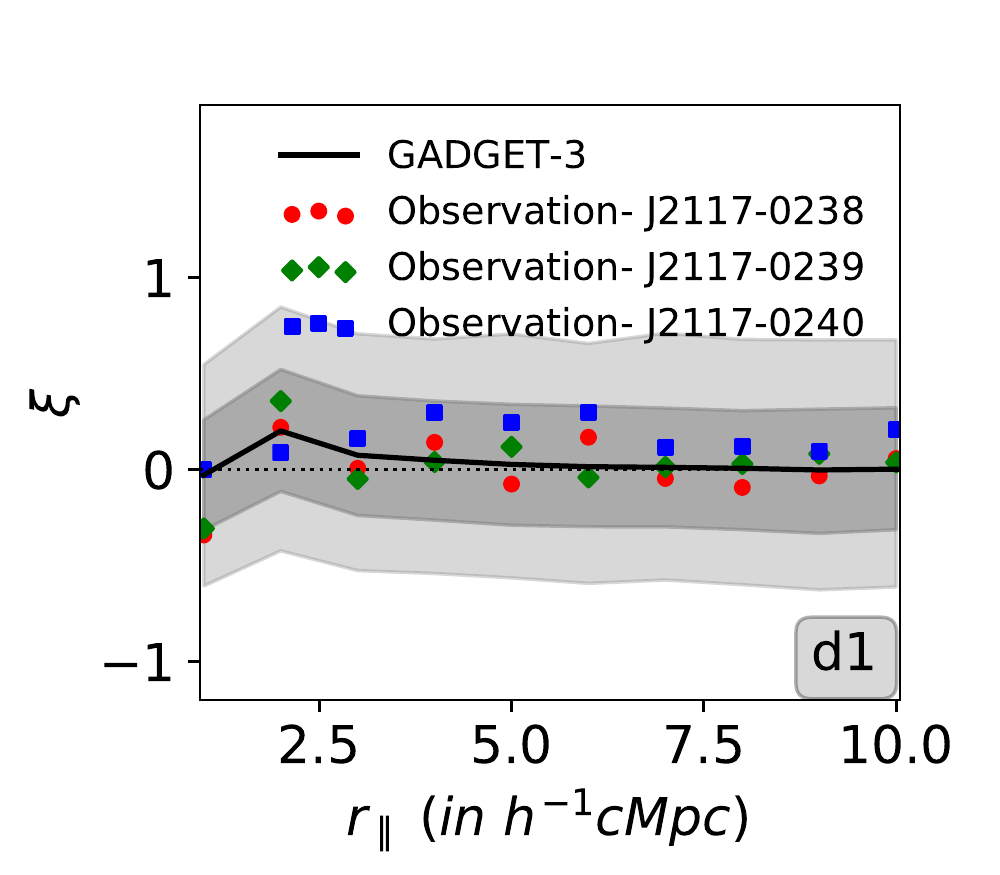}
		
		\end{minipage}

		\begin{minipage}{0.24\textwidth}
		    \includegraphics[width=4.5cm]{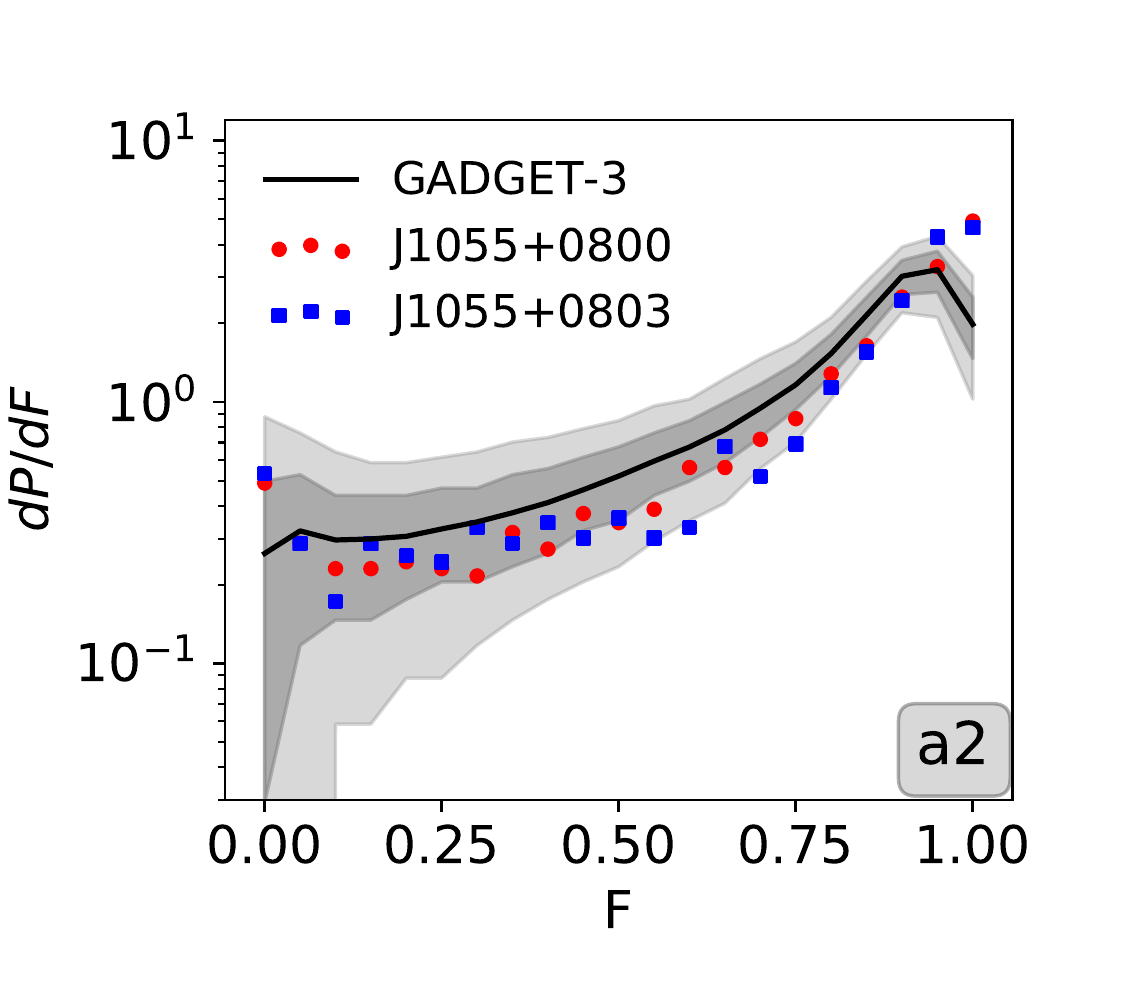}
		
		\end{minipage}%
		\begin{minipage}{0.24\textwidth}
			\includegraphics[width=4.5cm]{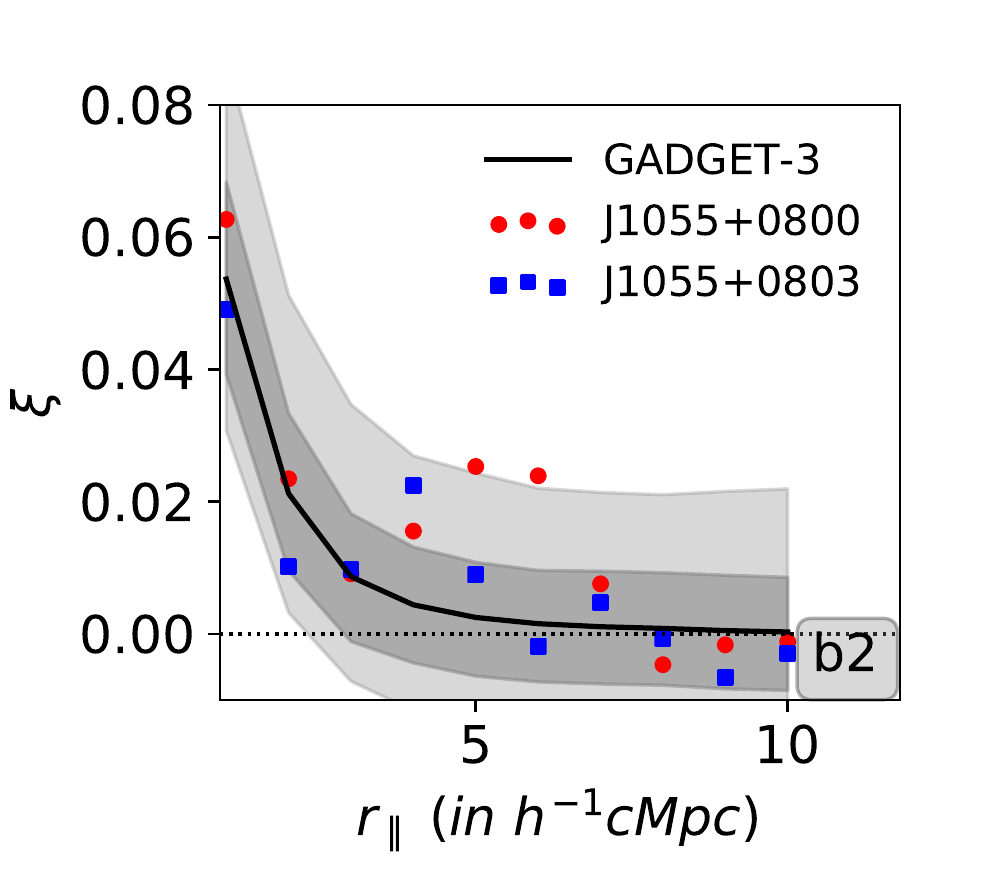}
		\end{minipage}
		\begin{minipage}{0.24\textwidth}
		    \includegraphics[width=4.5cm]{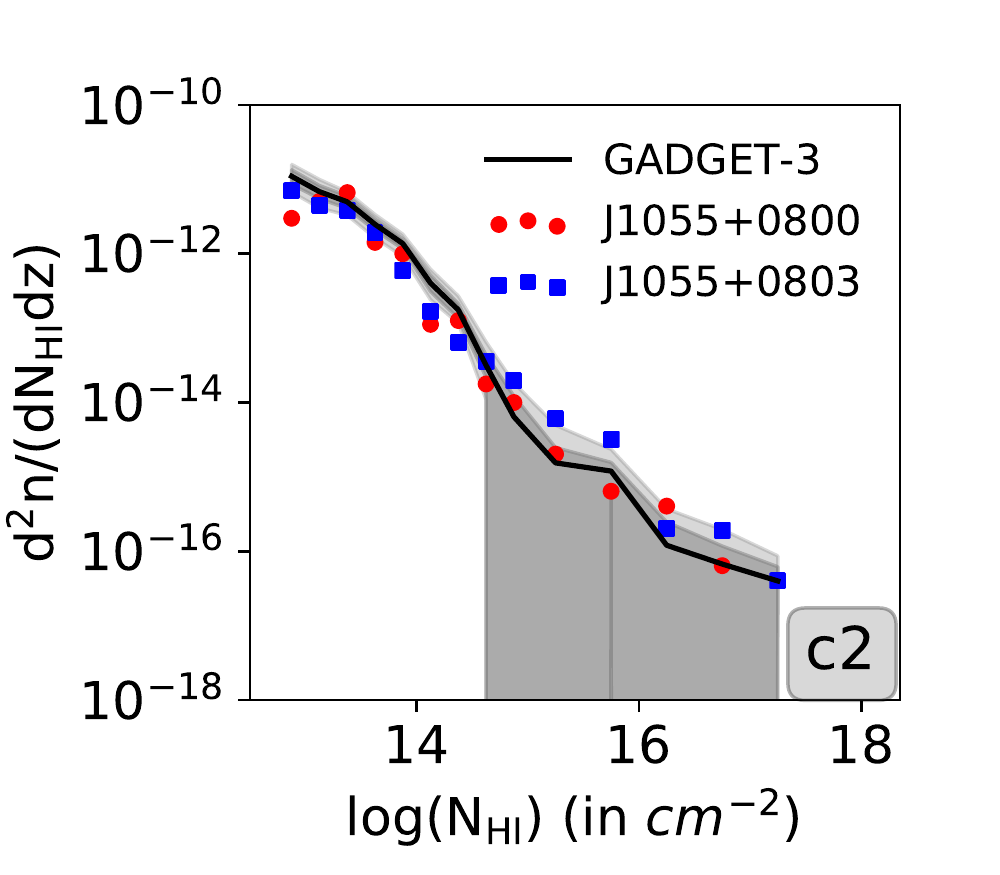}
			
		\end{minipage}%
		\begin{minipage}{0.24\textwidth}
			\includegraphics[width=4.5cm]{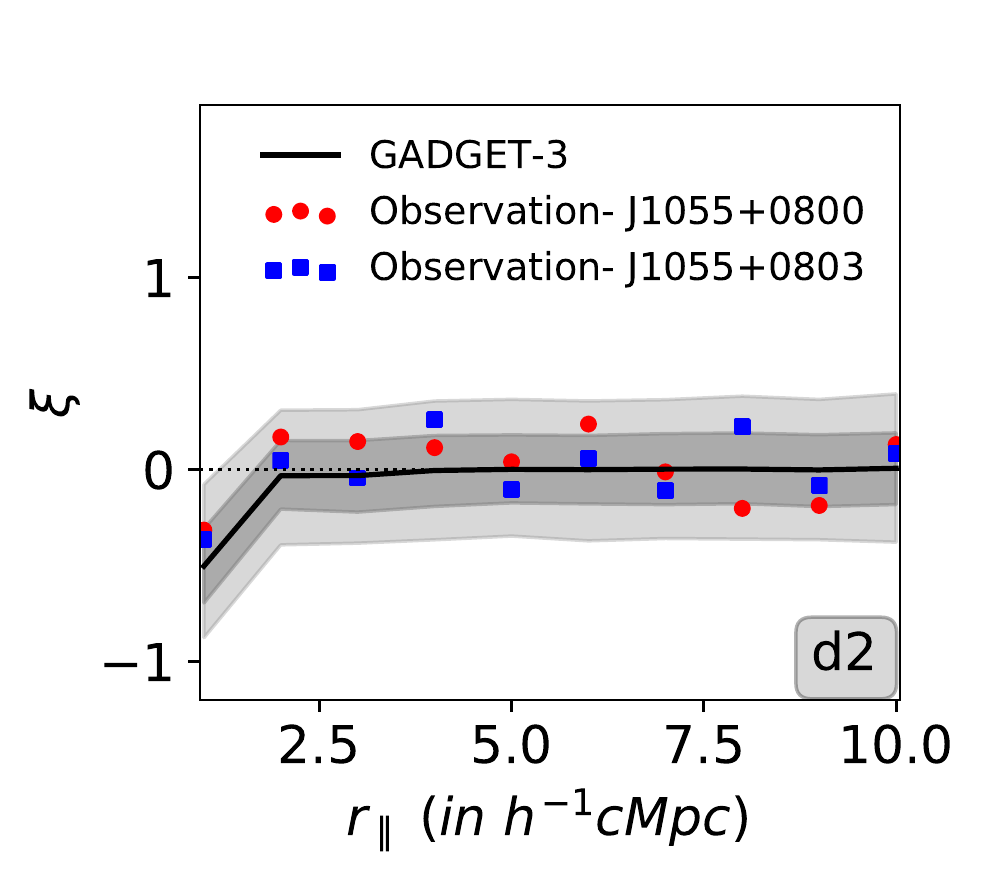}
		
		\end{minipage}

    	\begin{minipage}{0.24\textwidth}
    	    \includegraphics[width=4.5cm]{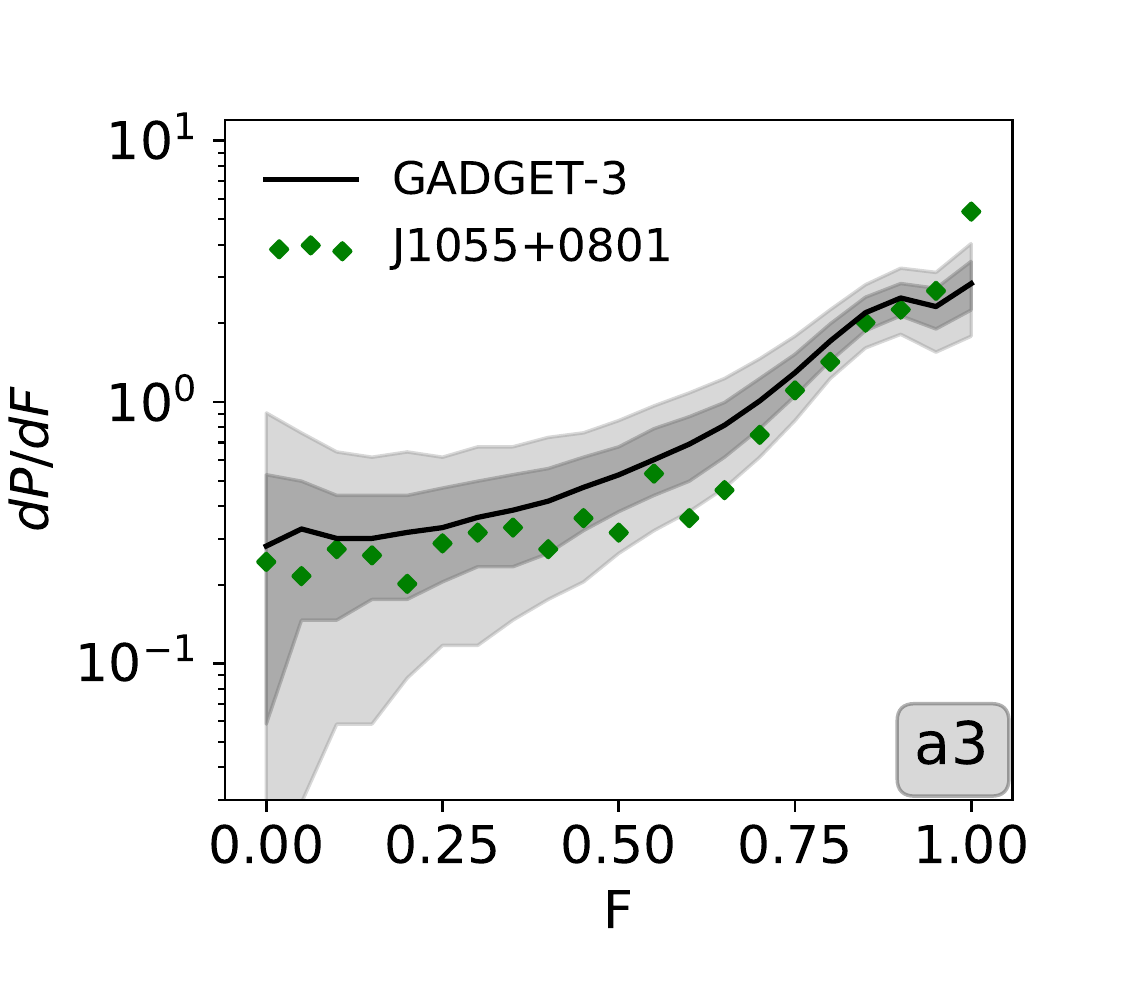}
    	
    	\end{minipage}%
    	\begin{minipage}{0.24\textwidth}
    		\includegraphics[width=4.5cm]{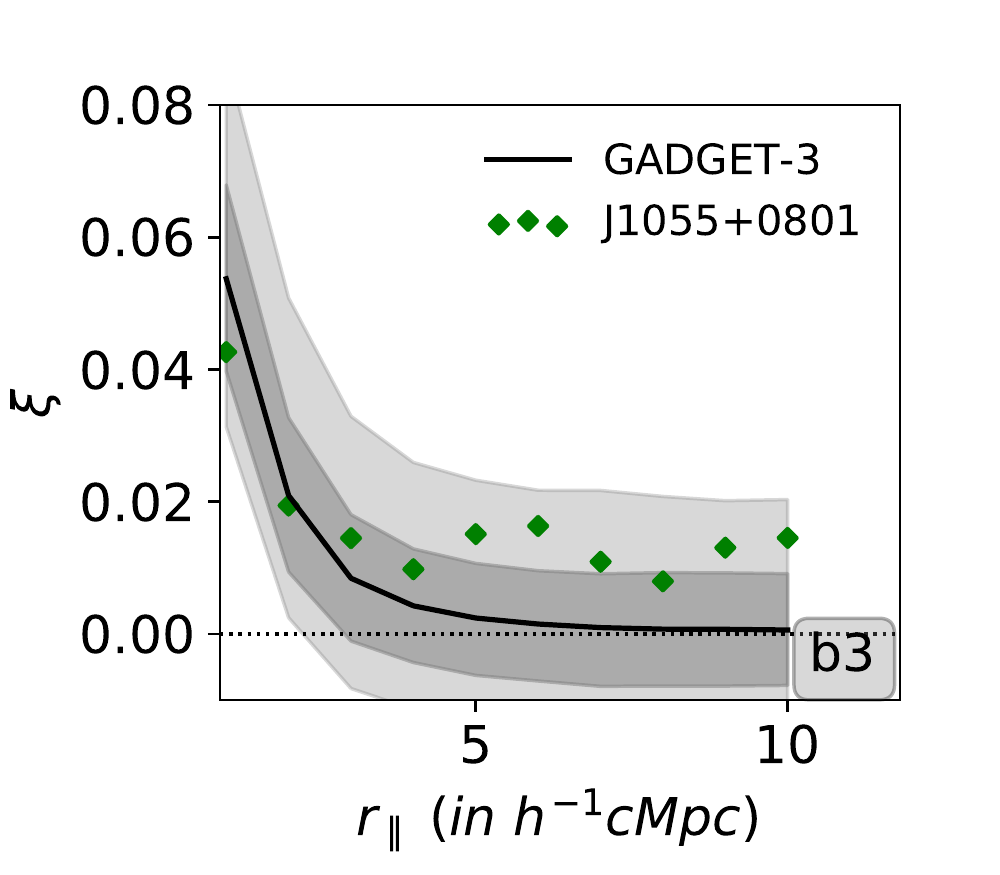}
    	
    	\end{minipage}
    	\begin{minipage}{0.24\textwidth}
    		\includegraphics[width=4.5cm]{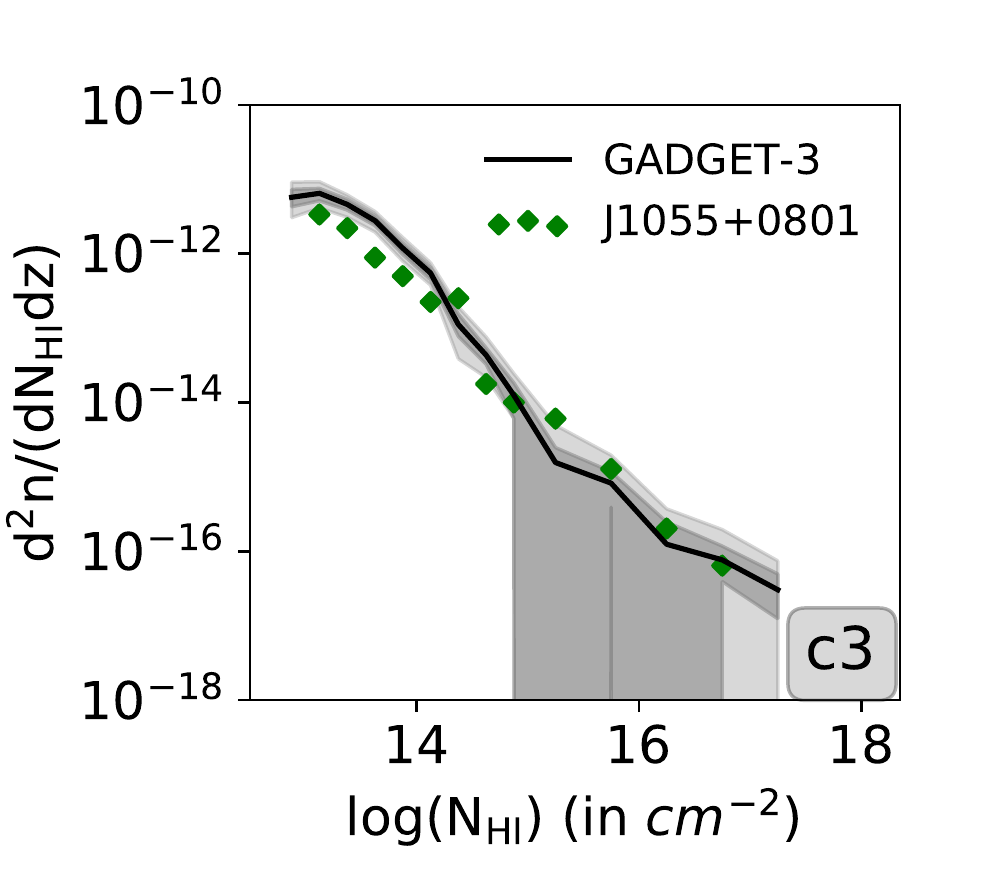}
    	\end{minipage}%
    	\begin{minipage}{0.24\textwidth}
    		\includegraphics[width=4.5cm]{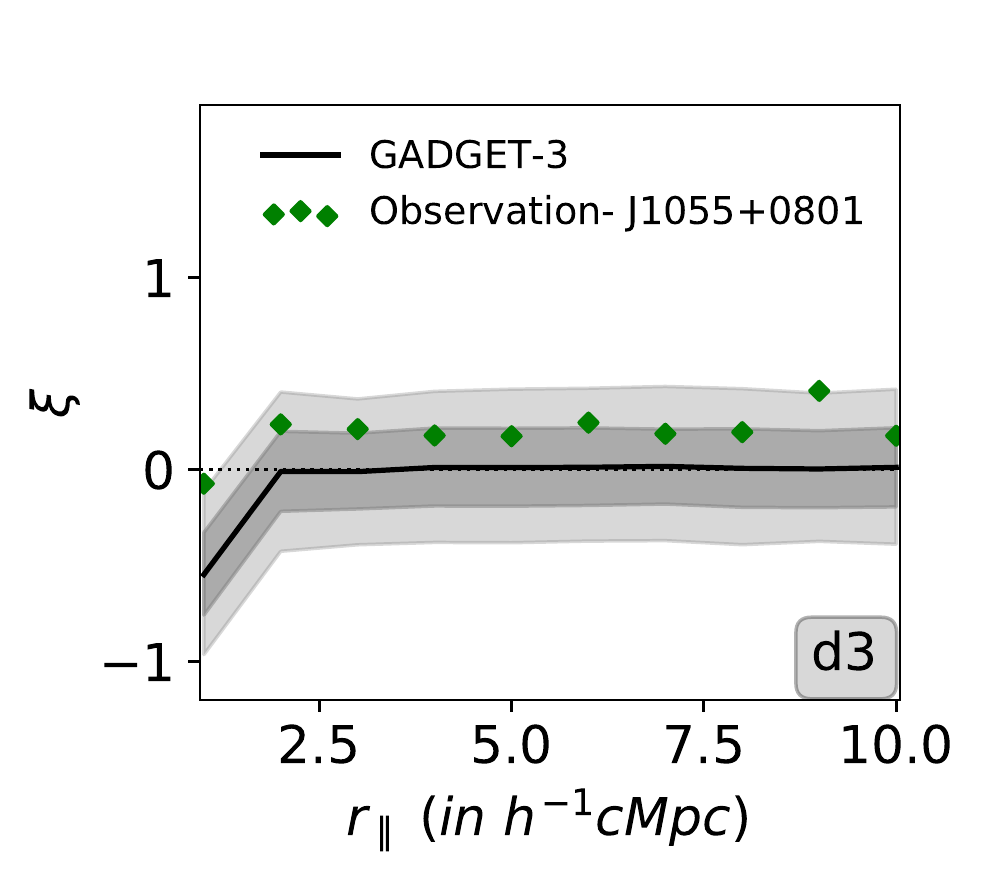}
    		
    	\end{minipage}
    	
    	\begin{minipage}{0.24\textwidth}
    		\includegraphics[width=4.5cm]{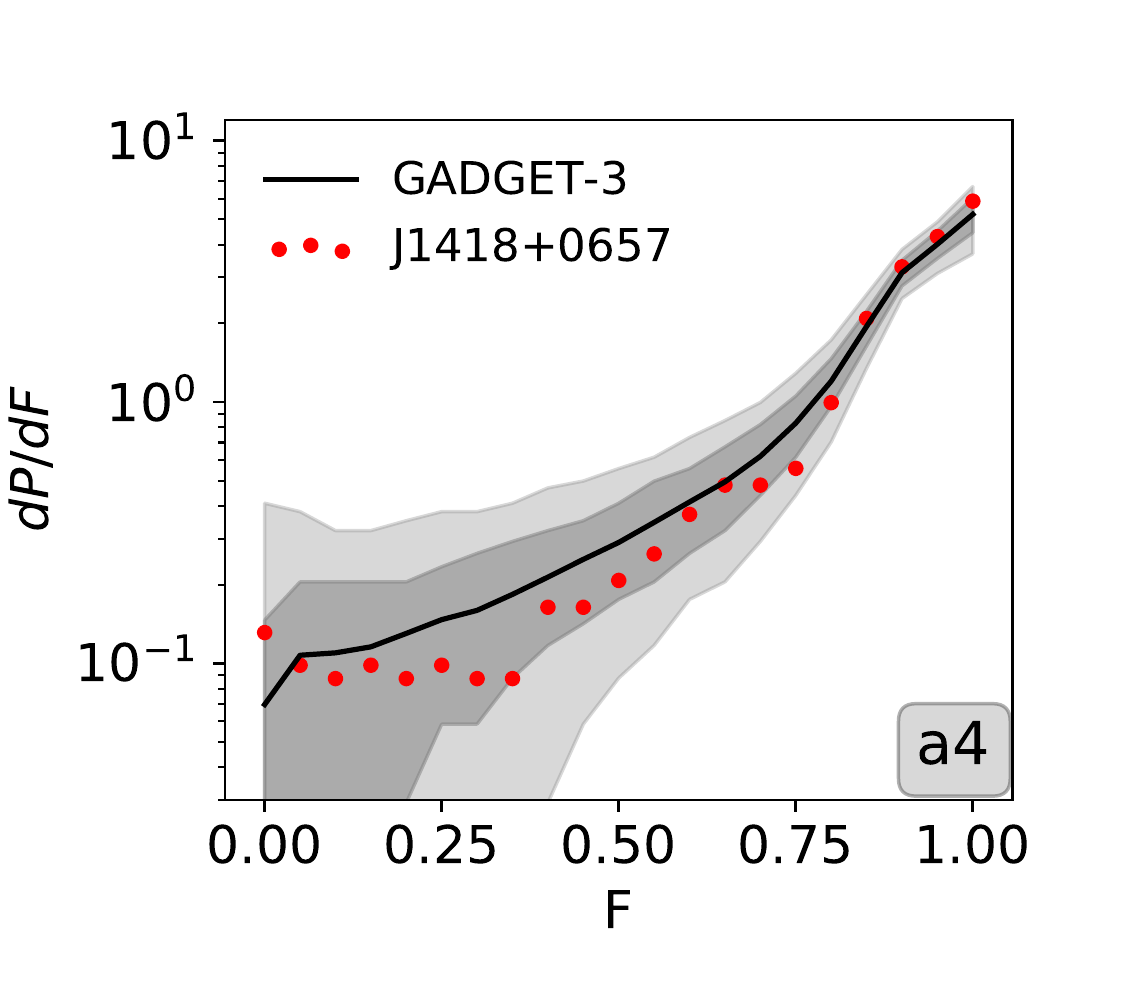}
    	\end{minipage}%
    	\begin{minipage}{0.24\textwidth}
    		\includegraphics[width=4.5cm]{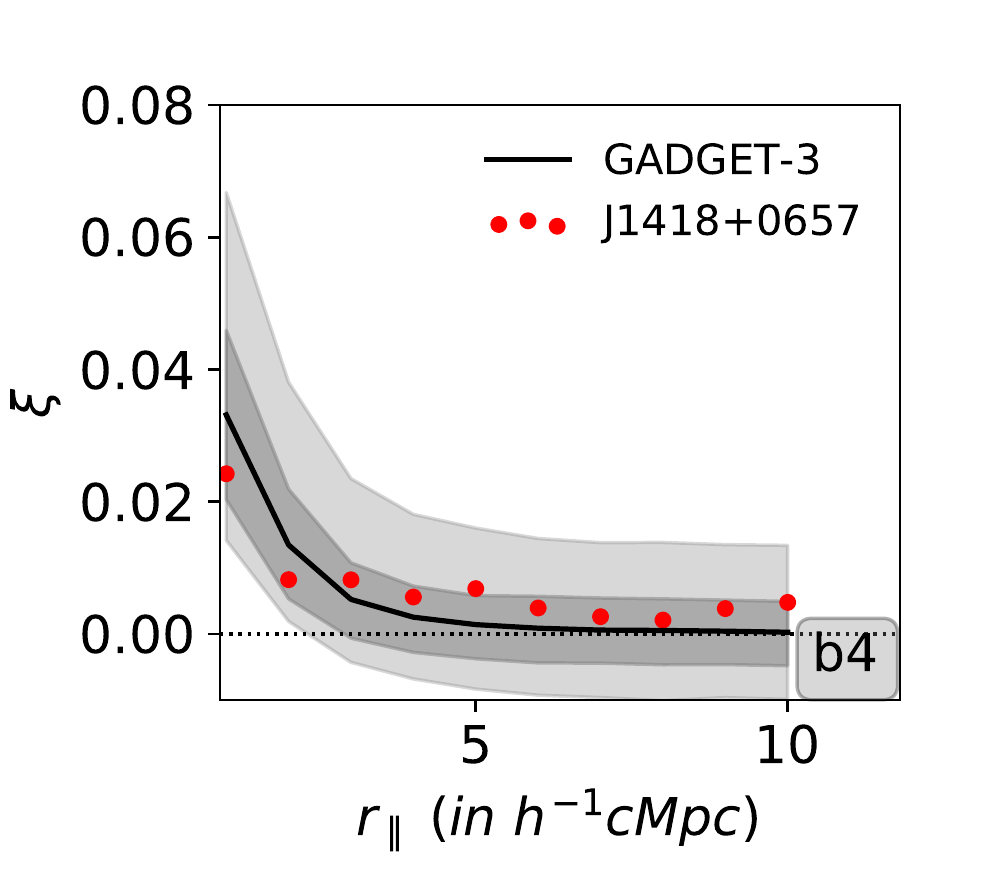}
    	\end{minipage}
    	\begin{minipage}{0.24\textwidth}
    		\includegraphics[width=4.5cm]{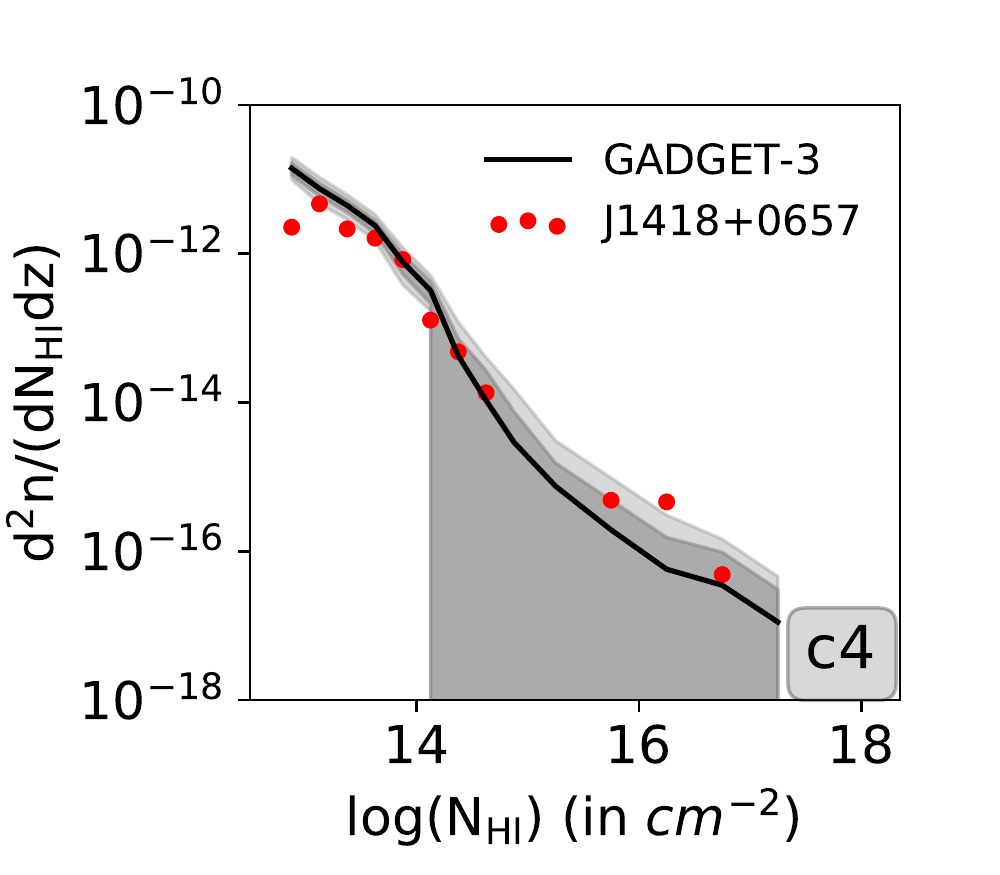}
    	\end{minipage}%
    	\begin{minipage}{0.24\textwidth}
    		\includegraphics[width=4.5cm]{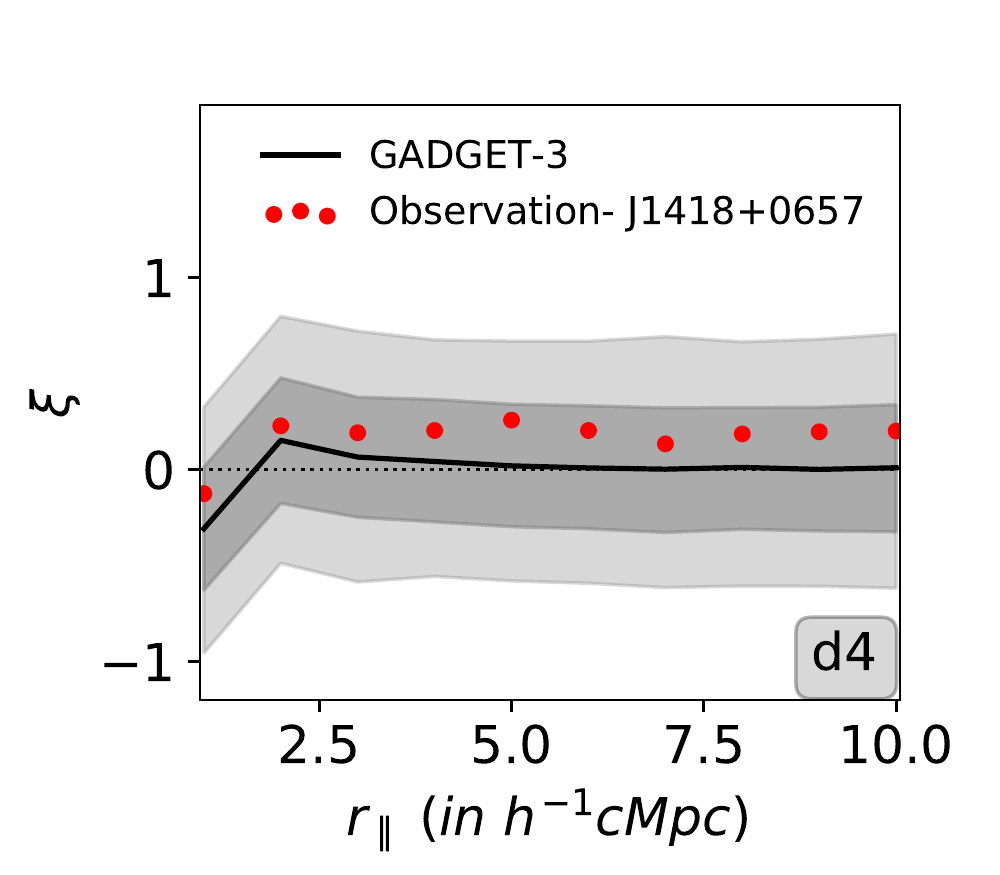}
    	\end{minipage}
    	
    	\begin{minipage}{0.24\textwidth}
    		\includegraphics[width=4.5cm]{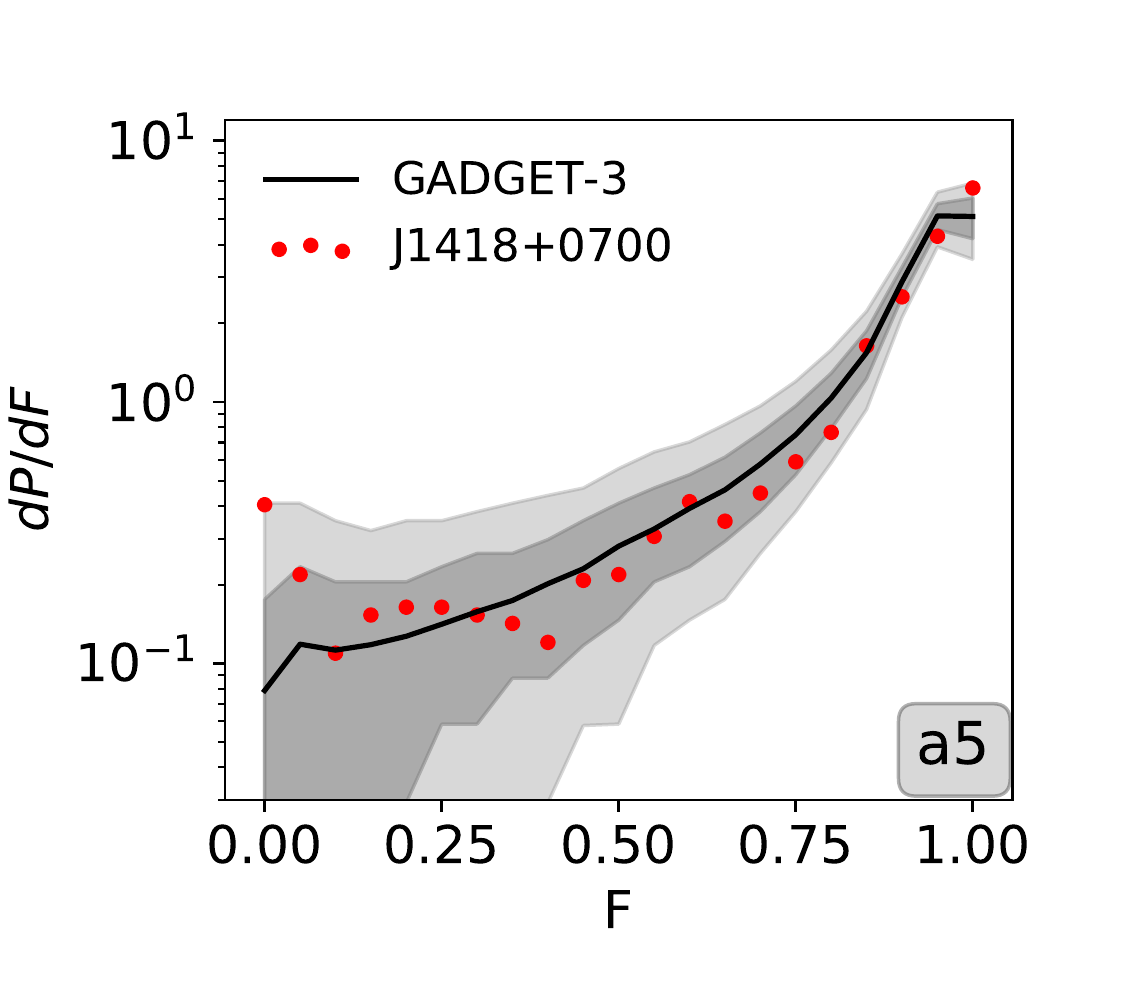}
    	\end{minipage}%
    	\begin{minipage}{0.24\textwidth}
    		\includegraphics[width=4.5cm]{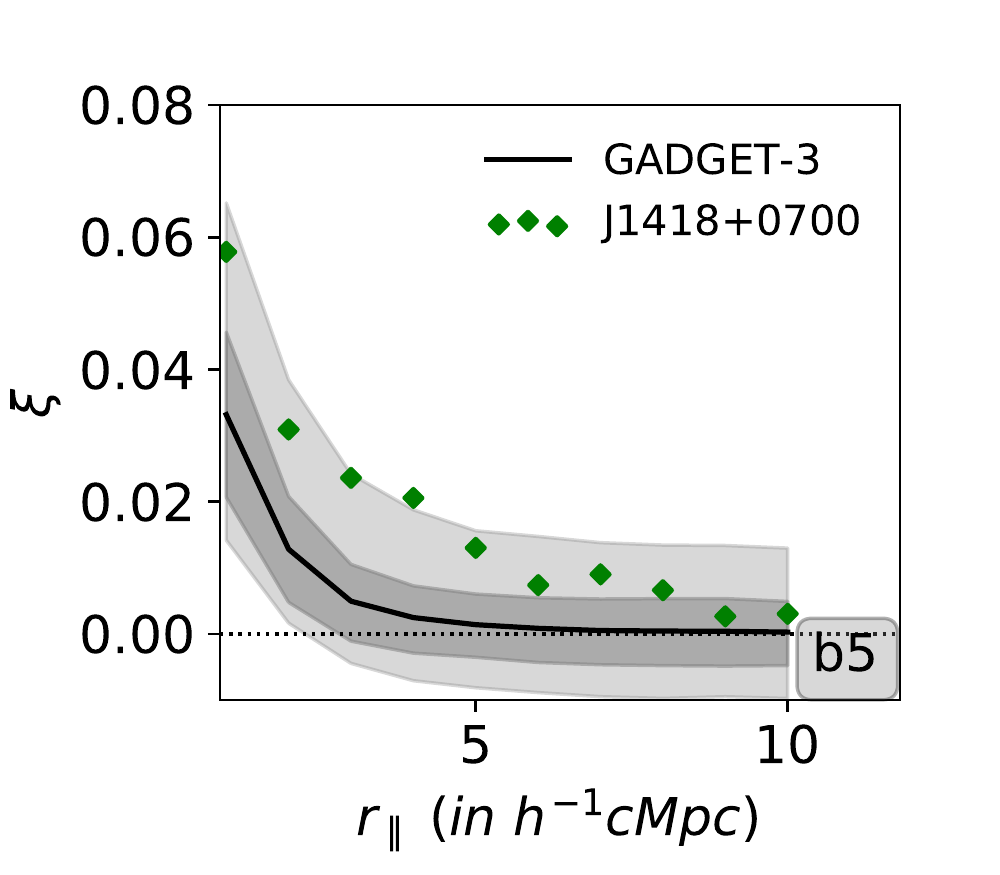}
    	\end{minipage}
    	\begin{minipage}{0.24\textwidth}
    		\includegraphics[width=4.5cm]{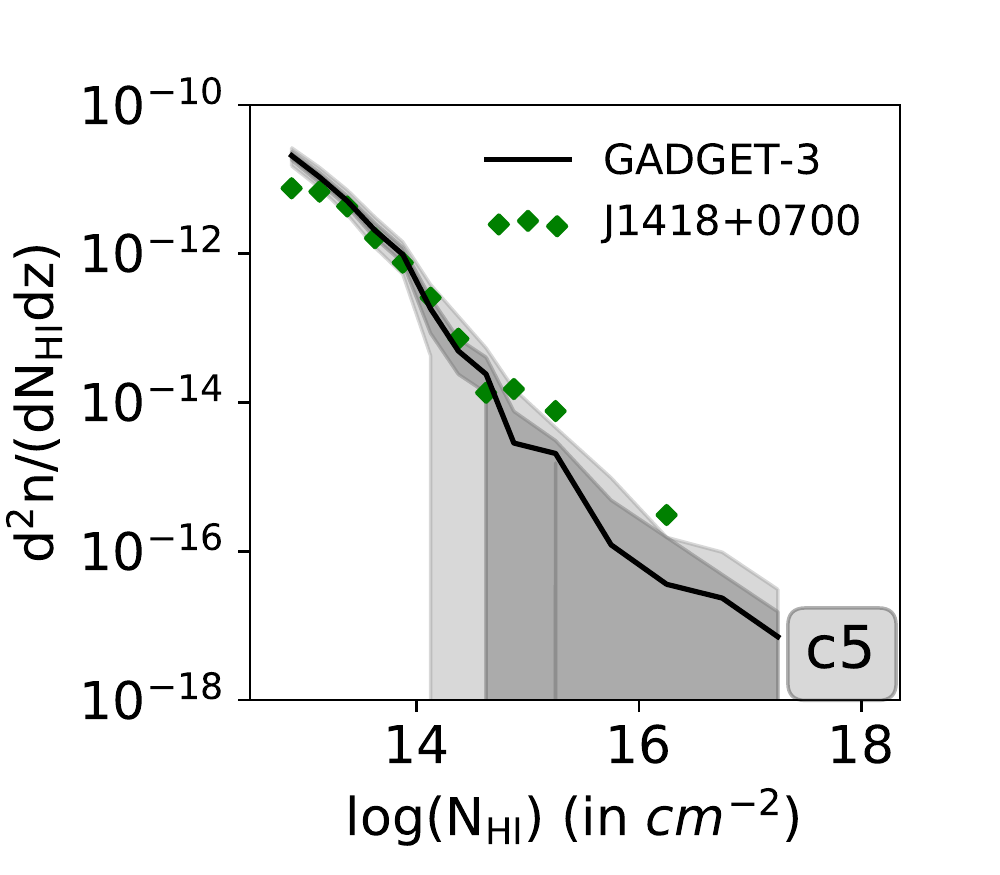}
    	\end{minipage}%
    	\begin{minipage}{0.24\textwidth}
    		\includegraphics[width=4.5cm]{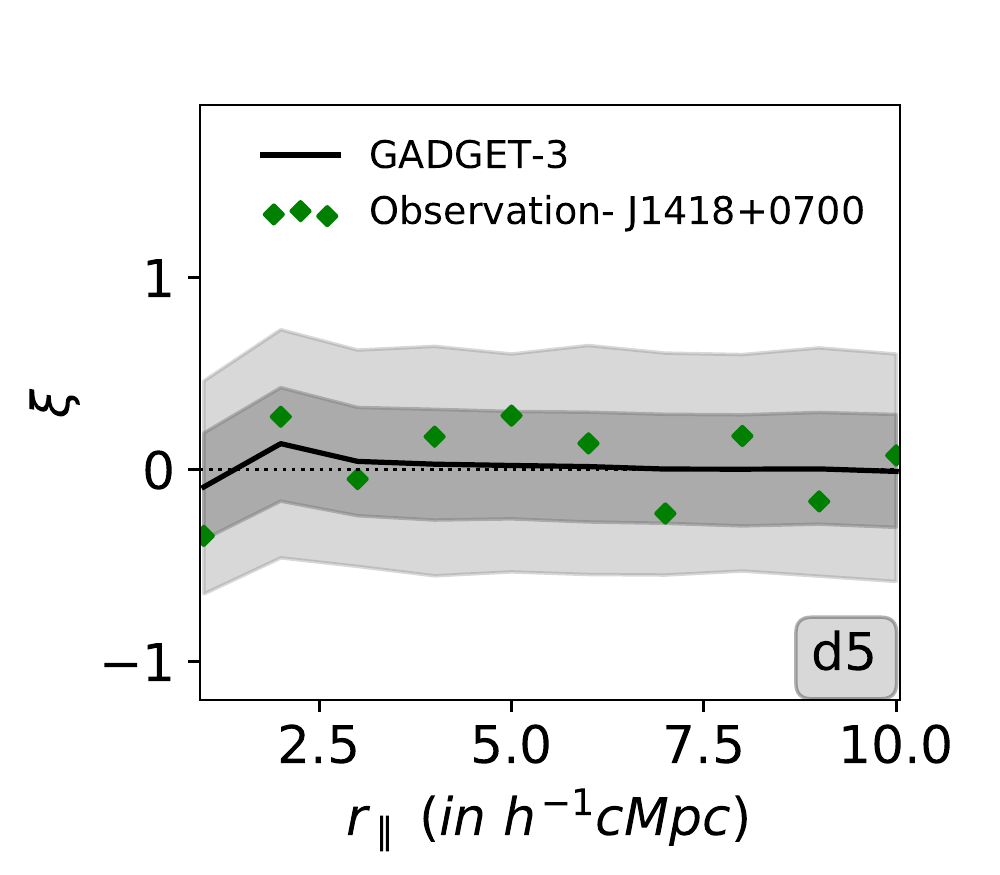}
    	\end{minipage}
    
		\caption{Statistics involving single \lya\ sight-line for the quasar triplets. The deep and light grey regions denote the $1\sigma$ and $2\sigma$ confidence interval from the simulations. Column-1: \lya\ transmitted flux probability distribution functions.
		Column-2: \lya\ transmitted flux longitudinal correlation function . 
		Column-3: Distribution of $N_{\rm HI}$ obtained from the Voigt profile decomposition. 
		Column-4: Longitudinal correlation function  for the Voigt profile decomposition of the \lya\ forest.
			}
\label{summary_fig}
	\end{figure*}

  First we compare the probability distribution function of the \lya\ transmitted flux (flux PDF) obtained in our simulations with the observed ones.
	The flux PDF is calculated for 20 bins in $F$ ranging from 0 to 1. As is usually done, values with $F<0$  goes in the first bin while values having $F>1$  goes in the last bin of the distribution. The first column in Fig.~\ref{summary_fig} shows the observed flux PDF.  In the top panel we show the results for three sightlines in "Triplet 1". As the spectral SNR is very similar we consider only one set of simulated results for comparison. In the case of "Triplet 2" SNR achieved along two sightlines are roughly a factor 2 higher than that along J1055+0801. Therefore we show comparison with the simulations in different panels (second and third from the top). In the case of the "Doublet", results for the two quasar sightlines are summarized in last two rows. 
	It is clear that by and large the observed distributions are consistent with the simulated one within 1$\sigma$
	confidence level.

	\subsection{Column density distribution function}
	
	\begin{figure}
        \centering
        \includegraphics[width=8cm]{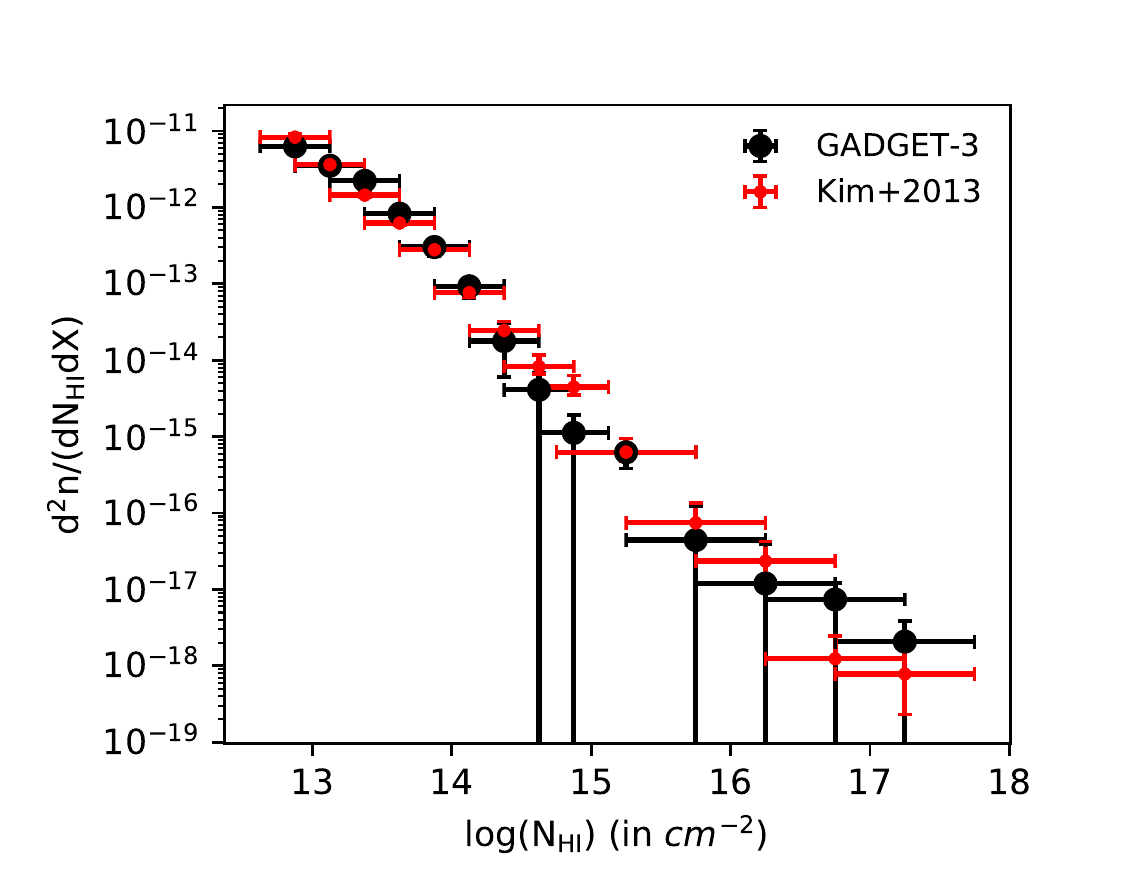}
        \caption{ Comparison of observed \citep[red dots from][]{kim2013} and simulated \HI\ CDDF
    (black dots).    }
        \label{Kim_NHI}
    \end{figure}
    
	 Next we compare the \HI\ column density distribution function (CDDF). The CDDF $f(N_{\rm HI},X)$, is defined as the number of \HI\ absorbers within absorption distance interval $X$ and $X+dX$ and column density interval $N_{\rm HI}$ and $N_{\rm HI}+dN_{\rm HI}$. 
	 The absorption distance is defined as ,
    \begin{equation}
        X(z)=\int \frac{H_0}{H(z)}(1+z)^2  \ ,
    \end{equation}
    by \citet{bahcall1969}.
 As a validation of our simulation, we compare the CDDF of our simulated spectra with that observed by \citet{kim2013} in Fig~\ref{Kim_NHI}. 
    The error plotted for \citet{kim2013} corresponds to 1$\sigma$ range. 
    For simulations, we show the 1$\sigma$ confidence interval due to multiple realizations. Our simulations match well within 1$\sigma$ of the CDDF from \cite{kim2013}.

    We compute the CDDF along each of our sightlines and compare them with the observations.
    These are summarized in panels shown in column 3 of Fig.~\ref{summary_fig}. Apart from CDDF measured along the line of sight to J1055+0801 (where we see less number of absorbers compared to the model predictions) the observed distribution along other sightlines are broadly reproduced. Given the small redshift path involved in each sightline such deviations are not unexpected.
    For the median SNR achieved in our spectra the typical
    limiting column density for an unresolved absorption component is log($N_{\rm HI}$) = 12.90 for quasars in "Triplet 1". In the case of "Triplet 2" the limiting column density is log($N_{\rm HI}$) = 13.10 for J1055+0801 and  log($N_{\rm HI}$) = 12.77 for the other two sightlines.  In the case of "Doublet", the limits are log($N_{\rm HI}$) = 12.95 and log($N_{\rm HI}$) = 12.74 repectively for J1418+0657 and J1418+0700. Therefore, for the full data set, we have a column density completeness close to $10^{13}$ cm$^{-2}$.
 
\section{Spatial correlations of \lya\ absorption:}	

The correlation properties of \lya\ forest absorption is usually studied using the statistics of transmitted flux \citep[see][for earlier work]{croft2002,viel2002,mcdonald2003,rollinde2003}. In this work 
we use statistics based on transmitted flux as well as individual "clouds" identified using  Voigt profile fitting.  
As we are dealing with hand full of sightlines our aim is to mainly quantify how probable are the observed properties in the framework of the simulations considered and not to  match the observations with simulated data by varying model parameters.
As we discussed before, unlike observations, our simulated spectra have finite wavelength coverage (limited by our box size). To make realistic comparisons, in what follows we divide the observed spectrum in to random chunks having wavelength intervals similar to that of our simulated spectra (1000 random chunks of the observation). We then compute the mean and probability distribution of the statistical quantity of our interest using these random chunks. These values will then be compared with the same obtained from the simulations (typically using 4000 random realizations).  
	
	\subsection{Transmitted flux based statistics}

			In this section, we investigate two-point correlation statistics $\xi(\textbf{r}_{\parallel},\textbf{r}_{\perp})$ of \lya\ transmitted flux as a probe of the spatial clustering of the IGM.
	
	\subsubsection{Longitudinal two-point Correlation}
	
We define the longitudinal two-point correlation ($\xi_{\parallel}(\Delta\textbf{r}_{\parallel})$) in transmitted flux
decrement, $D=F-\langle F\rangle$ as,
		\begin{equation}
		\xi_{\parallel}(\Delta\textbf{r}_{\parallel})=\langle  D(\textbf{r}^{\prime}_{\parallel}) D(\textbf{r}^{\prime}_{\parallel}+\Delta \textbf{r}_{\parallel}) \rangle \ .
		\end{equation}
		 Here $\langle F\rangle$ is the mean transmitted flux of the \lya\ absorption. Longitudinal correlation measures the flux correlation along the line of sight in the redshift space.

		 The two-point longitudinal correlation measured as a function of redshift separation are shown in the second column of Fig.~\ref{summary_fig}. 
		In the top panel we summarize the results for three sightlines in "Triplet 1". The dots are the average $\xi_\parallel$ measured at different values of $r_\parallel$ in observation. The solid curve is the mean value we find from our simulations. Deep and light grey shaded regions give the 1 and 2$\sigma$ confidence interval measured around the median values from our simulations. It is clear that the observed distribution is well within 1$\sigma$ range of our observations. As the spectral SNR achieved towards all three quasars are similar we plot them in the same figure.

		In the case of "Triplet 2" we notice that the longitudinal correlation function is consistent within 1$\sigma$ range seen in our simulations.
	The measured two-point longitudinal correlation is also consistent within 2$\sigma$ range predicted by our simulations for J1418+0657. However for J1418+0700 we find the measured correlations deviate by about 2$\sigma$. 
		From the figure it is also apparent that the longitudinal correlation found for sightlines in "Triplet 1" (that probe similar redshift range) are also lower than what is seen towards J1418+0700.
		This could imply a strong clustered absorption along this sightline compared to that of  J1418+0657.

\begin{figure}
    \centering
    \includegraphics[viewport=15 20 270 230, width=8cm,clip=true]{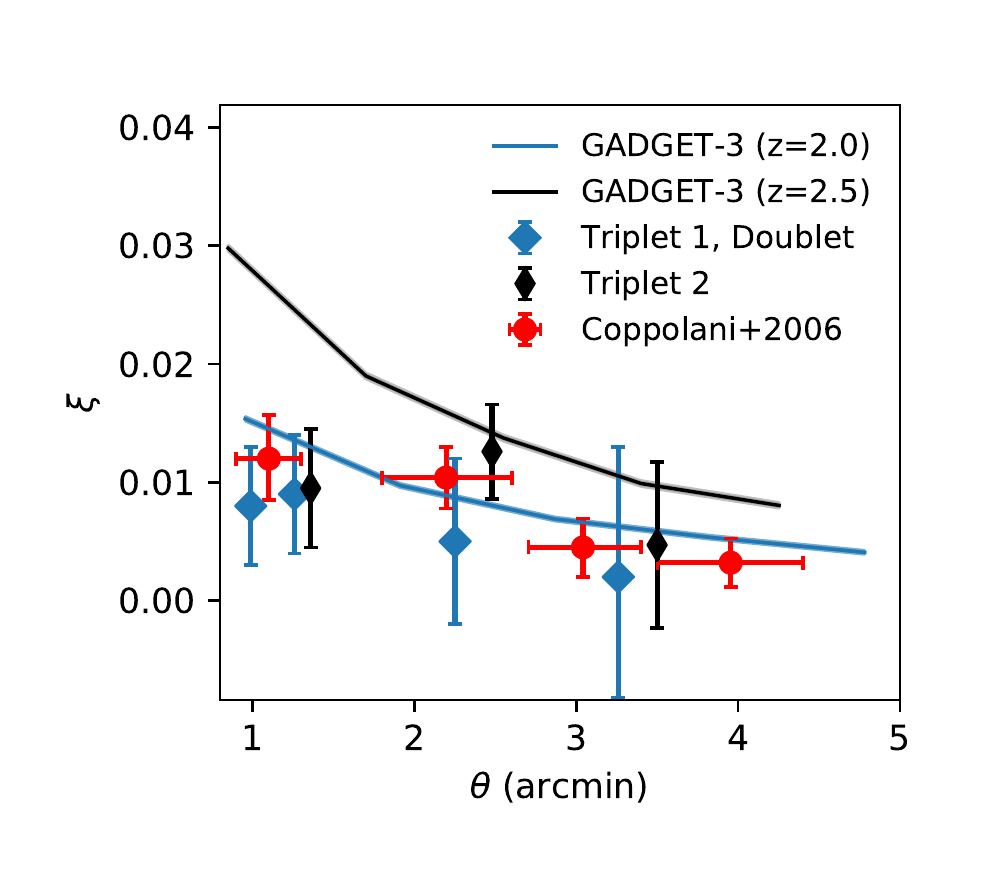}
    \caption{Transverse two-point correlation of transmitted flux as a function of angular separation. The red dots denote the mean measurements by \citep[][]{coppolani2006} computed from the values given in their Table~1. The error reflects the $1\sigma$ range. Measurements from the present sample are plotted with blue and black diamond symbols. The blue and black curves are from our simulated spectra at $z\sim2.0$ and $z\sim2.5$ respectively. }
    \label{fig_compare_t2p}
\end{figure}

    	\begin{figure*}
			\begin{minipage}{0.32\textwidth}
				\includegraphics[viewport=10 10 320 270,width=6cm, clip=true]{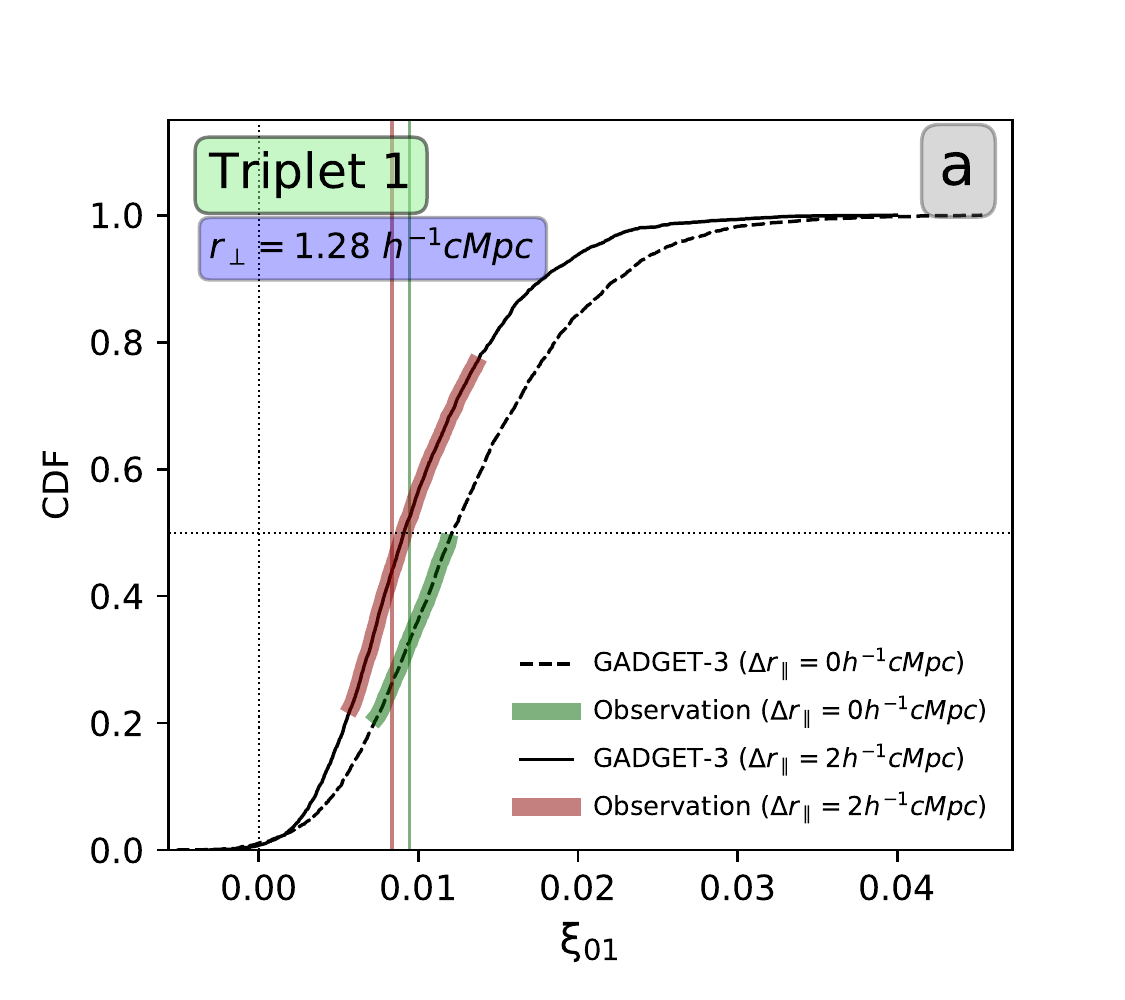}
			\end{minipage}%
			\begin{minipage}{0.32\textwidth}
				\includegraphics[viewport=10 10 320 270,width=6cm, clip=true]{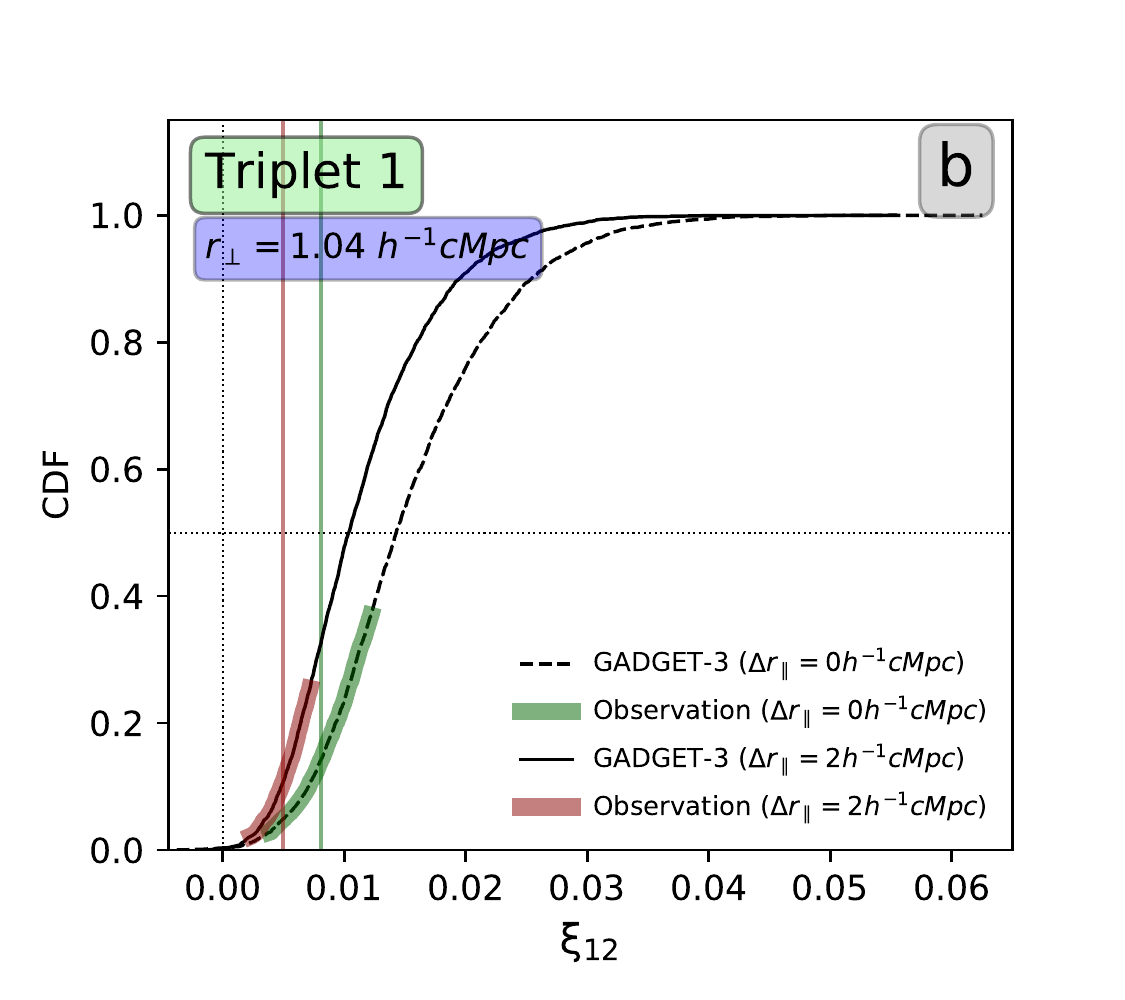}
			\end{minipage}
			\begin{minipage}{0.32\textwidth}
				\includegraphics[viewport=10 10 320 270,width=6cm, clip=true]{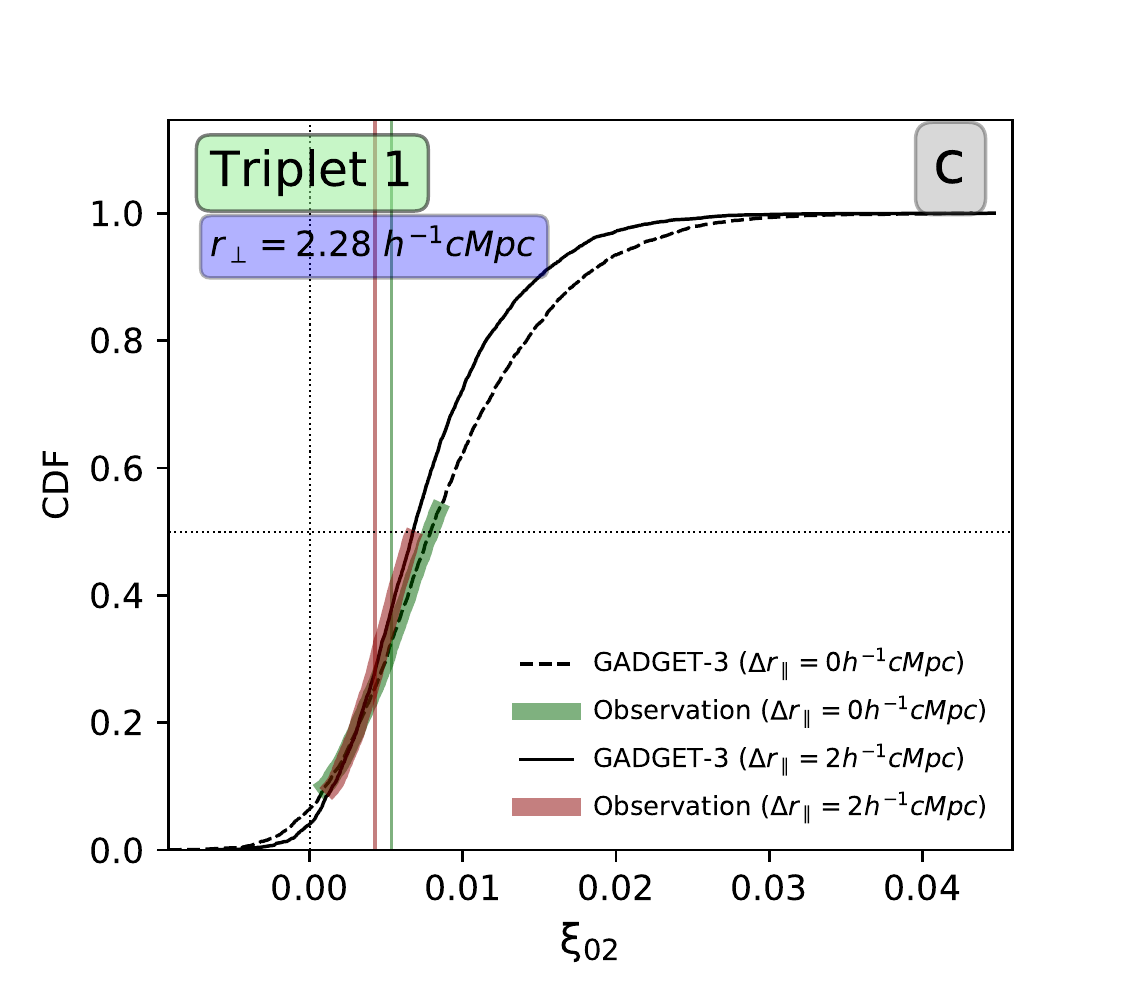}
			\end{minipage}%

			\begin{minipage}{0.32\textwidth}
				\includegraphics[viewport=10 10 320 270,width=6cm, clip=true]{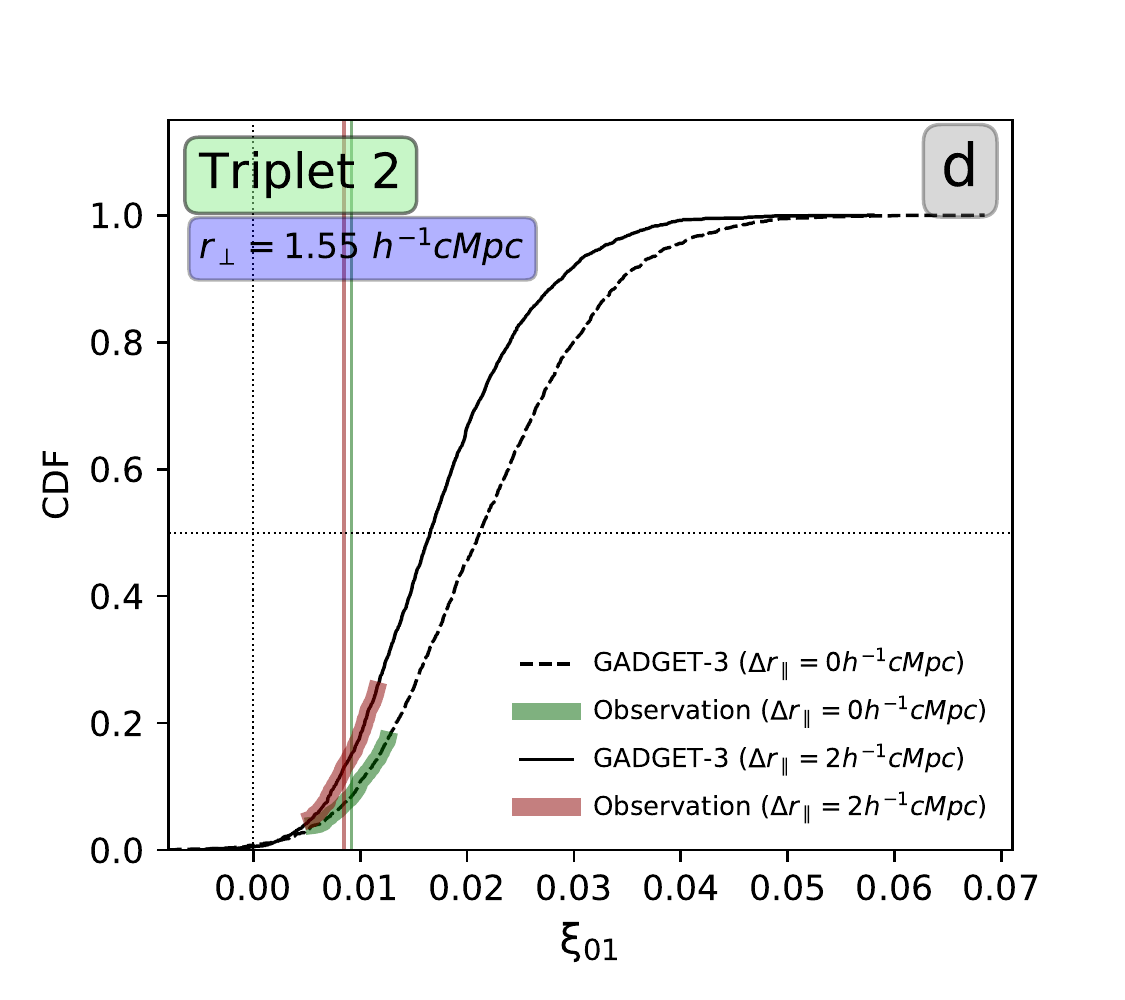}
			\end{minipage}%
			\begin{minipage}{0.32\textwidth}
				\includegraphics[viewport=10 10 320 270,width=6cm, clip=true]{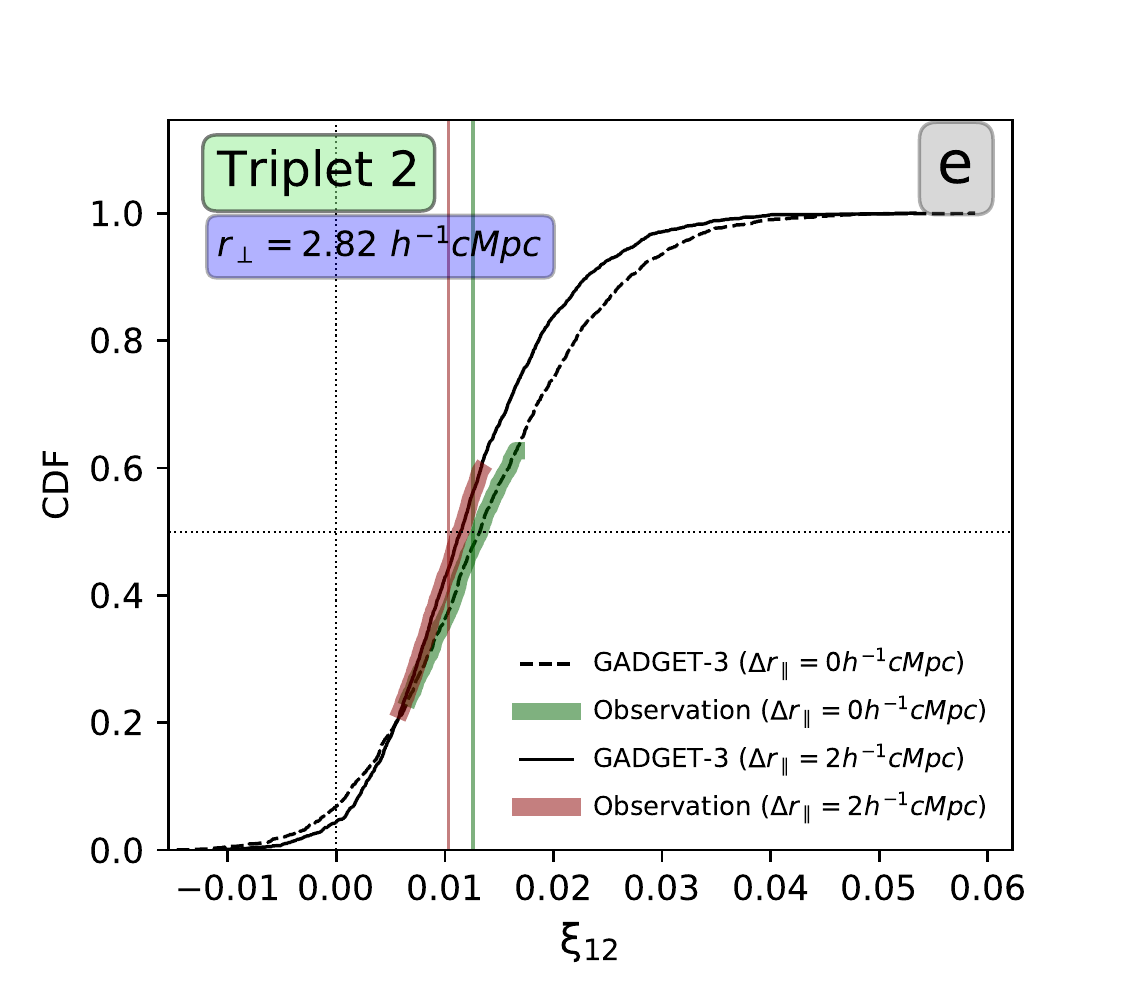}
			\end{minipage}
			\begin{minipage}{0.32\textwidth}
				\includegraphics[viewport=10 10 320 270,width=6cm, clip=true]{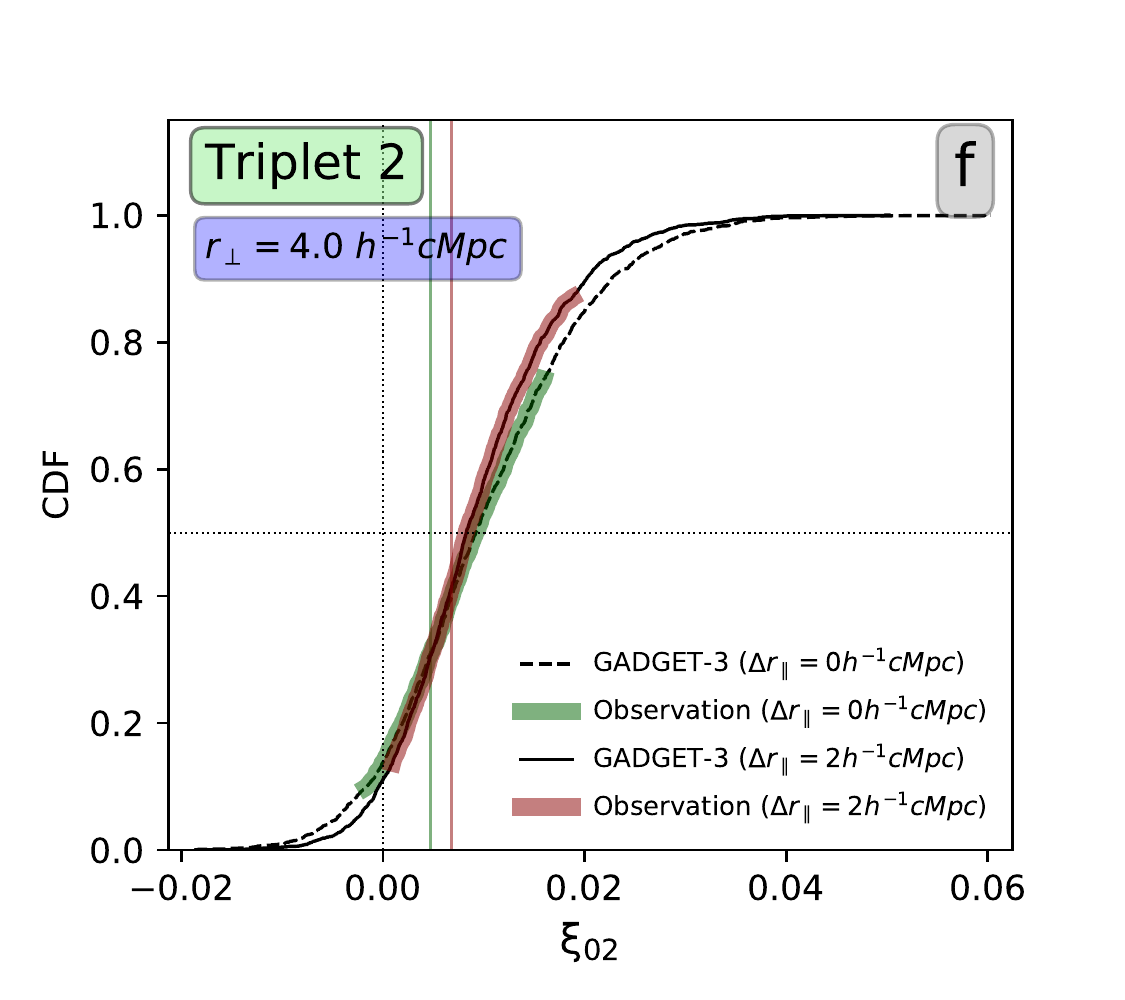}
			\end{minipage}%
			
			\begin{minipage}{0.32\textwidth}
				\includegraphics[viewport=10 10 320 270,width=6cm, clip=true]{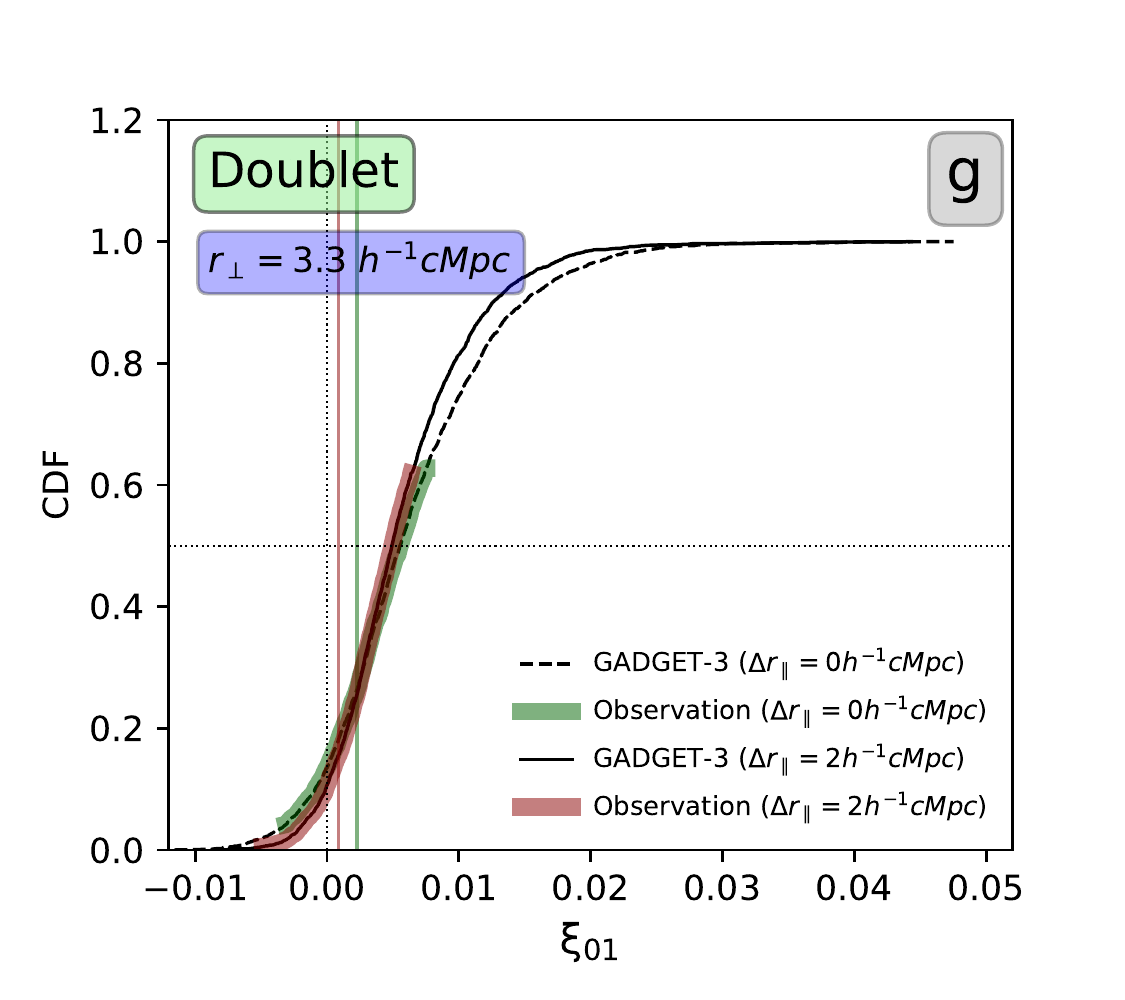}
			\end{minipage}%
			\begin{minipage}{0.32\textwidth}
				\includegraphics[viewport=10 10 320 270,width=6cm, clip=true]{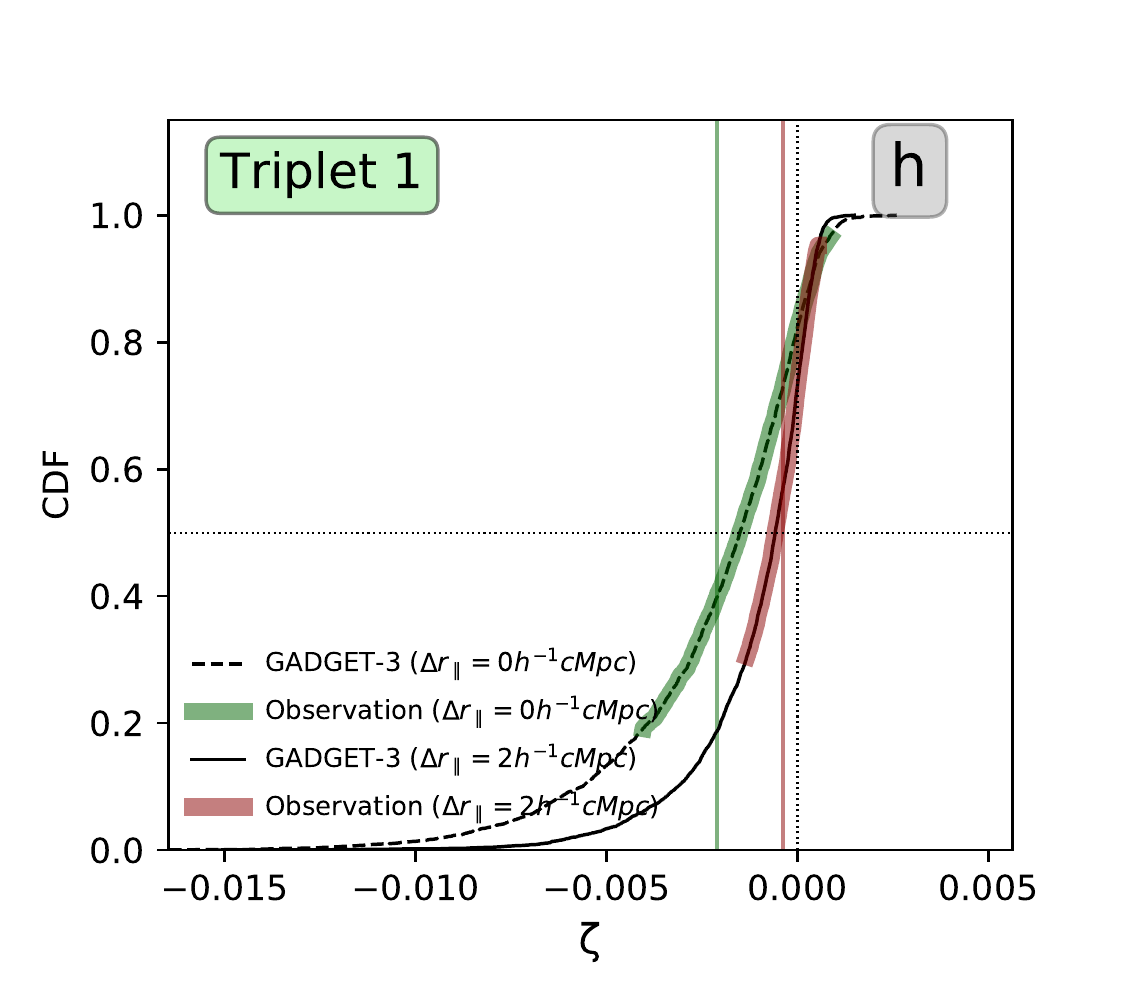}
			\end{minipage}%
			\begin{minipage}{0.32\textwidth}
				\includegraphics[viewport=10 10 320 270,width=6cm, clip=true]{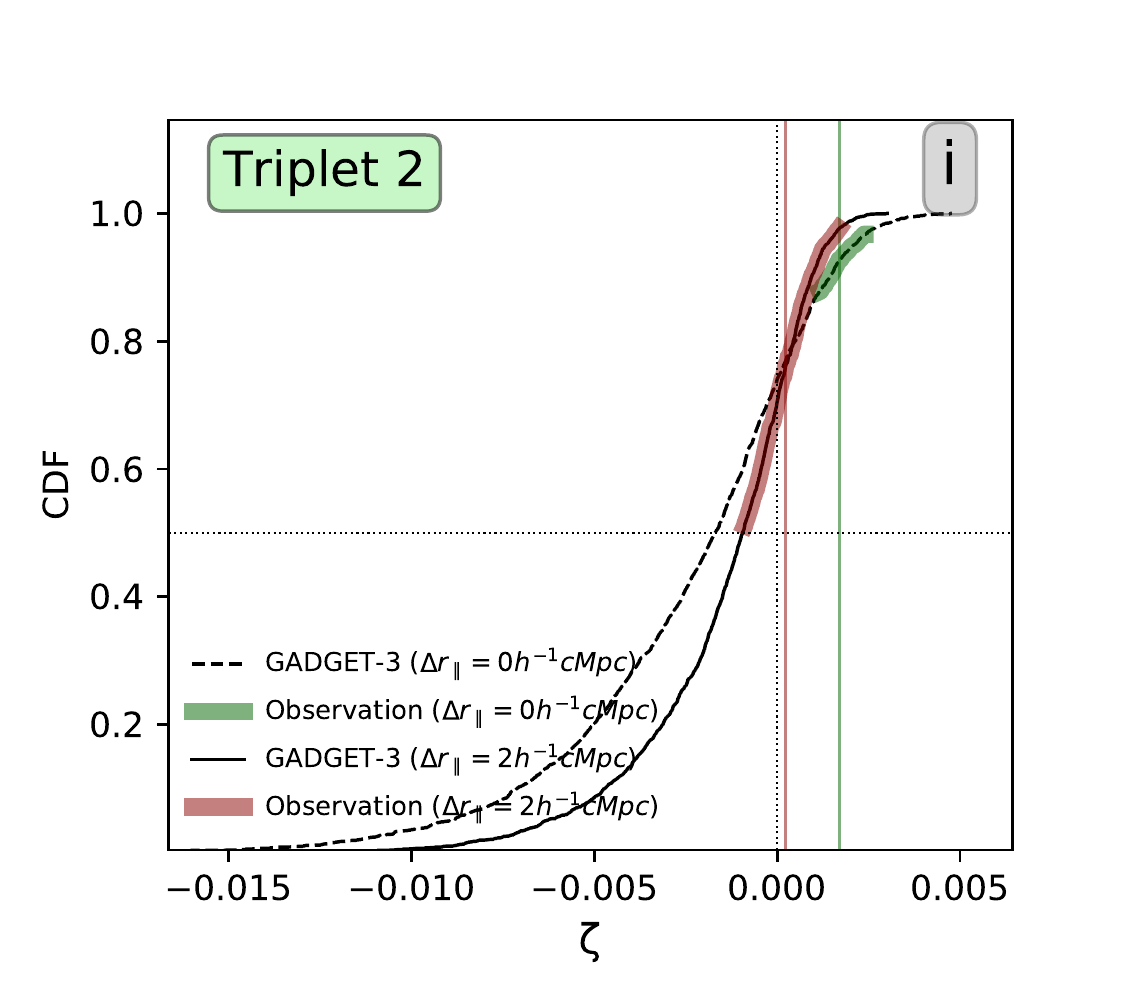}
			\end{minipage}%
			
			\caption{Cumulative distribution two- and three-point transverse correlation functions. The red and green vertical lines denote the mean of the observed correlation. The red and green regions overlaying the simulated curves represent the $1\sigma$ confidence interval of the observed triplet sightlines. 
			}
			\label{fig_f2p}
		\end{figure*}

\subsubsection{Transverse two-point Correlation}

Transverse correlation probes the clustering information between spatially separated two sightlines.
For a fixed pair separation $\Delta \textbf{r}_{\perp}$ between the sightlines, a two-point correlation can be defined as
		\begin{equation}
		\xi(\Delta \textbf{r}_{\parallel},\Delta \textbf{r}_{\perp})=\langle  D_1(\textbf{r}^{\prime}_{\parallel},\textbf{0}) D_2(\textbf{r}^{\prime}_{\parallel}+\Delta \textbf{r}_{\parallel},\Delta \textbf{r}_{\perp}) \rangle \ .
		\end{equation}
The transverse correlation function can be defined over the actual separation given by $\Delta r~=~\sqrt{\Delta r_{\parallel}^2+\Delta r_{\perp}^2}$. We will consider two-point correlation functions in the transverse direction for $\Delta r_{\parallel} = 0$ \citep[i.e correlating flux along two sightlines having same redshift as in,][]{coppolani2006}. In addition to this, we also calculate the transverse two-point-correlation by averaging over $\Delta r_{\parallel} = \pm 2 h^{-1}$ cMpc. This is done to average out the distortions due to peculiar velocity effects along the sightline. Subsequently, we will refer to this as $\Delta r_{\parallel} = 2 h^{-1}$ cMpc case.  Note in the flux based two-point correlation functions both the correlated over-dense and under-dense regions will have positive correlation amplitudes. However, as the mean transmission is close to the continuum, the mean of the flux two-point correlation will be more influenced by the over-dense (i.e strong absorption) regions. Uncorrelated regions will have negative correlation amplitudes.

\begin{table}
    \centering
    \caption{Results of Flux-based correlation analysis for $\Delta r_{\parallel}=2h^{-1}$ cMpc}
    \setlength\tabcolsep{2.5pt}
    \begin{tabular}{cccccc}
\hline
Sample & Correlation & $r_\perp$ & Obsereved & Probability & Percentile \\
       &             &(cMpc)   & values & &\\
\hline
\hline
Triplet 1 & $\xi_{01}$ & 1.28 & $+$0.008  & 0.54 & 44\\
          & $\xi_{12}$ & 1.04 & $+$0.005  & 0.23 & 11\\
          & $\xi_{02}$ & 2.28 & $+$0.004  & 0.39 & 28\\
          & $\zeta$    &      & $-$0.0004  & 0.65 & 57\\
%          \\
Triplet 2 & $\xi_{01}$ & 1.55 & $+$0.008  & 0.20 & 13\\
          & $\xi_{12}$ & 2.82 & $+$0.010  & 0.38 & 44\\
          & $\xi_{02}$ & 4.00 & $+$0.007  & 0.73 & 41\\
          & $\zeta$    &      & $+$0.0002  & 0.47 & 76\\
%          \\
Doublet & $\xi_{01}$ & 3.30 &  $+$0.001 & 0.62 & 16\\
\hline
    \end{tabular}
    \label{table_2p}
\end{table}
 
We compare the predictions from our simulation with the observed two-point transverse correlation ($\Delta r_{\parallel}=0$) as a function of angular separations by \citet[][]{coppolani2006}. Note the observations used in \citet{coppolani2006} were obtained with four times lower resolution (i.e FWHM$\sim$220 \kms) than the spectra used in our study and our simulated spectra (see column 5 in Table~\ref{Tab_obs}). It is clear from Fig.~\ref{fig_compare_t2p} that our simulated data for $z = 2$, reproduces the observed trend very well. This is expected as most of the data points with angular separations less than 4 arcmin in \citet{coppolani2006} sample the \lya\ forest in the range $1.9\le z\le2.3$.

Next we consider individual measurements of $\xi$ using all possible pairs of sightlines from our triplets. These points are also shown in Fig.~\ref{fig_compare_t2p}.  As explained before, we divided the observed \lya\ forest wavelength range into several segments (of size equal to that of the simulated spectrum) and estimated the mean $\xi$ and its $1\sigma$ confidence interval for each doublets. 
It is clear from Fig.~\ref{fig_compare_t2p} that the observed $\xi$ values for all the doublets corresponding to "Triplet 1" and "Doublet" are consistent with the simulated results for $z=2$. However, in the case of "Triplet 2" (where we use simulation data for $z\sim2.5$) while the observed values $\xi_{12}$ and $\xi_{02}$ are consistent with the models, $\xi_{01}$ measured between the sight lines along J1055+0800 and J1055+0801 is lower than the model predictions.

To quantify this further, in Fig.~\ref{fig_f2p} we compare our measurements (mean and $1\sigma$ confidence interval with thick shaded curves) with the cumulative probability distribution we get from our simulated sightlines (dashed and dotted curves).
The solid curves in Fig.~\ref{fig_f2p} are computed for $\Delta r_{\parallel}= 0 $ \citep[as done by][and for our data in Fig~\ref{fig_compare_t2p}]{coppolani2006}. The dotted curves in Fig.~\ref{fig_f2p} are obtained by integrating over $\Delta r_{\parallel}$ between $\pm 2 h^{-1}$cMpc.  Vertical dotted lines in each panel shows $\xi=0$ and the other two vertical lines provide the observed mean $\xi$ values for the two cases considered.
It is clear from the simulated curves that when we integrate $\xi$ along $r_\parallel$ we notice a decrease in the predicted value of the median and the scatter around the median. The difference between the two cumulative distributions are larger for the smaller pair separations. This trend is also roughly evident in the observed distributions as well.

In Table~\ref{table_2p} we quantify the comparison between our observations and simulations for the case where $\xi$ is obtained by integrating over $r_\parallel$. Fourth column of this table provides the mean values of $\xi$ and $\zeta$ ($\zeta$ is the transverse three-point correlation defined in section 5.1.3). Fifth column gives the probability that the observed mean and $1\sigma$ confidence interval is realised in our simulations. We obtain this by computing the fractional area of the shaded region in the cumulative distributions shown in Fig.~\ref{fig_f2p}. The last column in Table~\ref{table_2p} gives the percentile of the observed mean $\xi$ based on the cumulative distribution obtained from the simulations. 

In the top row we plot the transverse correlation function for the three possible doublets belonging to the "Triplet 1". 
It is clear that in the case of "Triplet 1" when we consider $\xi$ measurement with $r_\parallel =0$, the observed mean and $1\sigma$ confidence range tend to be in the lower end of the simulated distribution (i.e typically below the median). This  is also the case when we 
consider integration in $r_\parallel$ for two of the doublets. However, as can be seen from Table~\ref{table_2p}, the probability of realising the observed distributions of $\xi$ in our simulation box is not 
very low. 

Similarly in the case of "Triplet 2", the measured $\xi$ is lower than
our model prediction for the pair of sight lines towards J1055+0800 and J1055+0801 (i.e $\xi_{01}$ in Fig.~\ref{fig_f2p}). This doublet also 
shows the lowest probability in Table~\ref{table_2p}. For the other 
two pairs in this triplet the observed $\xi$ is consistent with the predictions within 1$\sigma$ level.  For the only pair for which we have X-shooter data in "Doublet", the measured $\xi$ is consistent with the model predictions. 

In summary, the transverse two-point correlation functions based on transmitted flux measured for all the doublets we could construct in our sample are not abnormal in the framework of our simulations. Next we look at the three-point correlation function based on the transmitted flux.

\subsubsection{Three-point Correlation in flux}

	If the matter density field is Gaussian then the two-point correlation would be sufficient to describe the spatial distribution of matter. However, due to non-linear nature of evolution of gravitational instabilities and structure formation, the density field is expected to be non-gaussian, necessitating the usage of three-point correlation $\zeta(\Delta \textbf{r}_{1\parallel},\Delta\textbf{r}_{2\parallel},\Delta\textbf{r}_{1\perp},\Delta\textbf{r}_{2\perp},\theta)$. For a fixed point in one of the LOS (say LOS 0), $\Delta \textbf{r}_{i\parallel}$ denotes the longitudinal separation between this reference point and points of our interest in $i^{th}$ ($i=1,2$) LOS , $\Delta \textbf{r}_{i\perp}$ denotes the transverse separation between the reference point and $i^{th}$ LOS and $\theta$ denotes the angle subtended by
	the lines joining the reference point and two other lines of sight in the sky plane. 
	Hence $(\Delta\textbf{r}_{1\perp},\Delta\textbf{r}_{2\perp},\theta)$ denotes the quasar triplet configuration in the sky-plane at the same redshift.

	For a given configuration of triplet source in the sky plane $(\Delta\textbf{r}_{1\perp},\Delta\textbf{r}_{2\perp},\theta)$, we define triple correlation $\zeta(\Delta \textbf{r}_{1\parallel},\Delta\textbf{r}_{2\parallel})$ as
	\begin{equation}
	\zeta(\Delta \textbf{r}_{1\parallel},\Delta\textbf{r}_{2\parallel})=\langle  D_0(\textbf{r}^{\prime}_{\parallel}) D_1(\textbf{r}^{\prime}_{\parallel}+\Delta\textbf{r}_{1\parallel}) D_2(\textbf{r}^{\prime}_{\parallel}+\Delta\textbf{r}_{2\parallel})\rangle \ .
	\end{equation}
	This gives the triple correlation as a function of two redshift space axis.  It is clear from the above equation that three-point correlation function will be negative when all three-points under consideration have strong absorption or only one of them have strong absorption. In all other cases the three-point function will be positive.
	
	In the last row in Fig~\ref{fig_compare_t2p} we plot the three-point correlation function ($\zeta$) measured with $\Delta r_\parallel = 0$ and $2 h^{-1} $cMpc for the Triplets 1 and 2. As we have seen in the case of $\xi$, $\zeta$ values also goes towards zero when we integrate it along the longitudinal directions.
	In the case of "Triplet 1", the observed distribution has large spread and hence the probability of realising the observed $\zeta$ distribution is also very high (see Table~\ref{table_2p}).  In the case of "Triplet 2" the observations trace the upper end of the probability distribution for both the cases considered.
	This indicates a slightly weaker clustering in the regions probed by "Triplet 2" in comparison to the simulations. As we discussed before at least one of the pairs in this triplet also shows weaker two-point correlation function.
	
	Thus when we consider individual two-point and three-point function measurements we find that the observed values are not statistically significant outliers in the framework of simulations considered here. However, what will be more important is to ask what is the probability for simultaneously reproducing the three $\xi$s and $\zeta$. This we will explore in section 5.3.

\begin{figure*}

	\begin{tabular}{lr}

		\includegraphics[viewport=10 10 300 270,width=6.5cm,clip=true]{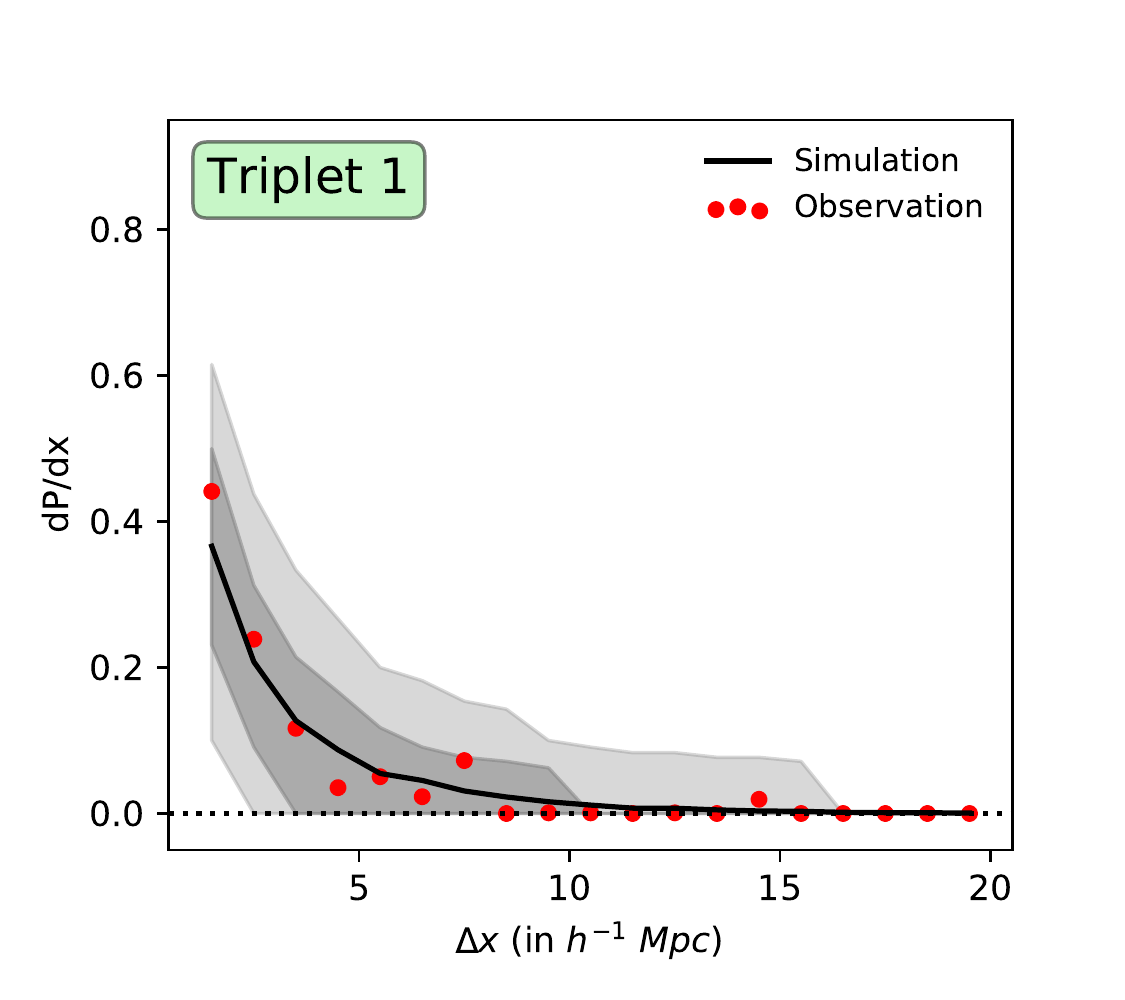}
	&	
		\includegraphics[viewport=10 10 300 270,width=6.5cm,clip=true]{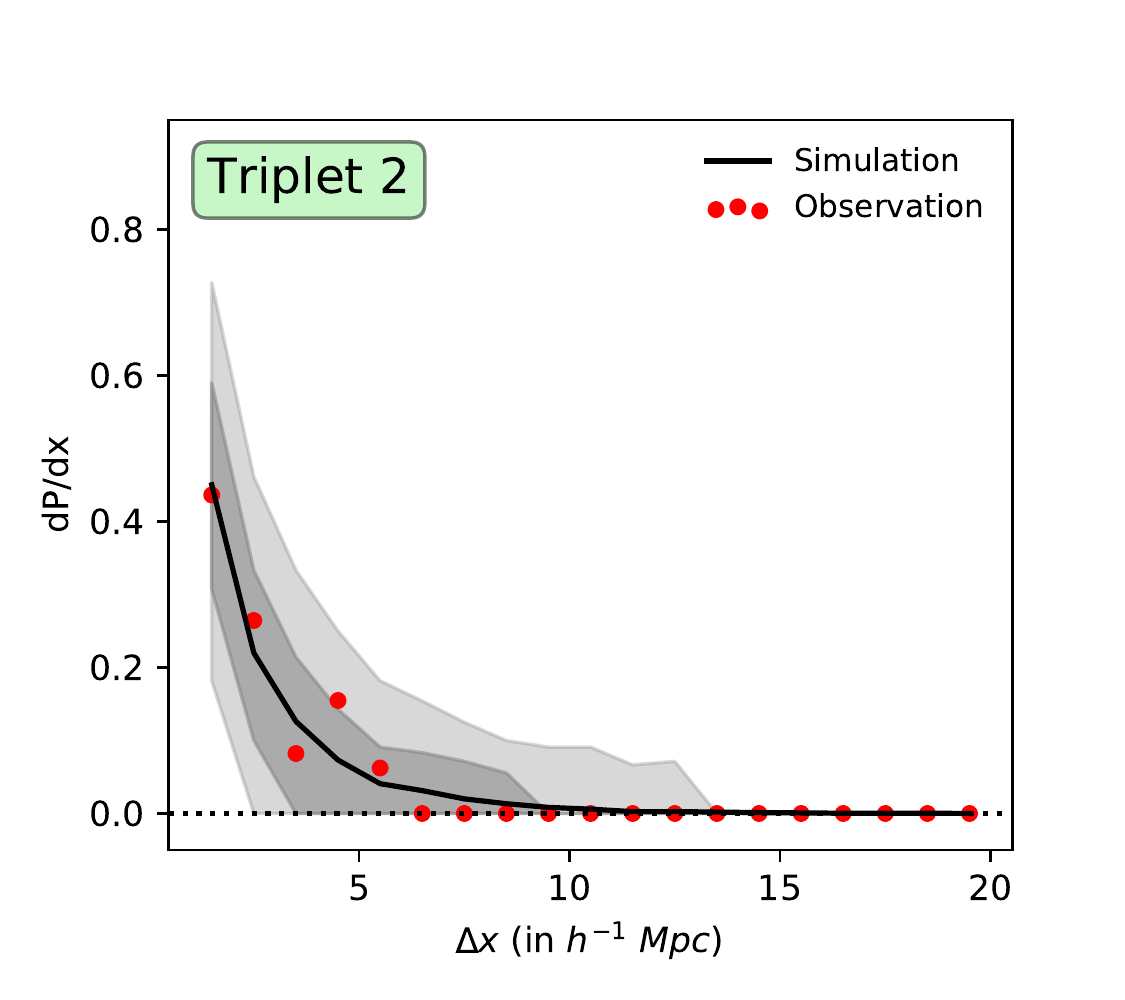}

	\\

   		\includegraphics[viewport=10 0 320 280,width=7cm,clip=true]{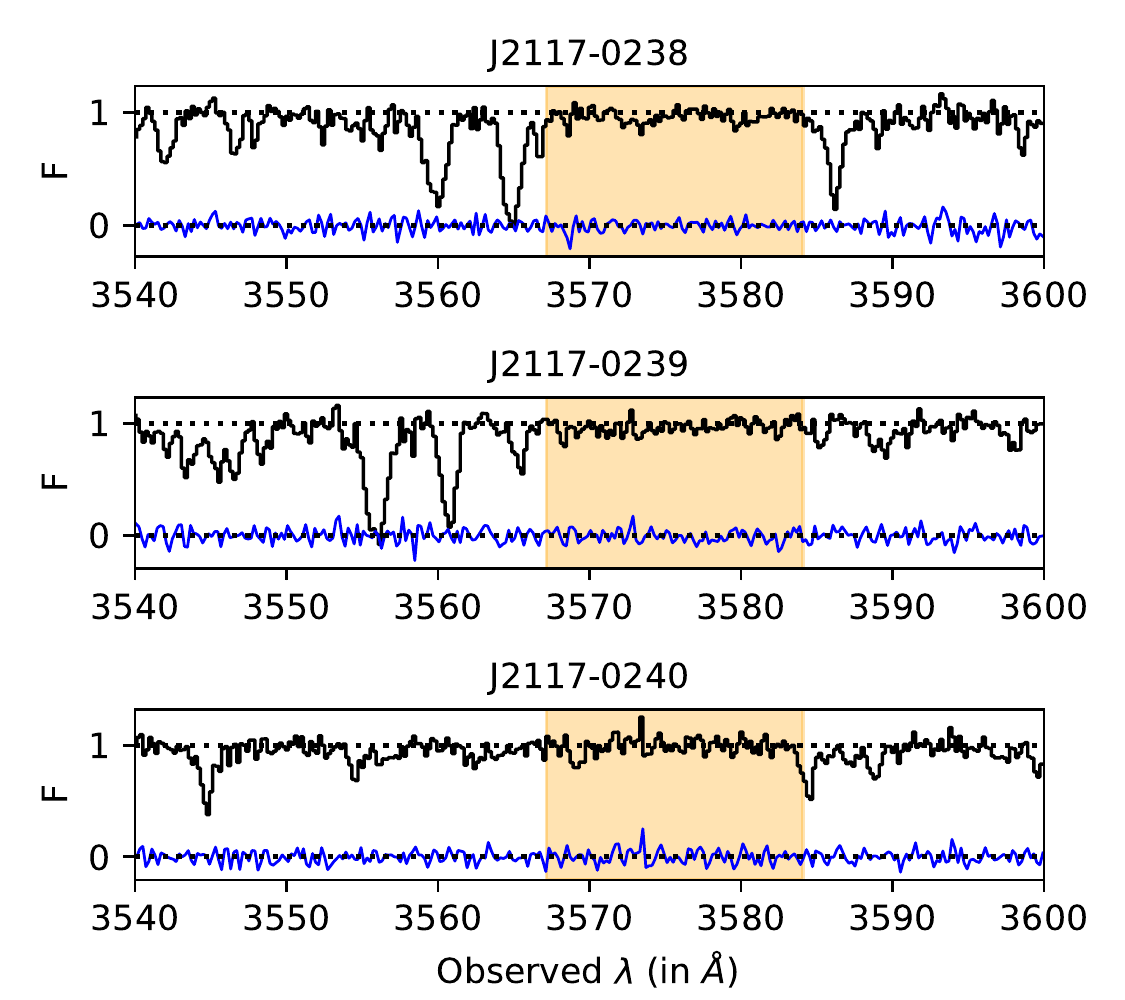}

  &

		\includegraphics[viewport=10 0 330 290,width=7cm, clip=true]{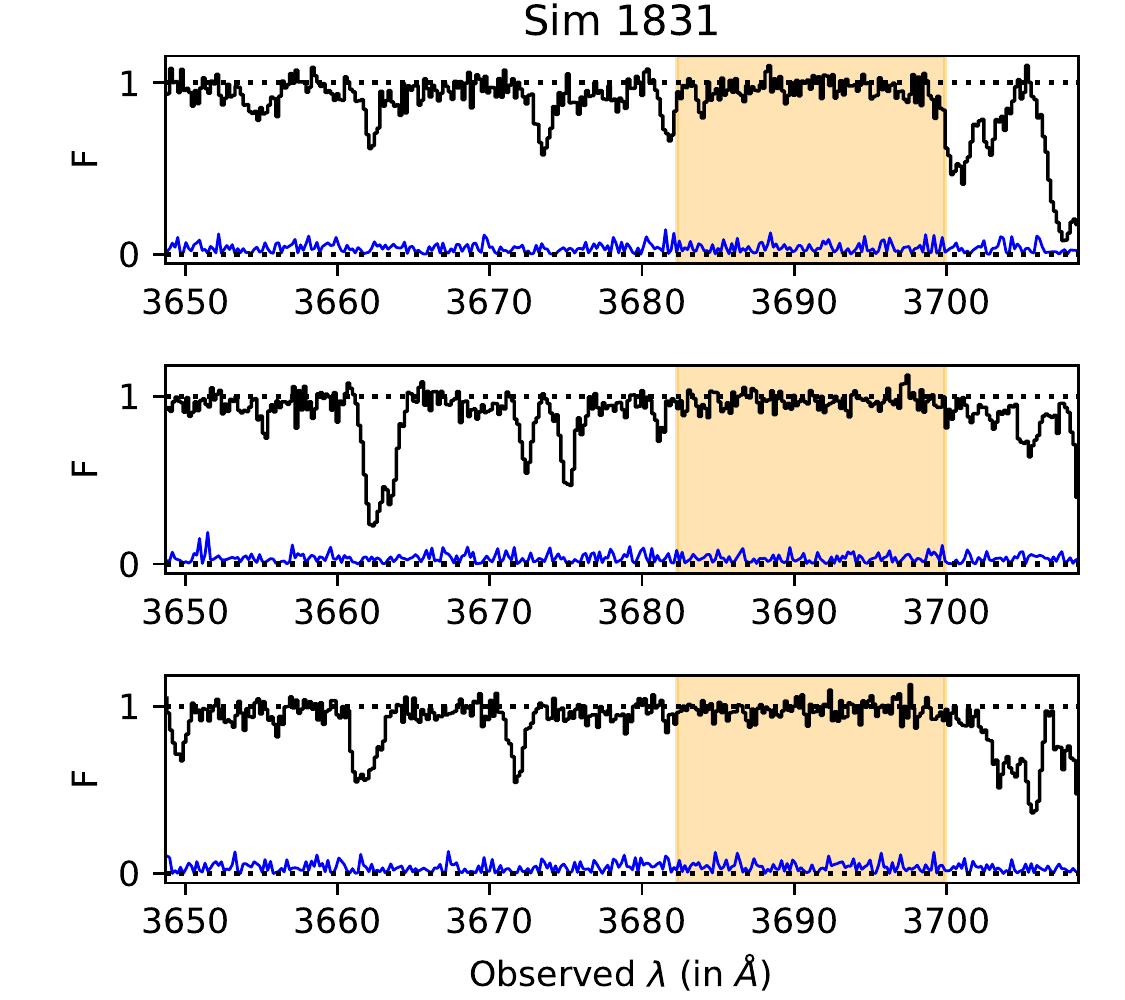}

	\\
	\end{tabular}
	\caption{$Top$: Probability distribution of the concurrent gap length in "Triplet 1" (left) and "Triplet 2" (right). 
	$Bottom-left$: Largest concurrent gap identified along "Triplet 1".
	$Bottom-right$: An example from our simulated spectra that resembles closest to the observed largest concurrent gap.}
	\label{gaps}

\end{figure*}

\subsubsection{Concurrent gap statistics in flux}

In this section, we probe the presence of large concurrent absorption gaps in \lya\ forest along our two triplets. Following \citep{rollinde2003}, for a transmitted flux $F$ in a wavelength bin to be a gap, it must satisfy the criteria
	\begin{equation}
	F>\Bar{F}-\sigma \ ,
	\end{equation}
	where $\sigma=(\sigma_{noise}^2+\sigma_{cont}^2)^{1/2}$ is the standard deviation due to the noise and error from continuum fitting. 
	We also do a running mean over 3 pixels in the spectra to remove spurious peaks due to noise. The $\sigma_{noise}$ is taken to be $1/\rm SNR_{min}$, where $\rm SNR_{min}$ is the minimum SNR value of the 3 sightlines.  We have performed independent continuum fits to get the residual flux error, $\sigma_{cont}$.

 A concurrent gap is a wavelength stretch over which our requirement for gap is satisfied along all three sightlines. 

For simulated sightlines, we simply record the continuous concurrent gap lengths for the three sightlines and then calculate the concurrent gap length probability distribution.
For the observed triplets, we select a random wavelength section whose size matches that of the corresponding simulated sightline and calculate the concurrent gap length probability distribution. We do this random selection 1000 times and average over them to get the observed gap length distribution.

In the top panel in Fig.~\ref{gaps}, we show the probability distributions of observed (red points) and simulated (solid line) concurrent gaps for "Triplet 1" (top left panel) and "Triplet 2" (top right panel). Thick and light shaded regions give 1 and 2 $\sigma$ range seen from the simulated spectra. In the case of "Triplet 1" the largest measured gap has a size of 17 \AA\ ($14.2h^{-1}$cMpc) around the observed wavelength of 3576\AA. 
 %A search for a similar gap in the simulation yielded 
 In our simulations we find $14.2\%$ of the total triplet sightlines have concurrent gap length  $\ge \rm 14.2h^{-1}$cMpc. In this case we have masked an intervening \CIV\ absorption present along the line of sight to J2117-0238 that also falls in the gap. For the individual sightlines, the underdense region corresponds to a gap of length 17.3 \AA\ ($\rm 14.5h^{-1}$cMpc) in J2117-0238, 22.8 \AA\ ($\rm 19h^{-1}$cMpc) in J2117-0239 and 29 \AA\ ($\rm 24.2h^{-1}$cMpc) in J2117-0240.  A spectra of a simulated triplet that closely resembles that of "Triplet 1" is also shown in the bottom panel in Fig.~\ref{gaps}. In passing we note that even when we consider the full sightline (not the small regions as we use for comparison with simulations) we find the same gap as the largest concurrent gap. 
 
 \citet{rollinde2003} have found a concurrent gap of 13\AA\ wide at $z=1.99$,  between four nearby sightlines that span an angular separations between 2.1 to 9.0 arc min. 
 \citet{cappetta2010} have reported two concurrent gaps along two triple QSO sightlines with the gap length of 10 and 15\AA\ with redshifts 1.82 and 1.93 respectively. 
 Thus we report the concurrent gap that is larger than those reported in the literaure using more than 3 sightlines. However, the transverse spatial scales probed by "Triplet 1" is smaller than those probed in the literature.

 In case of Triplet 2, such large concurrent gap length is not found (see top right panel in Fig~\ref{gaps}). The largest concurrent gap spreads over $8.7\ang$ corresponding to a length of $\rm 5.9h^{-1}$cMpc. The overall distribution of the gap is consistent with the predictions of our simulations. 

\subsection{Cloud based statistics}

In this section, we perform correlation analysis between Voigt profile components obtained along each line of sight using our automated Voigt profile fitting code {\sc viper} \citep{gaikwad2017b}. Unlike flux based statistics, in this case we will be able to probe the dependence of correlations on the $N_{\rm HI}$ (which is known to trace the underlying over density).  Moreover, there is an inherent degeneracy in flux based three-point correlation. Regions having an absorption in one of the sightlines and gap in the other two sightlines will give a negative three-point correlation in flux which is degenerate with regions having absorption in all the three sightlines. This degeneracy also makes it difficult to interpret the reduced three-point function (i.e Q defined below) measured using flux statistics.

\subsubsection{Longitudinal two-point correlation}
When we treat IGM being constituted by distinct clouds, we follow the standard procedure to compute correlation function (both for the simulated as well as observed spectra) with respect to a random distribution of clouds. 
The longitudinal two-point correlation is estimated using $Landy-Szalay$ estimator \citep{landy_szalay1993},
	
	\begin{equation}
		\xi(\Delta r_{\parallel})=\frac{DD-2DR+RR}{RR} \ .
	\end{equation}	
Here, "DD", "RR" and "DR" corresponds to data-data, random-random and data-random pair counts respectively measured at a separation of $\Delta r_\parallel$.
For a given sightline, we construct the random distribution of clouds 
using Poisson distribution with a mean number of clouds being equal to what is expected based on the observed redshift distribution of clouds having $N_{\rm HI}$ above a threshold value.

	Once we have the observed and random cloud distributions, 
	we consider all possible  combinations of pair separations between these clouds.	The $DD$ pair separation counts are normalized with the total number of pair combinations, (i.e, it is divided with $n_D(n_D-1)$ where $n_D$ is the number of clouds along a sightline). In a similar fashion, one can normalize the data-random pair separations $DR$ and random-random pair seperation $RR$.  We have used 100 random skewers for every data skewer. 

 The  measured $\xi(\Delta r_{\parallel})$ along our sightlines are compared with the expectations from the simulations in the last column of Fig.~\ref{summary_fig}. It is clear from this plot that the expected correlation is negative in the first bin corresponding to  transverse separations  $r_\parallel < 2h^{-1}$ cMpc. This is mainly because while there is no restriction on the minimum separations between the randomly generated clouds, thermal broadening together with instrumental resolution set a limit on the lowest measurable separation during the Voigt profile decomposition (i.e cloud exclusion). It is clear from the figure that expected transverse two-point correlation function of clouds is consistent within 1$\sigma$ range of our model predictions. Thus we do not encounter any abnormal line of sight clustering (with scales larger than 1$h^{-1}$cMpc) along the sightlines considered in this study.

\subsubsection{Transverse two-point Correlation}
\begin{figure*}

	\begin{minipage}{0.3\textwidth}
		\includegraphics[viewport=5 10 350 270,width=6.2cm, clip=true]{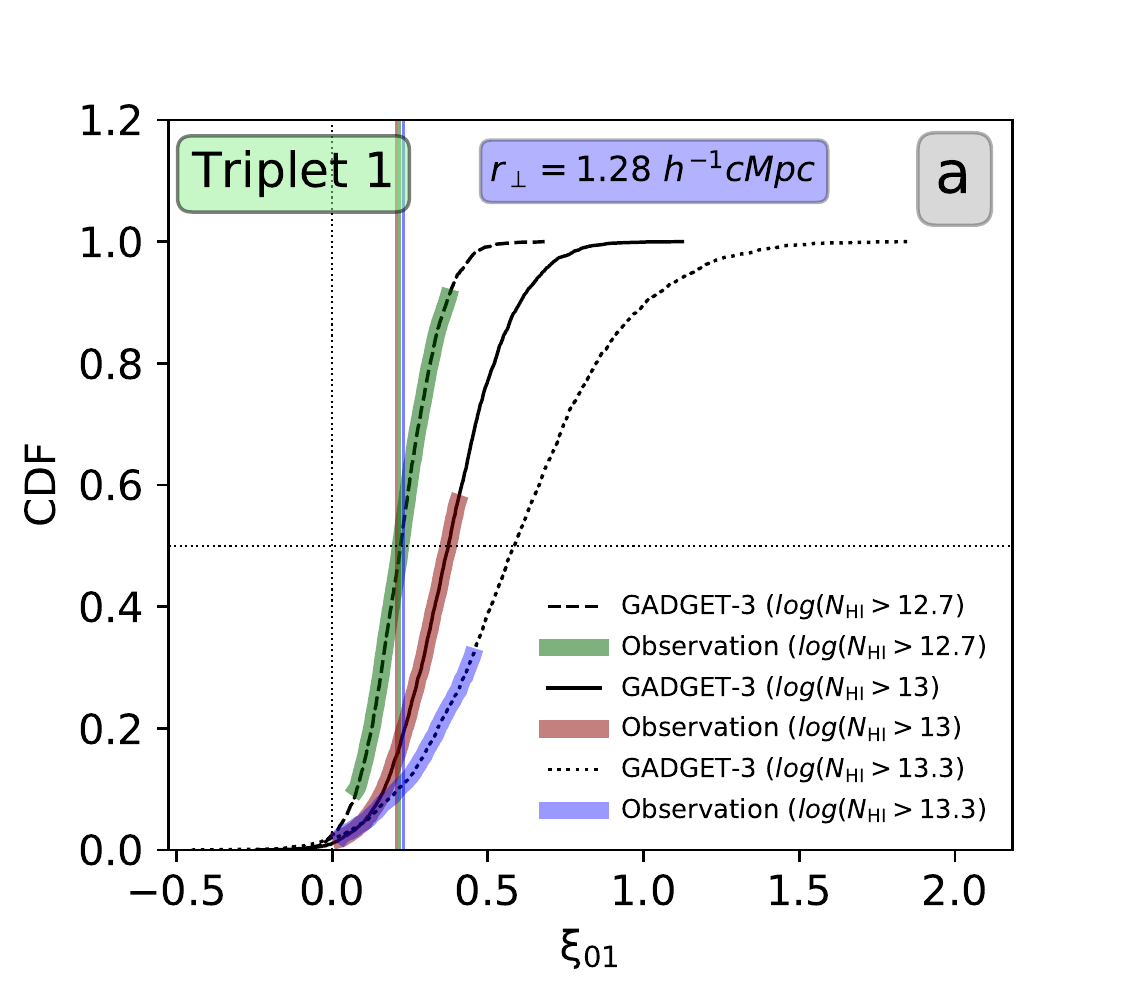}
	\end{minipage}%
	\begin{minipage}{0.3\textwidth}
		\includegraphics[viewport=5 10 350 270,width=6.2cm, clip=true]{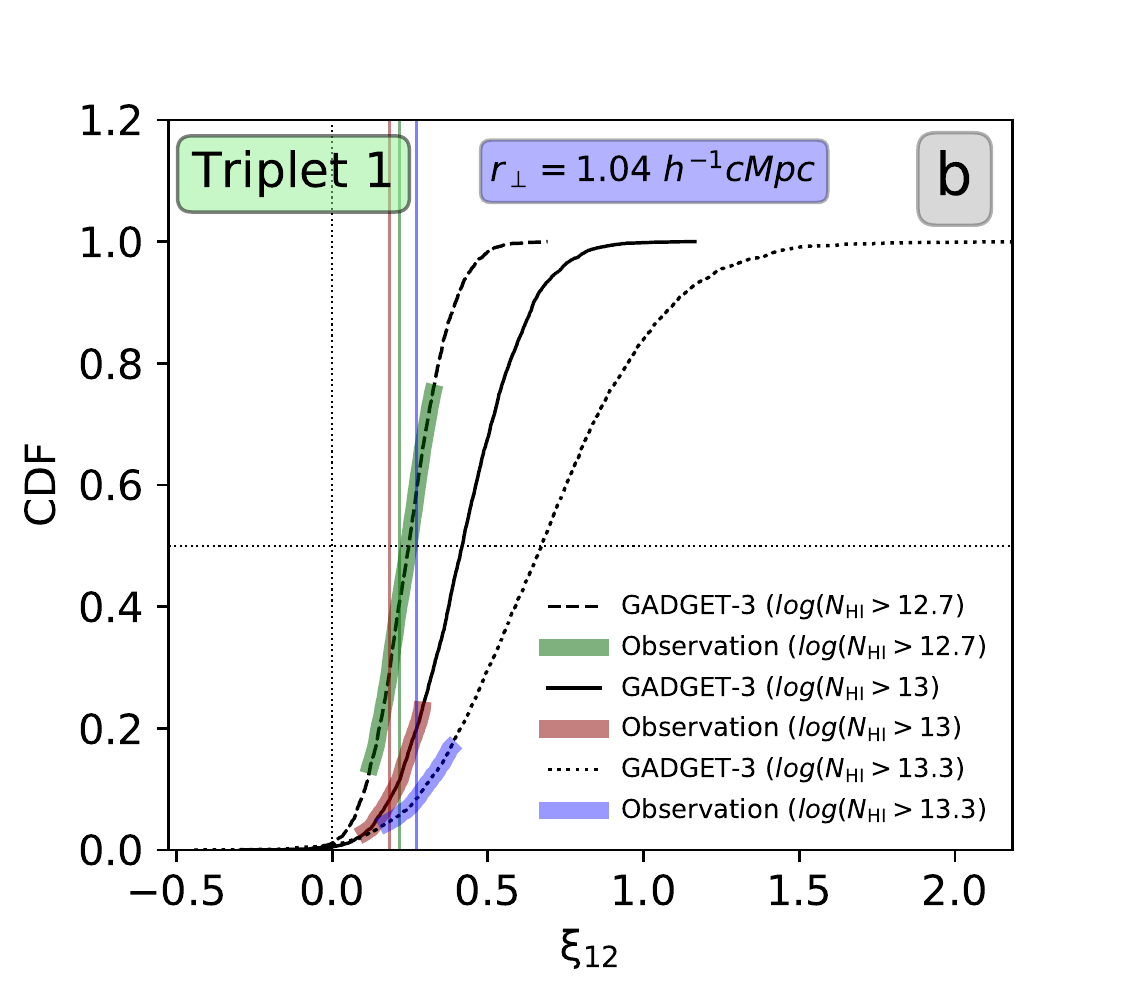}
	\end{minipage}
	\begin{minipage}{0.3\textwidth}
		\includegraphics[viewport=5 10 350 270,width=6.2cm, clip=true]{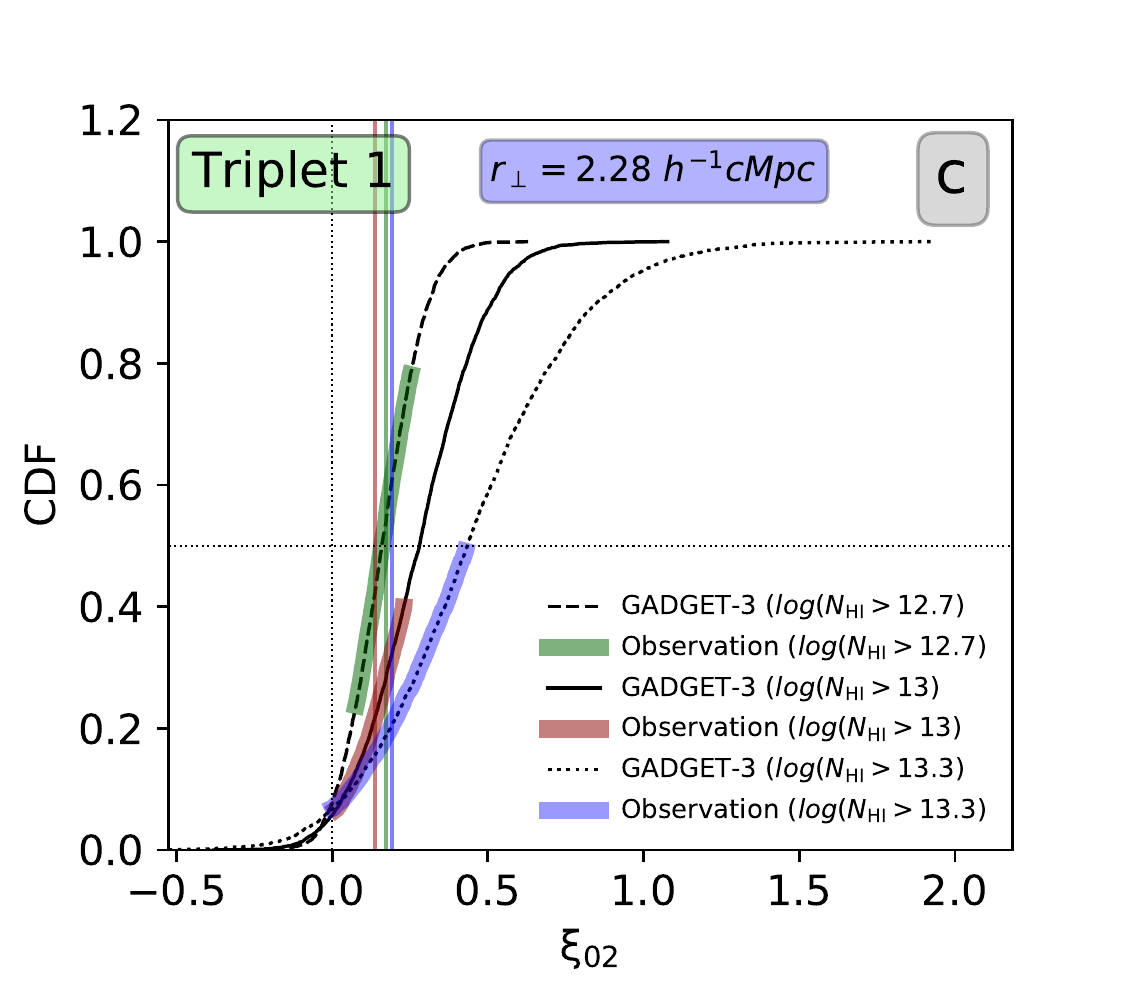}
	\end{minipage}%

	\begin{minipage}{0.3\textwidth}
		\includegraphics[viewport=5 10 350 270,width=6.2cm, clip=true]{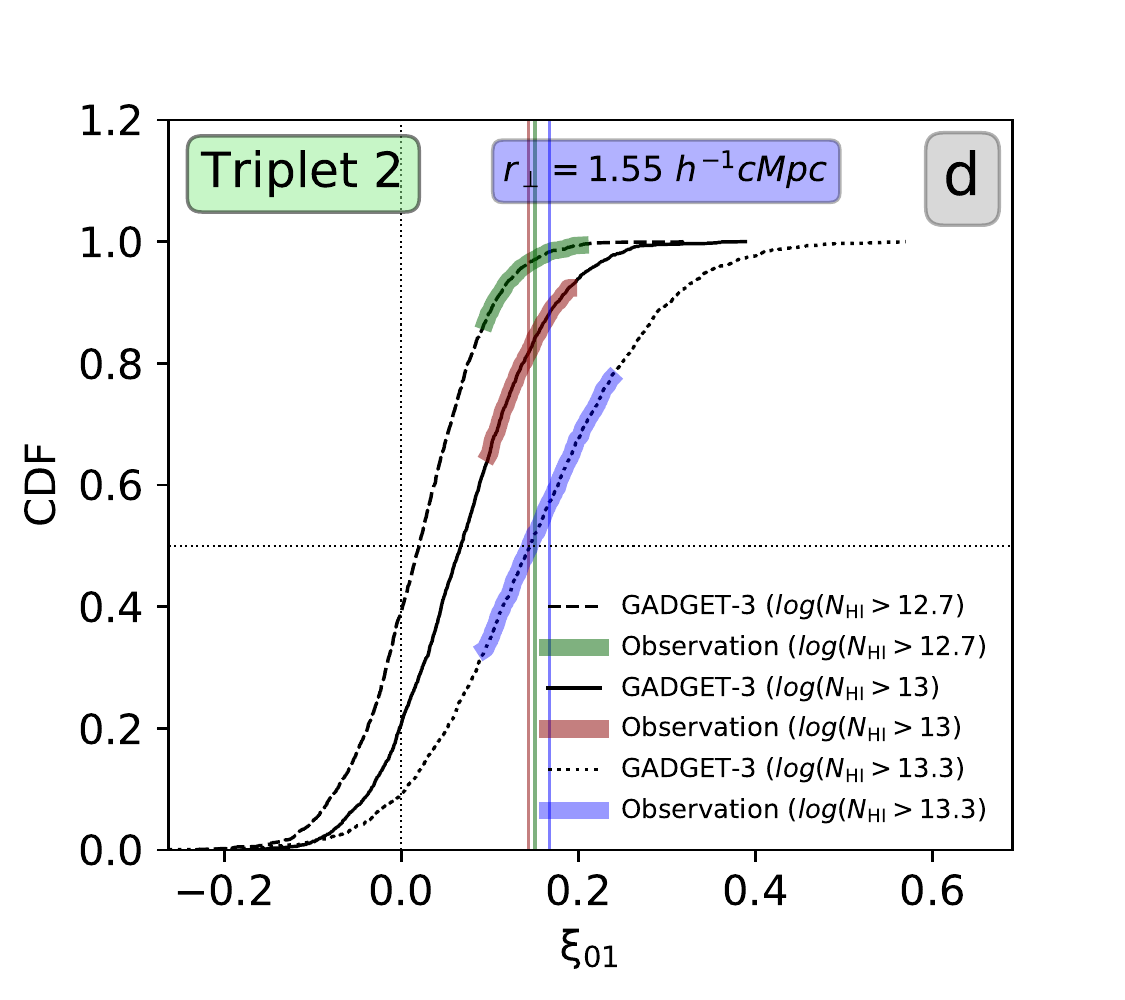}
	\end{minipage}%
	\begin{minipage}{0.3\textwidth}
		\includegraphics[viewport=5 10 350 270,width=6.2cm, clip=true]{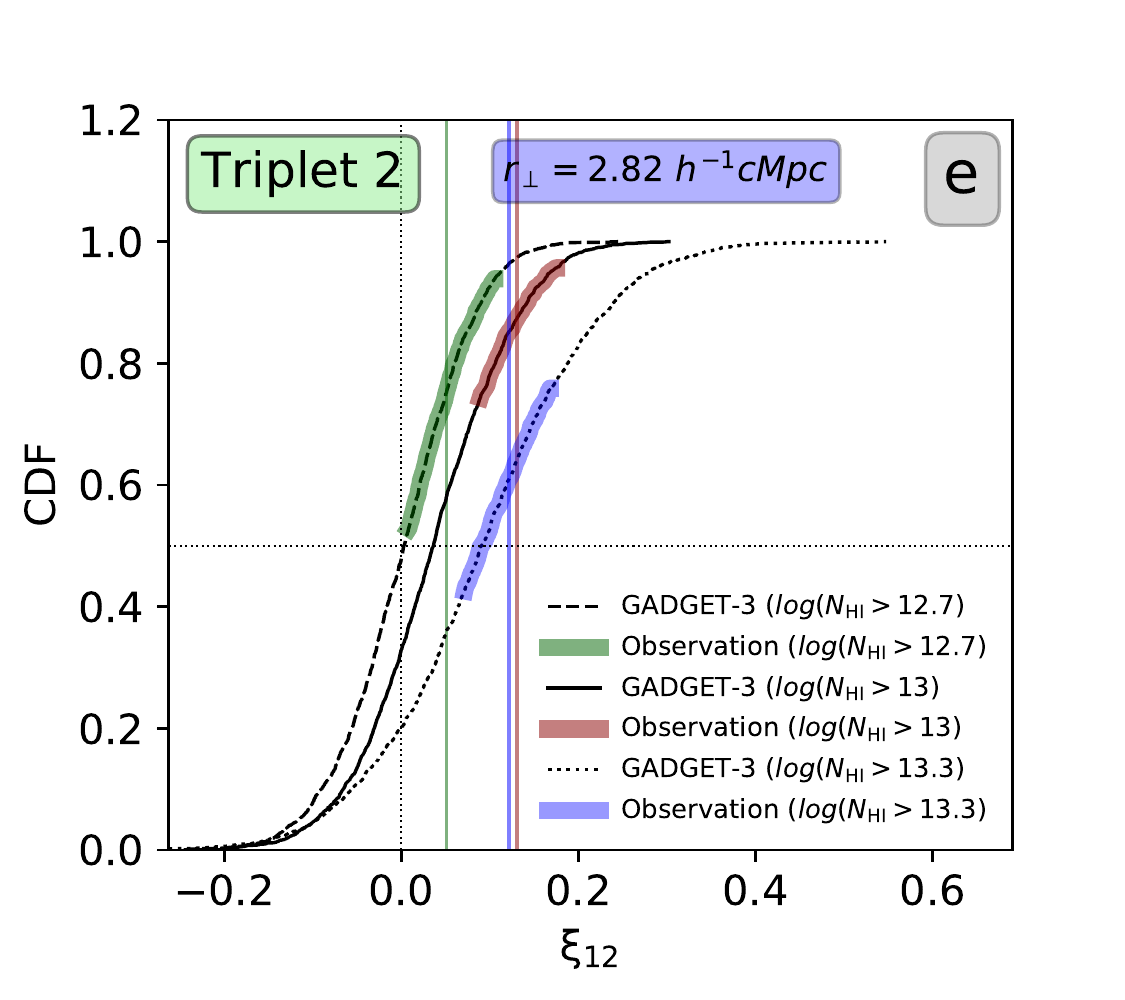}
	\end{minipage}
	\begin{minipage}{0.3\textwidth}
		\includegraphics[viewport=5 10 350 270,width=6.2cm, clip=true]{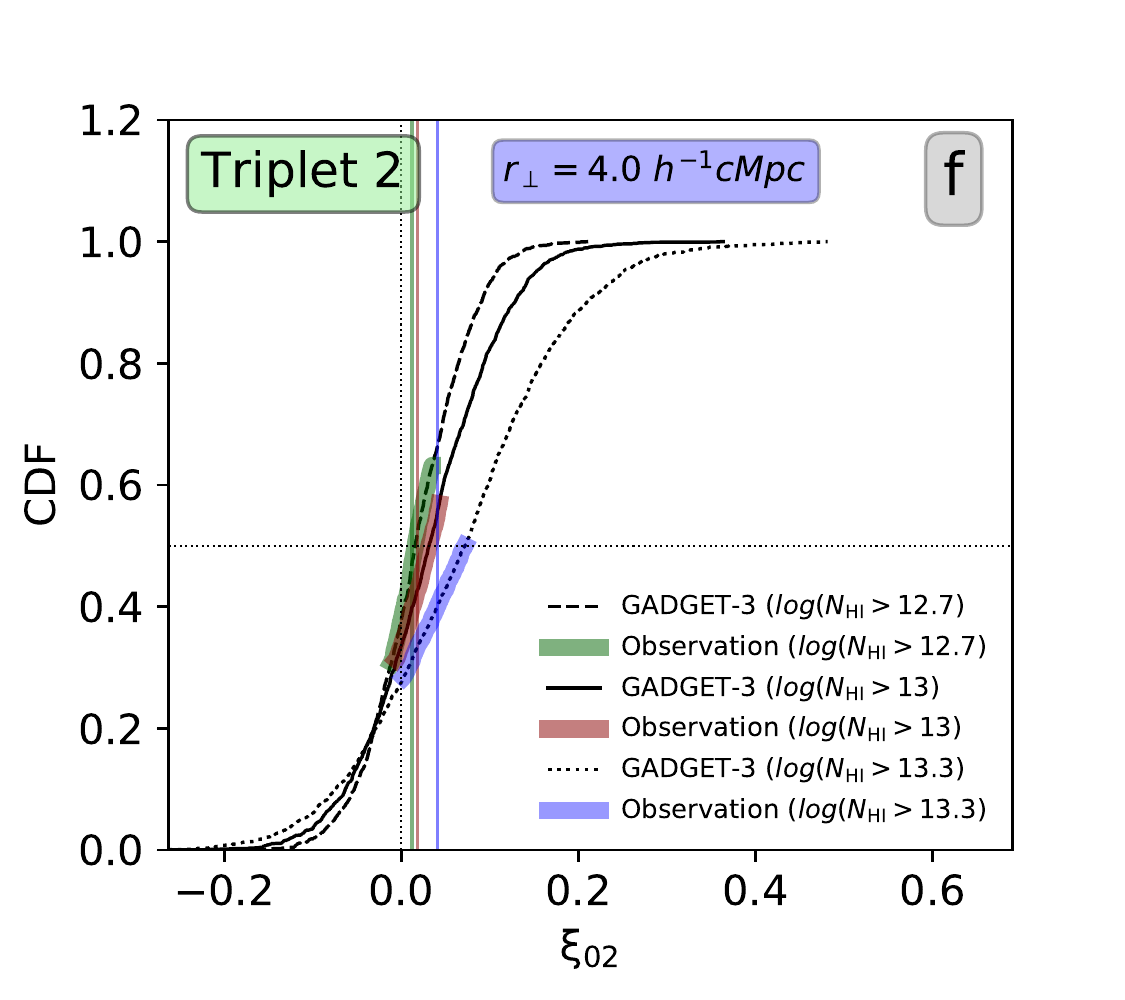}
	\end{minipage}%

    \begin{minipage}{0.3\textwidth}
		\includegraphics[viewport=5 10 350 270,width=6.2cm, clip=true]{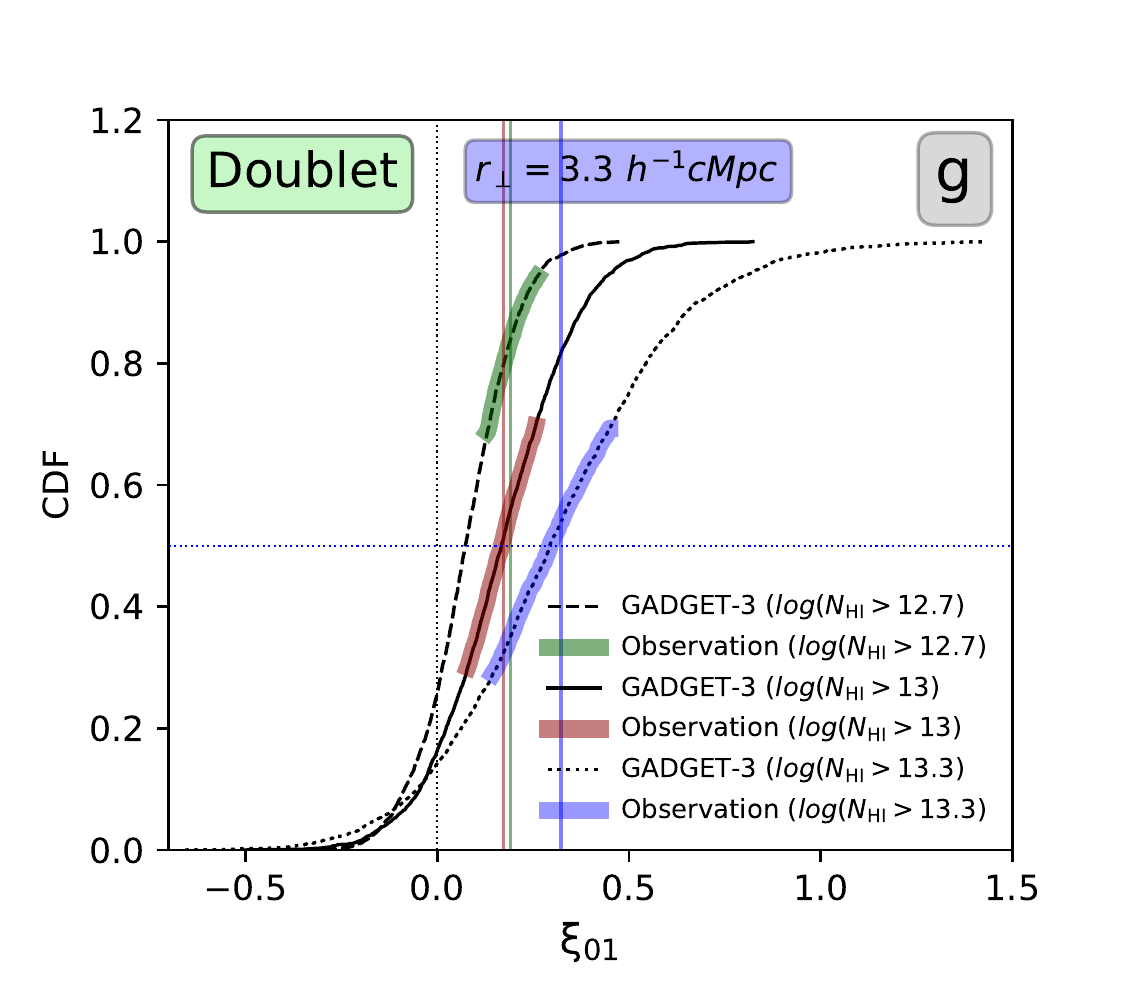}
	\end{minipage}%
	 \begin{minipage}{0.3\textwidth}
		\includegraphics[viewport=5 10 350 270,width=6.2cm, clip=true]{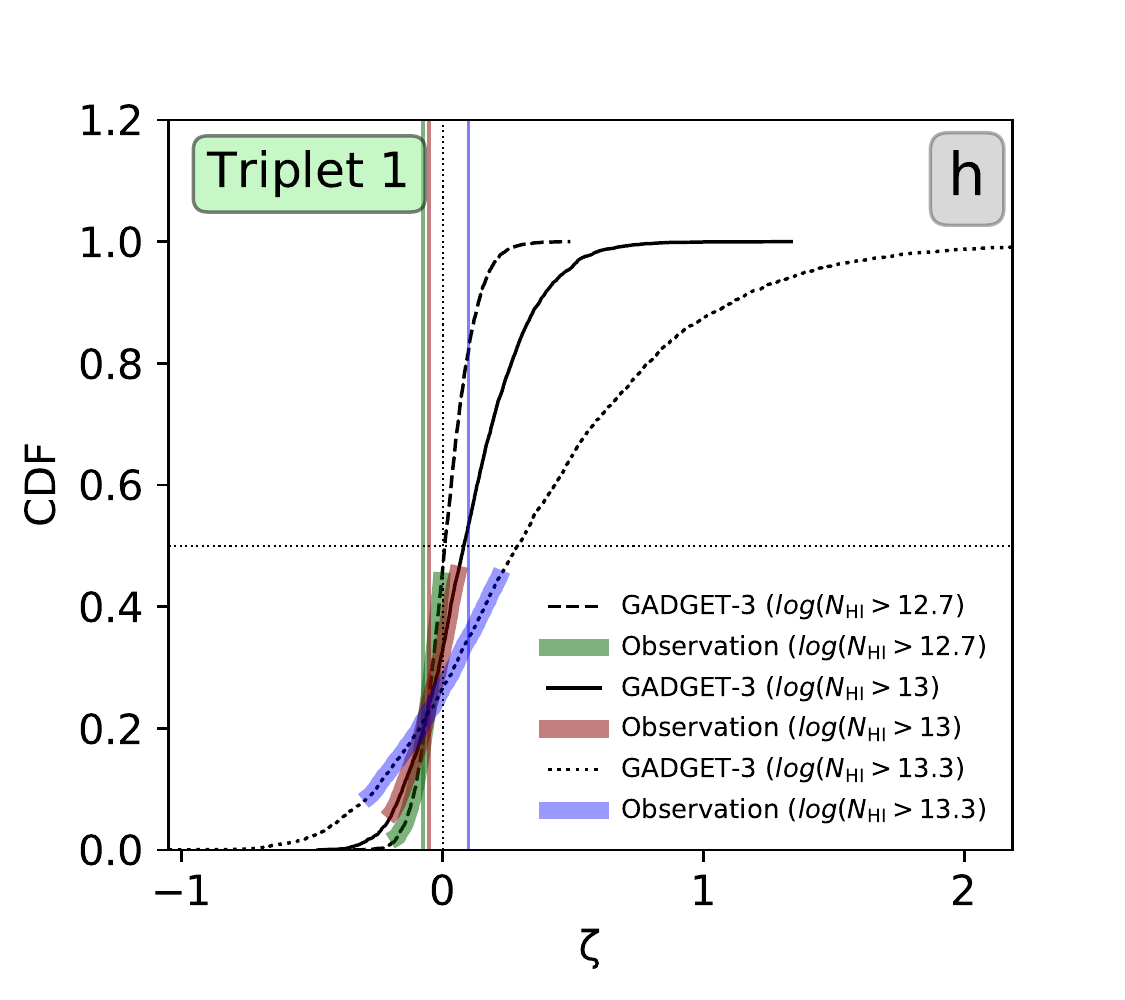}
	\end{minipage}%
	 \begin{minipage}{0.3\textwidth}
		\includegraphics[viewport=5 10 350 270,width=6.2cm, clip=true]{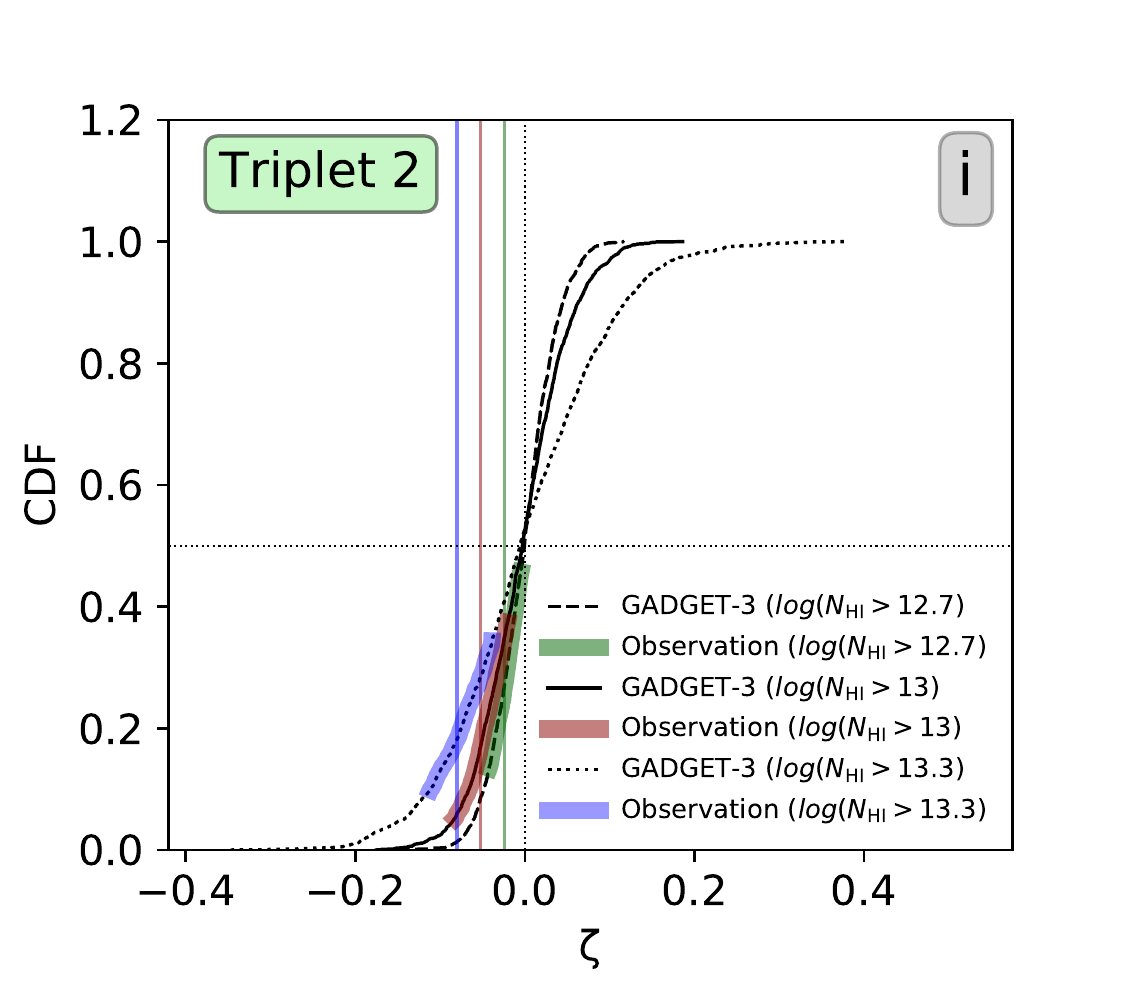}
	\end{minipage}
	
	\caption{Cloud based transverse two-point and three-point correlation cumulative distribution for the three triplets. The dashed, solid and dotted curves are the cumulative distribution functions for differnt neutral hydrogen column density cut-offs. The red, green and blue regions overlaying the simulated curve represents the $1\sigma$ confidence interval obtained from sub-sampling of the observed triplet sightlines for the two different longitudinal grids. The vertical lines with colors denote the mean of the corresponding sub-sampled observed correlation. 
	}
	\label{fig_cloud_stat}
\end{figure*}

To probe the transverse correlation between two data skewers $D1$ and $D2$ along two closely spaced sightlines, we generate two random skewers $R1$ and $R2$. The transverse correlation is then defined as 
\begin{equation}
\xi(\Delta r_{\parallel},\Delta r_{\perp})=\frac{D_1D_2-D_1R_2-R_1D_2+R_1R_2}{R_1R_2} \ .
\end{equation}
Similar to longitudinal correlation, the transverse correlation has been generated by averaging over 100 random skewers for every data skewers. We express the two-point correlation $\xi(\Delta r_{\parallel},\Delta r_{\perp})$ as a function of actual cloud separation $\Delta r =\sqrt{\Delta r_{\parallel}^2+\Delta r_{\perp}^2}$ in different plots. In Table~\ref{table_cloud_stat}, we summarize our measurement of transverse two-point correlation, of clouds having log~$N_{\rm HI}$$\ge$ 13.0, measured within the longitudinal separation bin of $\pm$ 2$h^{-1}$ cMpc (The effect of choosing a smaller longitudinal bin of $\pm$ 2$h^{-1}$ cMpc is shown in Fig.~\ref{fig_cloud_cdf_ap} in the Appendix). The entries in this table are similar to that of Table~\ref{table_2p}. 
In Fig~\ref{fig_cloud_stat}, we compare the predicted cumulative distributions 
from our simulated spectra with the observations (as in Fig.~\ref{fig_f2p})
for 3 different $N_{\rm HI}$ thresholds.

\begin{table}
    \centering
    \caption{Results of Cloud-based ($\rm log(N_{\rm HI})>13$) correlation analysis}
    \setlength{\tabcolsep}{2pt}
    \begin{tabular}{cccccc}
\hline
Sample & Correlation & r & Observed & Probability & Percentile \\
       &             &($h^{-1}$cMpc)   & values & &\\
\hline
\hline
Triplet 1 & $\xi_{01}$ & 1.28 & 0.207  & 0.55 & 15.5\\
          & $\xi_{12}$ & 1.04 & 0.184  & 0.20 & 8.3\\
          & $\xi_{02}$ & 2.28 & 0.137  & 0.33 & 21.8\\
          & $\zeta$    &      & -0.052 & 0.39 & 24.1\\
          & Q          &      & -0.122 & 0.83 & 26.6\\ 
 %         \\
Triplet 2 & $\xi_{01}$ & 1.55 & 0.143  & 0.27 & 81.7\\
          & $\xi_{12}$ & 2.82 & 0.131  & 0.21 & 87.4\\
          & $\xi_{02}$ & 4.00 &  0.018 & 0.26 & 42.7\\
          & $\zeta$    &      & -0.052  & 0.32 & 17.1\\
          & Q          &      & -2.246 & 0.68 & 31.7\\
  %        \\
Doublet & $\xi_{01}$ & 3.30 &  0.172 & 0.40 & 51.0\\
\hline
    \end{tabular}
    \label{table_cloud_stat}
\end{table}

In the case of "Triplet 1" we find  that the observed correlations is
slightly less (as suggested by the percentile) than the median predicted correlation for log~$N_{\rm HI}\ge$13.0 (see Table~\ref{table_cloud_stat} and Fig.~\ref{fig_cloud_stat}). This trend is consistent with what we are finding above based on statistics using transmitted flux. Our simulations also predict an increase in $\xi$ with increasing $N_{\rm HI}$. {While the observations confirm this trend, the dependence on $N_{\rm HI}$ is usually weaker than the predictions from the simulation in case of the two-point correlations} (see the vertical lines in Fig.~\ref{fig_cloud_stat}).

In the case of "Triplet 2" the observed $\xi$ for two of the closest separations (i.e $\xi_{01}$ and $\xi_{12})$ are found to be higher than our model predictions (for log~$N_{\rm HI}\ge 13$). This seems to be the case even
when we include lower $N_{\rm HI}$ clouds. However, the observed distribution is close to the simulation results when we consider log~$N_{\rm HI}\ge$13.3 for both the pairs.  In case of the longest separated pair in "Triplet 2" the measured $\xi_{02}$ for all three column density thresholds are consistent with the predictions of our simulations. 
This seems also to be the case  for "Doublet".

%Next we consider the three-point correlation between clouds along three lines of sights.

\subsubsection{Three-point correlation in clouds}

Here, we study the three-point correlation function between three data skewers $D1$, $D2$ and $D3$ using three random skewers $R1$, $R2$ and $R3$ generated by the same approach discussed above. For a fixed sightline configuration (i.e $\Delta r_{1\perp},\Delta r_{2\perp},\theta$), three-point correlation is a function of two redshift space separations along two sightlines with respect to the reference sightline, i.e., $\zeta=\zeta(\Delta r_{1\parallel},\Delta r_{2\parallel})$. Hence, we generate data-data-data triplet separations in 2D logarithmic bins since there are two redshift space separation axis involved.
This gives us $D_1D_2D_3$ which we normalize by dividing with $n_D(n_D-1)(n_D-2)$. Similarly, we generate the data-data-random, data-random-random and random-random-random triplet separations. The three-point correlation is then given, following \cite{szapudi-szalay1998}, as 

\begin{equation}
\zeta(\Delta r_{1\parallel},\Delta r_{2\parallel})=\frac{D_1D_2D_3-DDR_{(123)}+DRR_{(123)}+R_1R_2R_3}{R_1R_2R_3} \ ,
\end{equation}  
where $DDR_{(123)}=D_1D_2R_3+D_1R_2D_3+R_1D_2D_3$ and $DRR_{(123)}=D_1R_2R_3+R_1D_2R_3+R_1R_2D_3$. Here also, we use 100 random skewers for every data skewers.

\begin{figure}
	\includegraphics[viewport=10 10 385 360,width=7cm, clip=true]{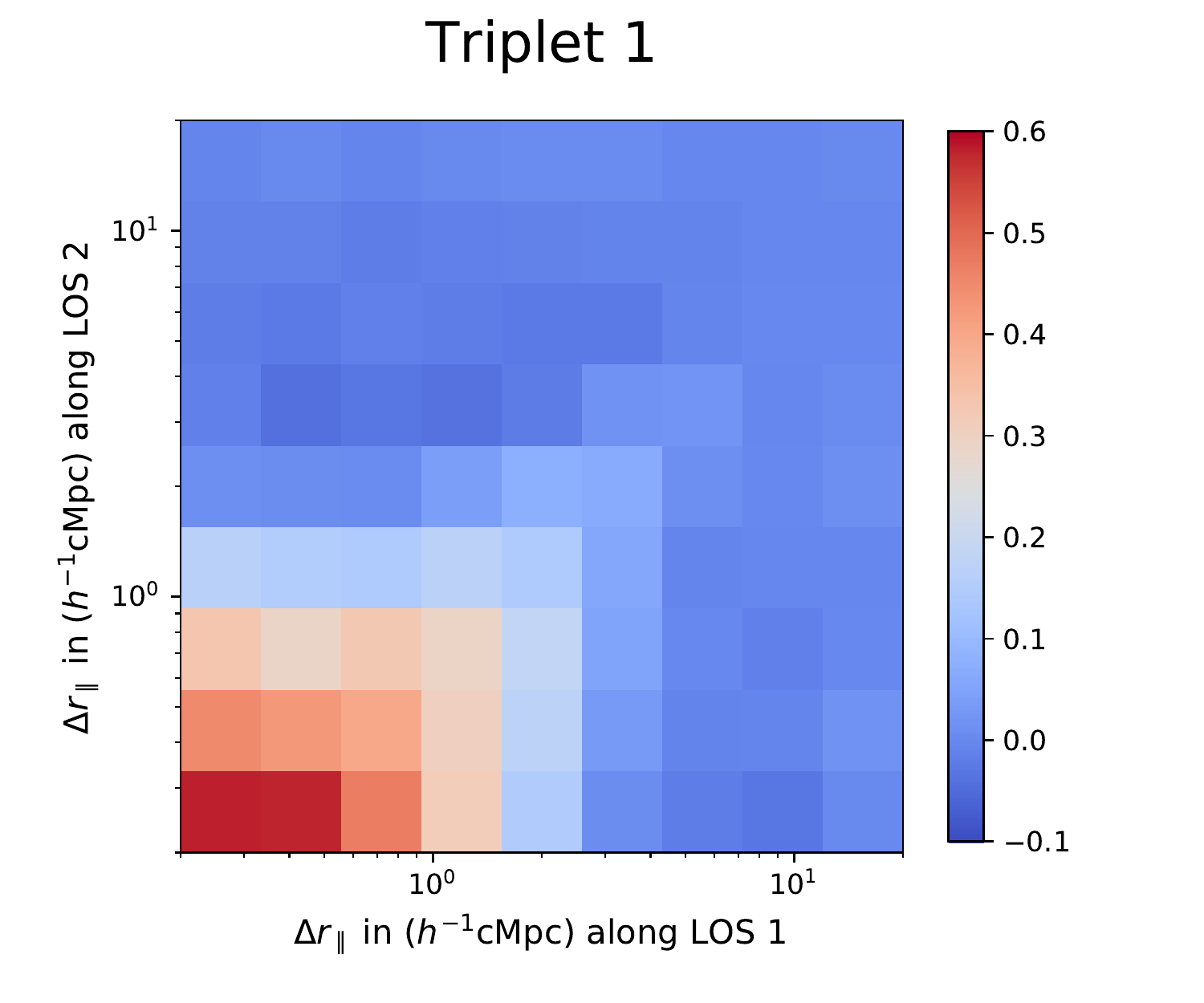}
	\includegraphics[viewport=10 10 385 360,width=7cm, clip=true]{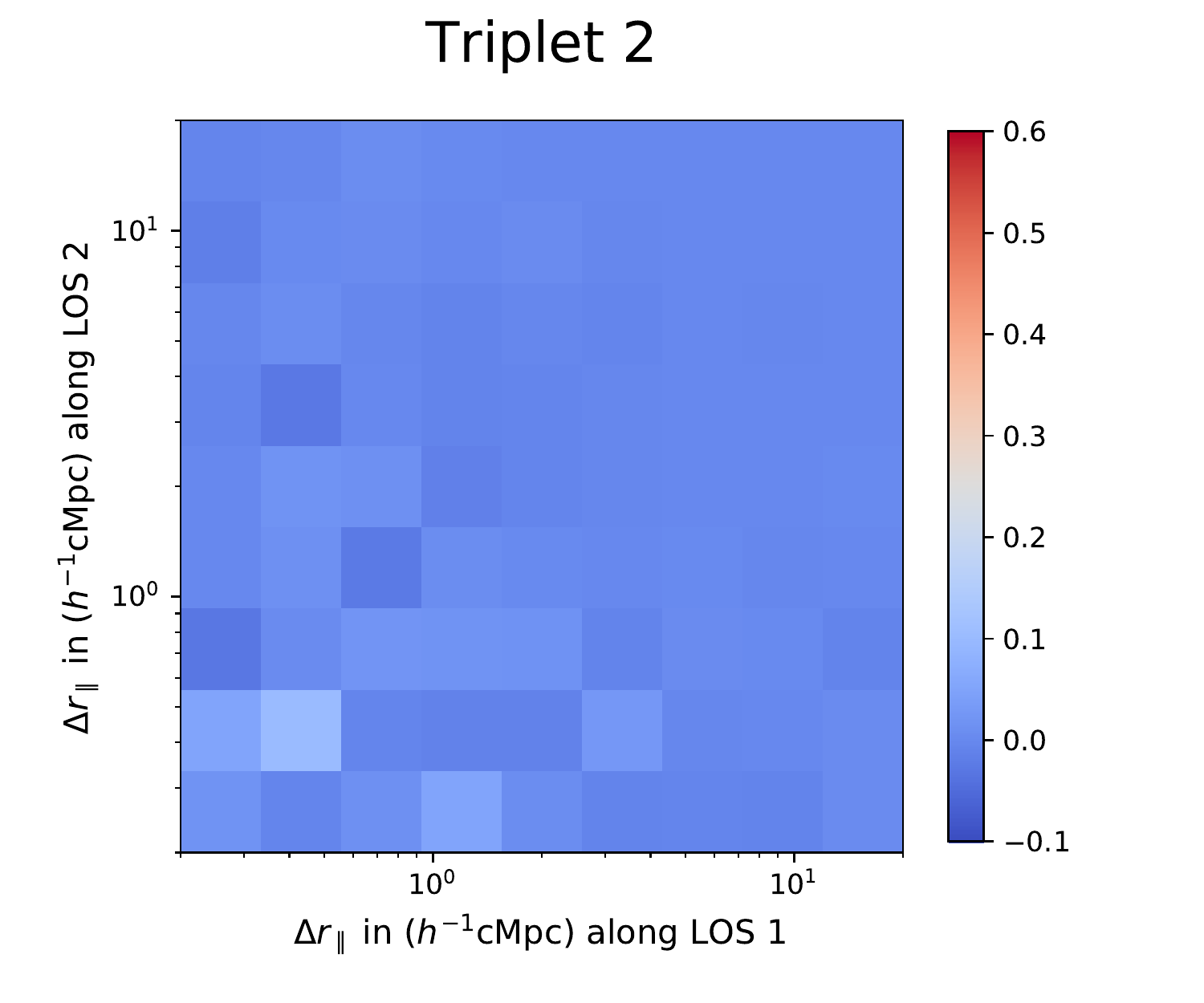}
	\caption{Cloud based ($\rm log(N_{HI})>13$) three-point correlation function from simulation as a function of redshift space separation of clouds along 2nd and 3rd sightline  of the triplet from the 1st sightline. The color of a bin represents triple-correlation for that bin. }
\label{t3_ind}
\end{figure}

In Fig.~\ref{t3_ind} we plot the triple correlation measured  in our simulations for configuration similar to the observed triplets among clouds along LOS~1 and LOS~2 with respect to those along LOS~0 (see  Fig.~\ref{config} for reference). 
Strong correlations are seen up to  the longitudinal separations of 2$h^{-1}$cMpc along both these sightlines in "Triplet 1". So while computing the three-point function we integrate  the correlations over 2$h^{-1}$cMpc along each sightlines.  However, in the case of "Triplet 2" we do not see strong correlations. This is not surprising given the larger angular separation between the sightlines. The measured values of the three-point correlations (summed over $\pm 2h^{-1}$cMpc along each of the longitudinal directions) for both the triplets  for log $N_{\rm HI}$$\ge$13 are given in Table~\ref{table_cloud_stat}. The cumulative distribution are shown in the bottom panels of Fig.~\ref{fig_cloud_stat}. 

In the case of configurations similar to "Triplet 1", our simulations predict positive values of $\zeta$ for 
cases log($N_{\rm HI})\ge13$ and 13.3. However when we include weak
$N_{\rm HI}$  clouds $\zeta$ becomes close to zero. The measured values of $\zeta$ and its 1$\sigma$ range from the observations are lower than the median values obtained in the simulations. Even though the increase in $\zeta$ with the increase in $N_{\rm HI}$ is evident for the observations the actual $\zeta$ value obtained even for the higher $N_{\rm HI}$ clouds are lower than those realised in the simulations. In the case of configuration similar to "Triplet 2", our simulations predict median $\zeta$ close to zero for the three $N_{\rm HI}$ thresholds considered. 
The observed values are less than the median from the simulations and mostly negative but follow the expected trend of the cumulative distributions (see Fig.~\ref{fig_cloud_stat}),that is,
$\zeta$ becomes more negative with increasing $N_{\rm HI}$ in the lower end of the cumulative distribution.

In the case of galaxies, it is usual procedure to define the reduced three-point function Q through the \textit{hierarchical ansatz} suggested by \citet[][]{peebles1980}. In this case
the observed three-point correlation function can be written in term of the cyclic combination of respective two-point correlation function as 
\begin{dmath}
\zeta(\textbf{r}_{01\perp},\textbf{r}_{12\perp},\textbf{r}_{02\perp})=Q\ [\xi(\textbf{r}_{01\perp})\xi(\textbf{r}_{12\perp})+\xi(\textbf{r}_{12\perp})\xi(\textbf{r}_{02\perp}) +\xi(\textbf{r}_{01\perp})\xi(\textbf{r}_{02\perp})] \ . 
\end{dmath}
For clouds with $N_{\rm HI}$ above a threshold, one can construct a similar reduced three-point correlation function Q.
\begin{figure*}
	\begin{minipage}{0.95\textwidth}
		\center

		\includegraphics[viewport=20 10 400 270,width=8.3cm, clip=true]{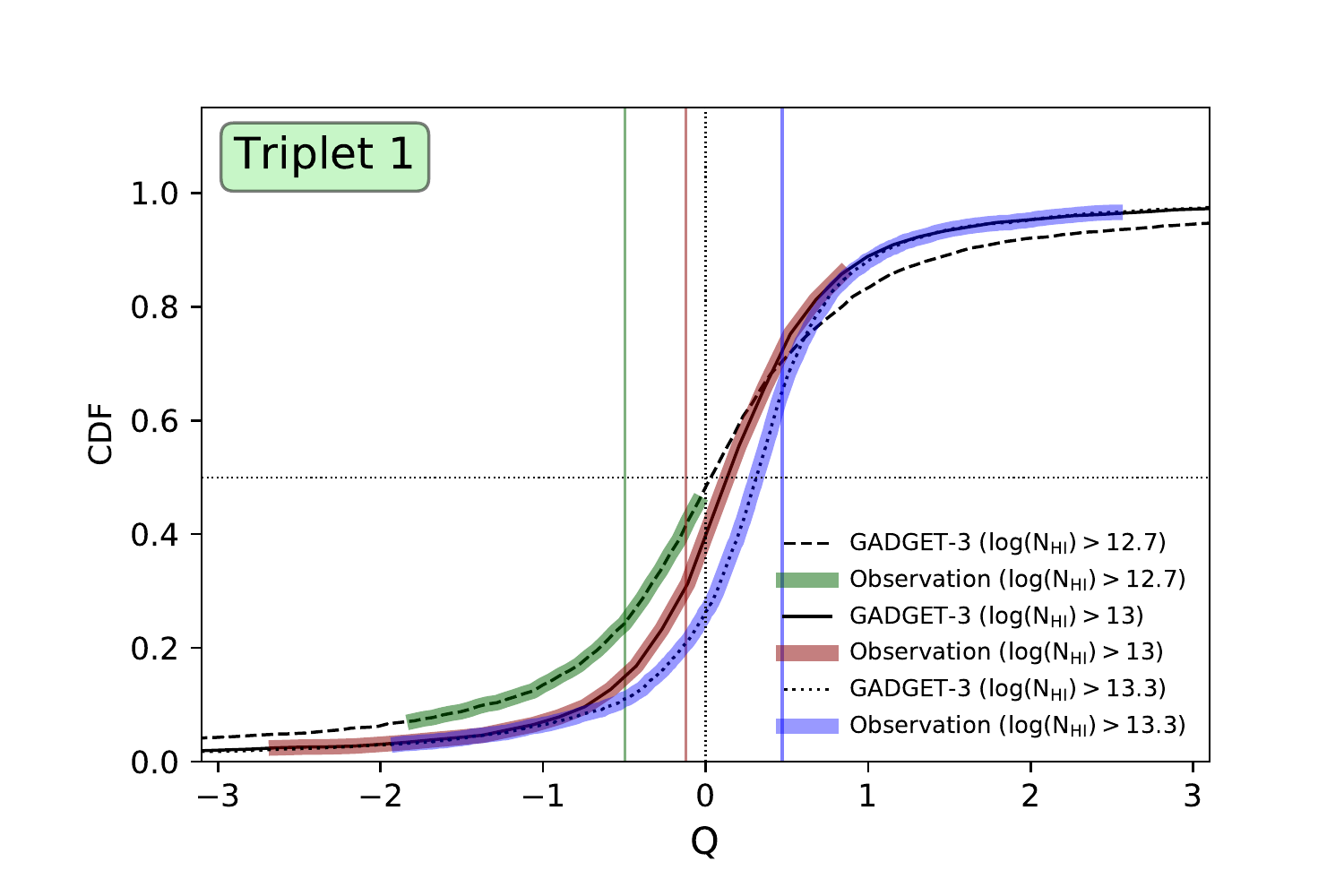}%
		\includegraphics[viewport=20 10 400 270,width=8.3cm, clip=true]{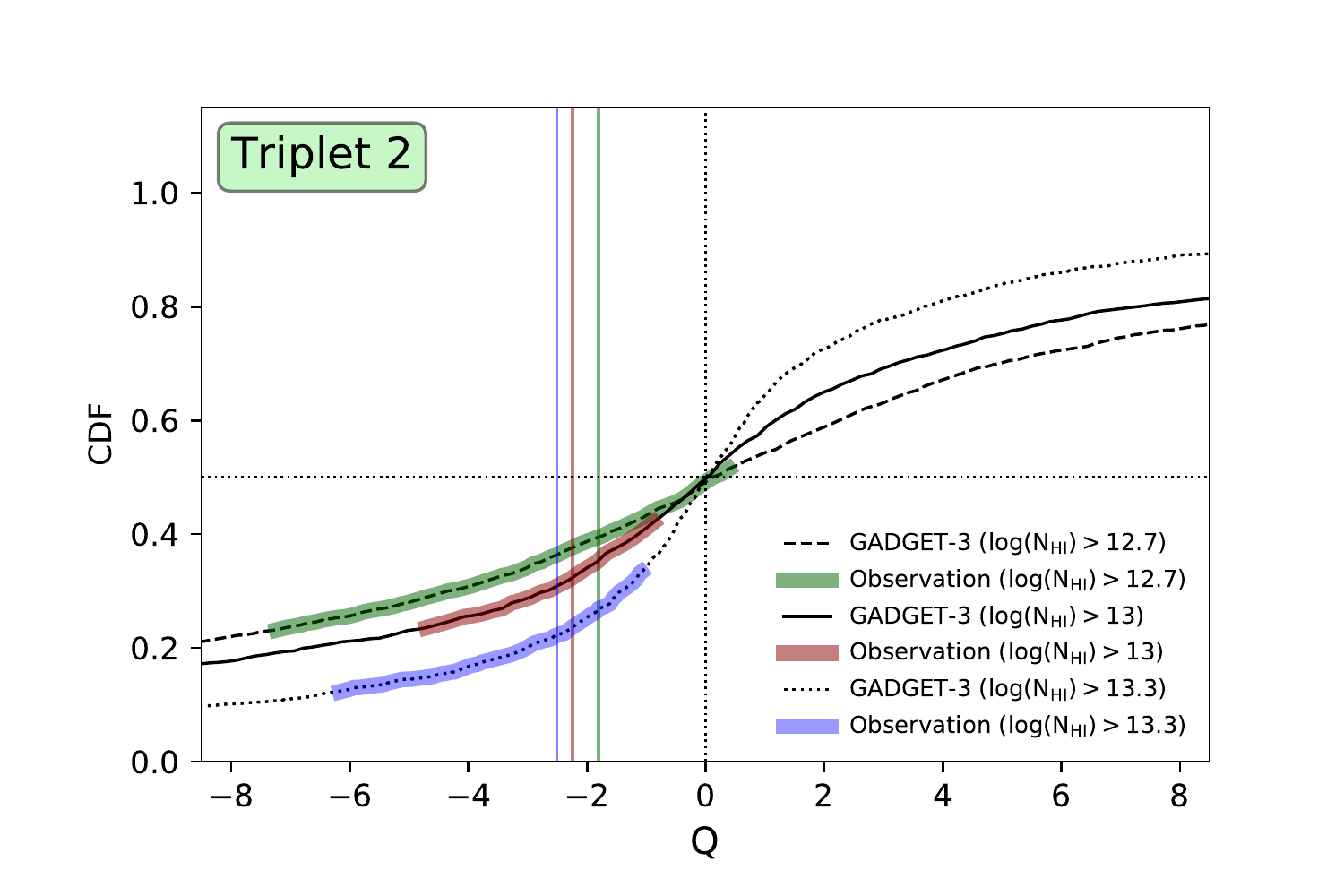}
	\end{minipage}%
	
	\caption{
	Reduced three-point correlation (Q) for the two triplets. The dashed, solid and dotted black lines represent the cumulative distribution function for different \HI\ column density cut-offs in simulations. The red, green and blue regions overlaying the simulated curve represents the $1\sigma$ confidence interval obtained from sub-sampling of the observed triplet sightlines. The vertical lines with colors denote the median observed Q value.
	}
	\label{fig_q}
\end{figure*}

In case of galaxies where the two-point correlation is high, one expects Q to be a positive quantity \citep[see][]{peebles1980}. This
need not be the case for IGM, where the clustering is weak.
In Fig~\ref{fig_q} we plot the cumulative distribution of Q for different $N_{\rm HI}$ thresholds.
The probabilities and percentiles associated with the observed Q values are given in Table~\ref{table_cloud_stat}.	Note that the observed value of Q mentioned in Table~\ref{table_cloud_stat} is taken to be the median value out of all the 1000 randomly sub-sampled skewers. This is done since Q can diverge when two-point correlations approach zero and thereby artificially boost
the mean value.

In the case of configuration similar to "Triplet 1", our simulations predict median Q value to be close to zero for clouds having log$(N_{\rm HI})>12.7$ with a large scatter. It is also evident that median Q value increases with increasing $N_{\rm HI}$. Also cumulative distribution for high $N_{\rm HI}$ clouds show smaller scatter. Increasing Q values with increasing $N_{\rm HI}$ is clearly evident even in observations. It is also evident from the Table~\ref{table_cloud_stat} that the observed Q distributions can be realised with high probability in our observations.

As expected based on the $\zeta$ values obtained above the median Q values are zero in the case of "Triplet 2" irrespective of whatever threshold column density we adopt.  Also the observed points are most of the time negative underlying the lack of strong correlation across the line of sights considered.

\subsubsection{Concurrent gap statistics in clouds}
\begin{table}
    \centering
    \caption{Results of concurrent gap statisitics in cloud}
    \setlength\tabcolsep{2.5pt}
    \begin{tabular}{ccccc}
\hline
Sample & $\rm log(N_{HI})$  & Largest gap & Wavelength  & \% of simulated\\
       &    threshold                         & ($h^{-1}$cMpc) &     range (\AA)     &  triplets      \\
\hline
\hline
Triplet 1 & 13 & 11.1 & 3596.6-3610.0  &  20.7\\
          & 13.5 & 25.4 & 3584.0-3614.2  &  5.8\\
          & 14 & 36.4 & 3614.7-3653.1  & 20.6\\
%          \\
Triplet 2 & 13 & 5.6 & 4150.0-4158.1  &  9.1\\
          & 13.5 & 7.6 & 4255.0-4267.1  &  41.1\\
          & 14 & 17.4 & 4148.7-4174.2  & 25.0\\
\hline
    \end{tabular}
    \label{table_cloud_gap}
\end{table}
In the cloud treatment of IGM, the definition of gap is more straightforward. The region between two clouds in a sightline is considered to be a gap. 
We searched for concurrent gaps as all the common regions in the three neighboring sightlines devoid of any clouds.

\begin{figure}
    \centering
    \includegraphics[viewport=10 23 330 260,width=8.5cm,angle=0,clip=true]{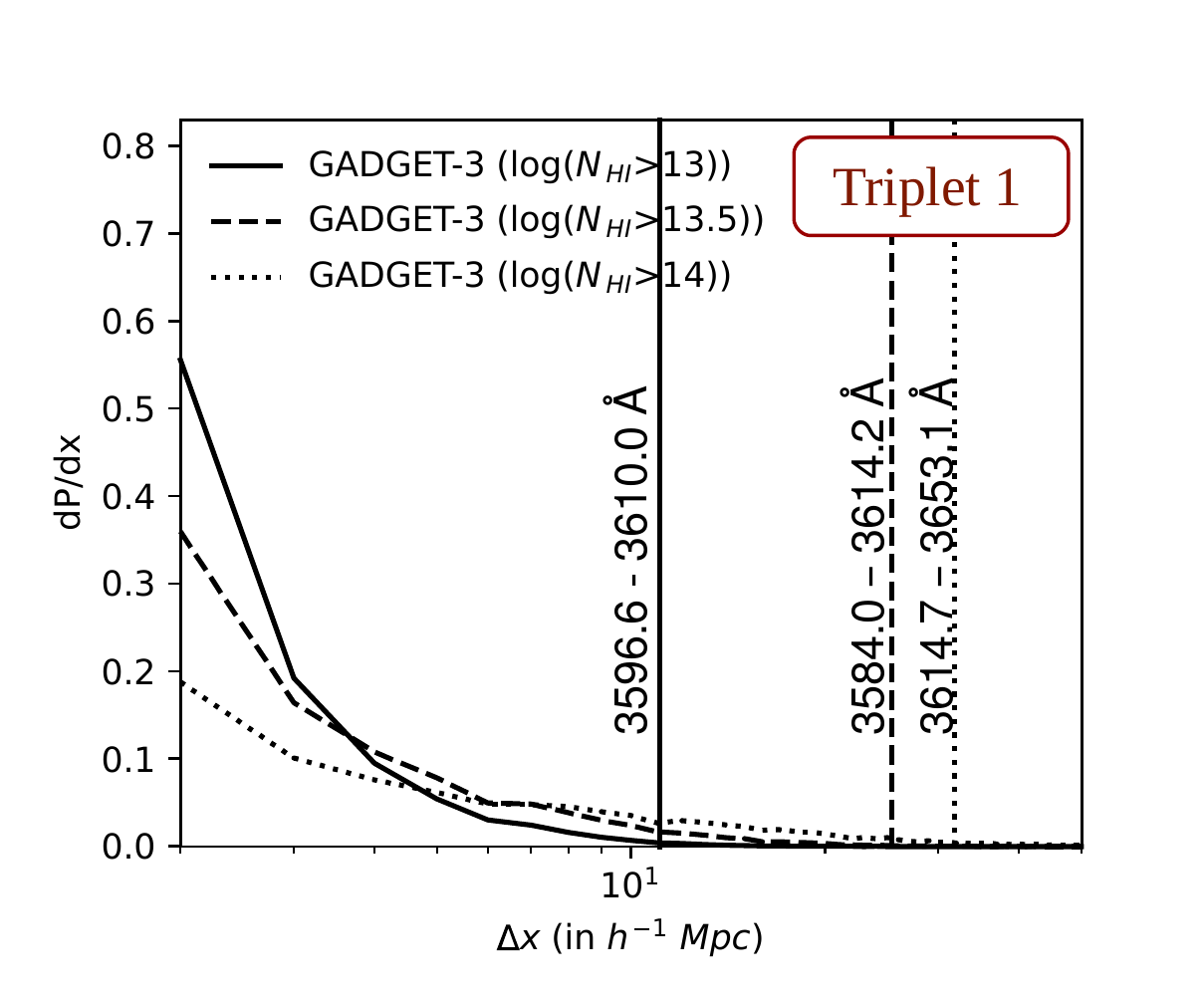}
    \includegraphics[viewport=10 10 330 260,width=8.5cm,angle=0,clip=true]{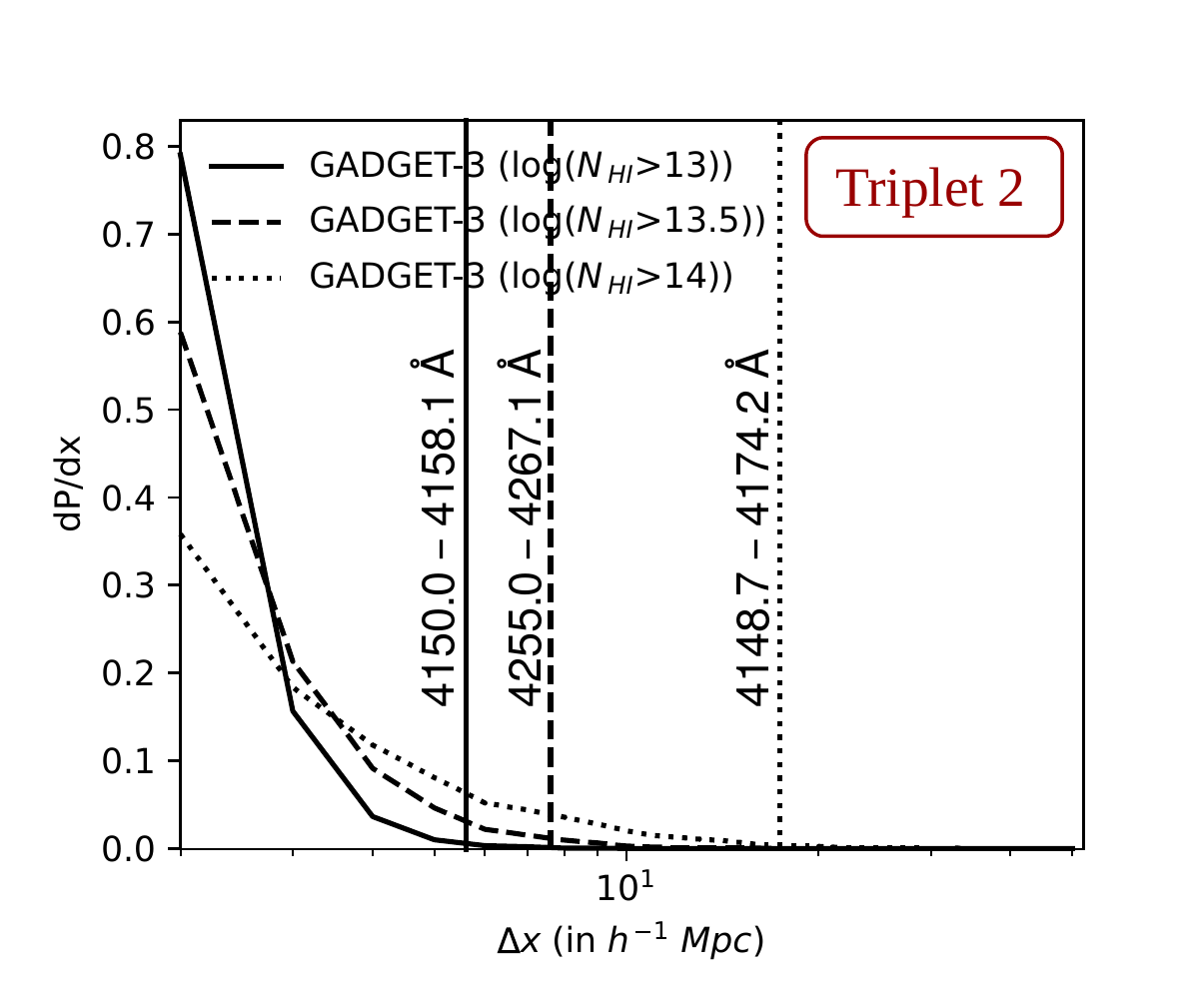}
    \caption{Concurrent gap distribution between clouds for different $N_{\rm HI}$ thresholds. The vertical lines with linestyle same as the gap distribution curves represent the largest concurrent gap (its wavelength range is provided) in the observed triplet for that corresponding $N_{\rm HI}$.}
    \label{Cloud_gappdf}
\end{figure}

Fig~\ref{Cloud_gappdf} shows the concurrent gap distribution between clouds obtained using 4000 simulated triplet sightlines for Triplet 1 and 2. 
These figures also show the gap probability distribution from our simulations for three different $ N_{\rm HI}$ threshold values. We also plot the largest concurrent gap sizes using vertical lines for different $ N_{\rm HI}$ thresholds. As expected the distribution shifts towards larger gap sizes with increasing $N_{\rm HI}$ thresholds and one picks up larger concurrent gaps more and smaller ones less. One can also see that Triplet 1 (which has smaller spatial separation in the projected space) picks up larger concurrent gaps more and smaller ones less as compared to Triplet 2.
We also mention the wavelength range of the largest concurrent gaps that one gets for different $ N_{\rm HI}$ thresholds. 
%These results are tabulated in 
In Table~\ref{table_cloud_gap}, we summarize the details of the largest gaps and associated  probability of occurrence. The probability given in the last of the table is the percentage of simulated triplet sightlines which have atleast one concurrent gap greater than the largest concurrent gap in the observations.
In case of Triplet 1, it is that the largest gap that one picks up for $log(N_{\rm HI})>13$ (11.1$h^{-1}$cMpc) is a subset of the gap that one picks up for $log(N_{\rm HI})>13.5$ (25.4$h^{-1}$cMpc). For $log(N_{\rm HI})>14$ , we pick the largest gap  (36.4$h^{-1}$cMpc) just adjacent to the previous gap. 
 In case of Triplet 2,  the largest concurrent gap that one picks up with $ log(N_{\rm HI})>13$ (5.6$h^{-1}$cMpc) is a subset of the gap that one picks up with $log(N_{\rm HI})>14$ (17.4$h^{-1}$cMpc). For $log(N_{\rm HI})>13.5$ , we pick the largest gap  (7.6$h^{-1}$cMpc) roughly 80\ang\ redward of the previous gap. 

\subsection{Correlation statistics}

Till now comparison of the three two-point and one three-point correlations for each triplet with simulations were done independent of each other.
Next we explore how closely our simulations can simultaneously reproduce all the three $\xi_s$ and $\zeta$ of a triplet.
In this exercise, instead of assigning average statistics for the entire observed triplet sightlines, we assign statistics to definite sections. These sections are obtained by uniformly selecting 10 regions from the observed sightline with redshift path length similar to the simulation. For each section we measure all the four correlation parameters $(\zeta,\xi_{01},\xi_{12},\xi_{02})$ which we denote by $x^i$ where $i=0,1,2$ and 3.  

We identify sightlines in the simulation which are closest to the observed sightlines in terms of these correlation parameters. For this, we define a \textit{closeness factor} $\rm C_4$ of the simulated sightlines with the observed sightlines using all the four correlation parameters as,
\begin{equation}
{\rm C_4}= \left(\sum_{i=0}^{i=3} \frac{(x^i-x^i_{obs})^2}{4 {\sigma}_i^2}\right)^{\frac{1}{2}} \ ,
\end{equation}
where $\sigma_i$ is the standard deviation of $i^{\rm th}$ parameter in the simulation. 
The simulated triplet which gives the smallest $\rm C_4$ with respect to the observations can be identified. 
We also define {\it closeness factor} ${\rm C_3}$ based only on the three
two-point correlation functions, i.e.,
\begin{equation}
{\rm C_3}= \left(\sum_{i=1}^{i=3} \frac{(x^i-x^i_{obs})^2}{3 \sigma_i^2}\right)^{\frac{1}{2}} \ .
\end{equation}
For each "Triplet", we identify the sections (based on flux- and cloud based statistics) with highest and lowest three-point function in our observations (given by id O-Max or O-Min in column 1 of Table~\ref{Table_C}).
The wavelength range for these observed sections are given in the second column of Table~\ref{Table_C}. In columns 3 to 6 of Table~\ref{Table_C}, we provide the three- and two-point correlation values. Next, we identify simulated sightlines with smallest $C_3$ and $C_4$ values (column 7 in Table~\ref{Table_C})  separately. These simulated sightlines are identified as Triplet-x-S3 and Triplet-x-S4 respectively in the table.
We also estimate the percentage of simulated sightlines having statistics similar to (i.e within 1 $\sigma_i$ for all statistics considered) the observations in the chosen section. 
These are provided for two cases; one considering only two-point statistics and the other considering both two- and three-point statistics (columns 9 and 10 in Table~\ref{Table_C}).

First we consider the flux based statistics for the segment with minimum $\zeta$ value (i.e having strong three-point correlation). It is evident that ${\rm C_3}$ and ${\rm C_4}$ based selection picked completely different simulated sightlines. We find that about 22.6\% of 4000 simulated triplet sightlines produce all the three $\xi$ values that matches with the observed values within their corresponding $1\sigma$ levels. When we demand similar matching also for $\zeta$ the percentage falls to 12.3\%. Even when we consider the segment with maximum $\zeta$ (or weak three-point correlation) the simulated sightline picked by ${\rm C_3}$ and ${\rm C_4}$ are not the same. However, the probability of sighlines producing the $\xi$ and $\zeta$ close to the observed values are higher than what we found for the segment with strong $\zeta$.

In the case of "Triplet 2" the segment with the strongest $\zeta$ is between 4167.5-4318.9 \AA\ (see Fig.~\ref{fig_J1055_spectra} in the Appendix). It is interesting to note from the table that $\xi_{02}$ (measured between the largest angular separation in the configuration) is larger than the other two $\xi$s.  The simulated triplet sightline picked based on ${\rm C_4}$ is also picked by ${\rm C_3}$. It is clear in our simulations that this configuration is very rare (i.e only 2\%  of the simulations produce all the correlations within their corresponding 1$\sigma$). As before when we consider the maximum $\zeta$ (i.e less correlated) segment ${\rm C_3}$ and ${\rm C_4}$ pick different simulated sightlines. We also find the probability of producing the high $\zeta$ sightlines are higher than low $\zeta$
sighlines.

Next we repeat the analysis for the cloud based (for log$(N_{\rm HI}) \ge$ 13) statistics. 
In the case of "Triplet 1" same segment has strongest three-point correlation function whether we use flux or cloud based statistics. However, unlike flux based statistics the probability of realising all four correlation function values within 1$\sigma$ of the observed values in our simulation turns out to be very low (i.e $\sim 3$\%). The segment with lowest three-point correlation function values is slightly shifted with respect to what we found based on flux statistics. However there is a considerable overlap between the two segments. Here also we find that ${\rm C_3}$ and ${\rm C_4}$ pick different simulated sightlines. Moreover the probabilities are less than what we find when we use flux statistics.

In the case of "Triplet 2", we do not obtain similar segments between flux and cloud approach based on our criteria of strongest three-point correlation. This is interesting since the segment which has the highest negative three-point correlation in flux (4167.5-4318 \AA) is supposed to be strongly correlated. But, when we go to cloud based analysis, we do not pick that region to have the strongest three-point correlation. Rather, in cloud based analysis, this particular region has a three-point correlation of -0.07. This discrepancy is probably seen because of a strong absorption present in one of the sightlines ($\lambda\sim 4290 \ang $) of "Triplet 2". In case of flux, presence of a strong absorption in one of the sightlines will make the three-point correlation largely negative when it correlates with either gaps or absorption in both the adjacent sightlines. This is not true for clouds as it gives positive three-point correlation only with coherent absorption in all the three sightlines.

\begin{table*}
\caption{Correlation Statistics}
	\begin{tabular}{lccccccccc}
		\hline
		\textbf{Triplet Id} & Wavelength ($\ang$) & $\zeta$     & $\xi_{01}$ & $\xi_{12}$ & $\xi_{02}$ & $\rm C_3$ & $\rm C_4$ &\multicolumn{2}{c}{Percent with} \\
		 & & & & & & & & $1\sigma_{i=1,2,3}$ & $1\sigma_{i=0,1,2,3}$ \\ 
		\hline
		\hline
		\multicolumn{10}{c}{\underline{\textbf{Flux based} }}\\
		Triplet1-O-Min & 3778.2-3900.2 & -0.0018 &  0.013          & 0.006                     & 0.004           & -           & -    & 22.6  &  12.375                       \\
		Triplet1-S4 &  & -0.0017 &   0.014         & 0.004                     & 0.003           &0.18          & \textbf{0.16} &   &                                     \\
		Triplet1-S3 &  & 0.0005 &    0.013        & 0.006                    & 0.004 		& \textbf{0.09}          & 0.71         &     &         \\

		Triplet1-O-Max & 3539.5-3661.5 & 0.0007 &  0.009          & 0.005                     & 0.004           & -           & -         & 37.5 &       33.5                 \\
	Triplet1-S4 &  & 0.0006 &   0.009        & 0.005                    & 0.003           & 0.13       & \textbf{0.11}  & &
		\\
		Triplet1-S3  & & -0.0003 &  0.009         & 0.004                     & 0.005           & \textbf{0.03}       & 0.31  & &
                               \\
		\\
%		\\
		Triplet2-O-Min & 4167.5-4318.9 & -0.0014 &  0.003          & 0.003                     & 0.019           & -           & -            & 2.0 &  1.9       \\
		
		Triplet2-S4-S3  & & -0.0017 &   0.003         & 0.007                    & 0.019           & \textbf{0.30}       & \textbf{0.27}  & &
		\\
		
%		\\
		Triplet2-O-Max & 4055.0-4206.4 & 0.0012 &  0.009          & 0.013                     & 0.005           & -           & -       & 31.6 &   24.2           \\
		Triplet2-S4 &  & 0.0009 &   0.008         & 0.013                     & 0.004           & 0.07       & \textbf{0.09}  & &
		\\
		Triplet2-S3  & & -0.0012 &  0.010         & 0.013                     & 0.005           & \textbf{0.06}       & 0.52  & &
		\\

%		\\
%		\\
		\multicolumn{10}{c}{\underline{\textbf{Cloud based} }}\\
	    Triplet1-O-Max & 3778.2-3900.2 & 0.05 &  0.11          & 0.10                     & -0.04           & -           & -      &  3.4  &     2.825                \\
		Triplet1-S4-S3 & - & 0.05 &   0.17         & 0.11                     & -0.05           &\textbf{0.19}          & \textbf{0.17} &    &                                       \\
%		\\
		Triplet1-O-Min & 3500-3622 & -0.24 &    0.30        & 0.37                    & 0.29 		& -          & -        &  33.2   &     9.825           \\
        Triplet1-S4 & - & -0.21 &    0.31        & 0.39                    & 0.27 		& 0.12          & \textbf{0.13}  & &
        \\
        Triplet1-S3 & - & 0.05 &    0.29        & 0.37                    & 0.30 		& \textbf{0.05}          & 0.70    &     &                    \\

		Triplet2-O-Max & 4069-4220.4 & -0.01 &  0.12          & 0.15                     & 0.00           & -           & -          & 12.3 & 8.5              \\
		Triplet2-S4-S3 & - & -0.01 &   0.12         & 0.16                     & -0.01           &\textbf{0.05}          & \textbf{0.12}  &        &                                   \\

		Triplet2-O-Min & 4139.4-4290.8 & -0.08 &    0.15        & 0.14                    & 0.05 		& -          & -       &  14.0   &   4.15              \\
        Triplet2-S4 & - & -0.07 &    0.16        & 0.15                    & 0.07 		& 0.20          & \textbf{0.22}  & &
        \\
        Triplet2-S3 & - & 0.04 &    0.14        & 0.13                    & 0.05 		& \textbf{0.08}          & 1.23  &      &                     \\

\hline
	\end{tabular}
\label{Table_C}
\end{table*}

\section{Correlations between \lya, \CIV, DLAs and Quasars}

\subsection{Quasar -\lya\ absorption transverse correlations}
	
\begin{figure}
    \centering
    \includegraphics[viewport=30 60 545 800,height=8.9cm,angle=270,clip=true]{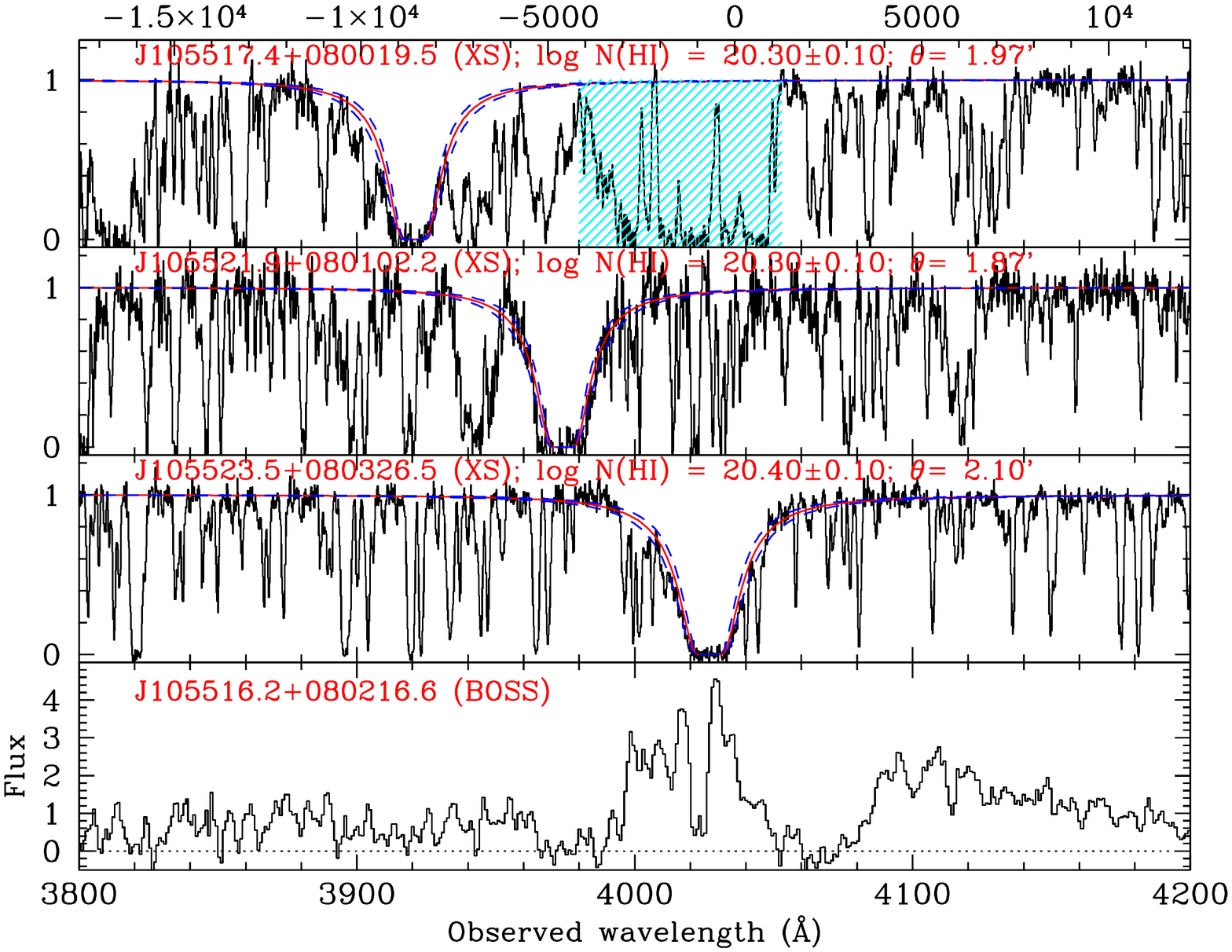}
    \includegraphics[viewport=30 30 560 800,height=9cm,angle=270,clip=true]{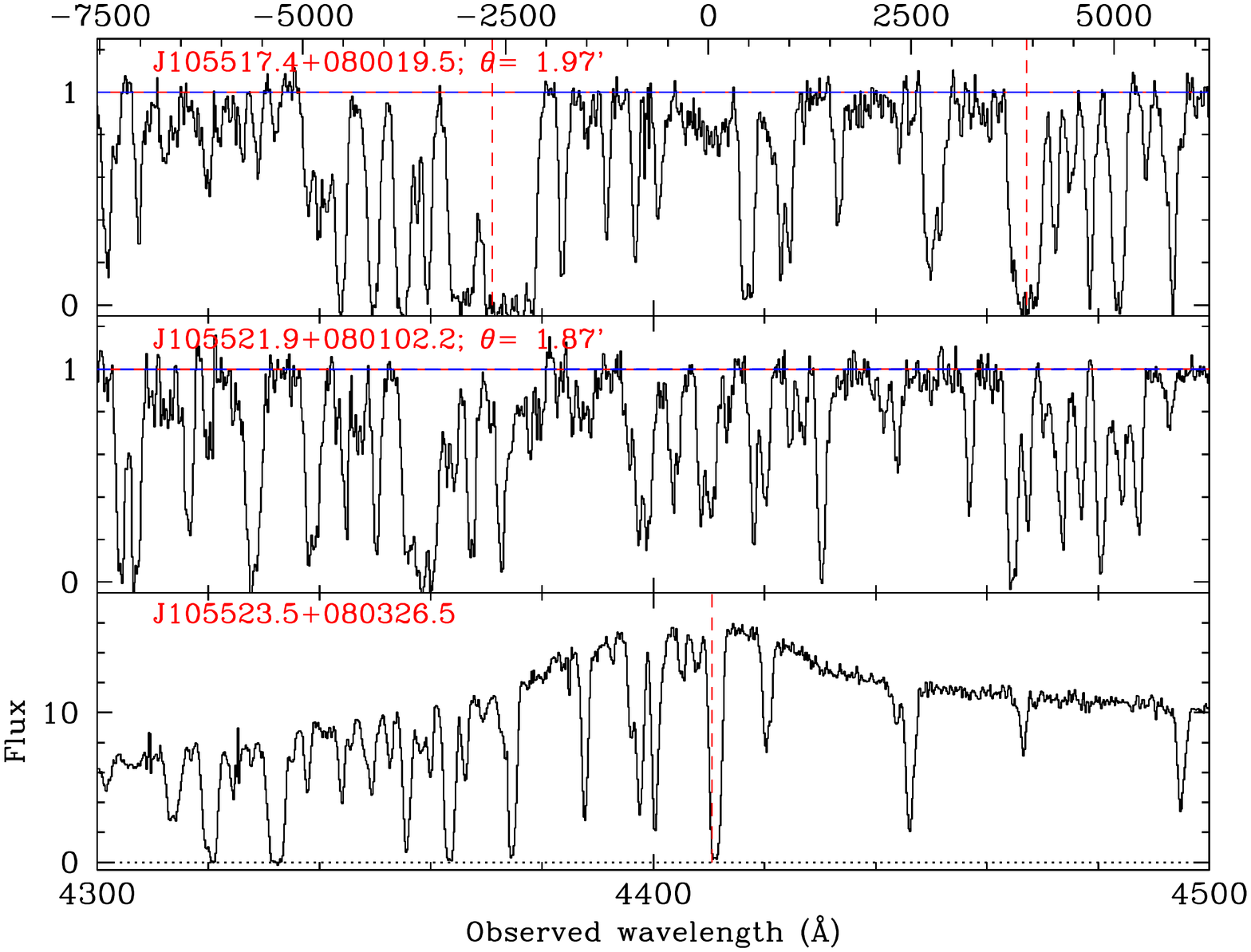}
\includegraphics[viewport=30 30 600 800,height=9cm,angle=270,clip=true]{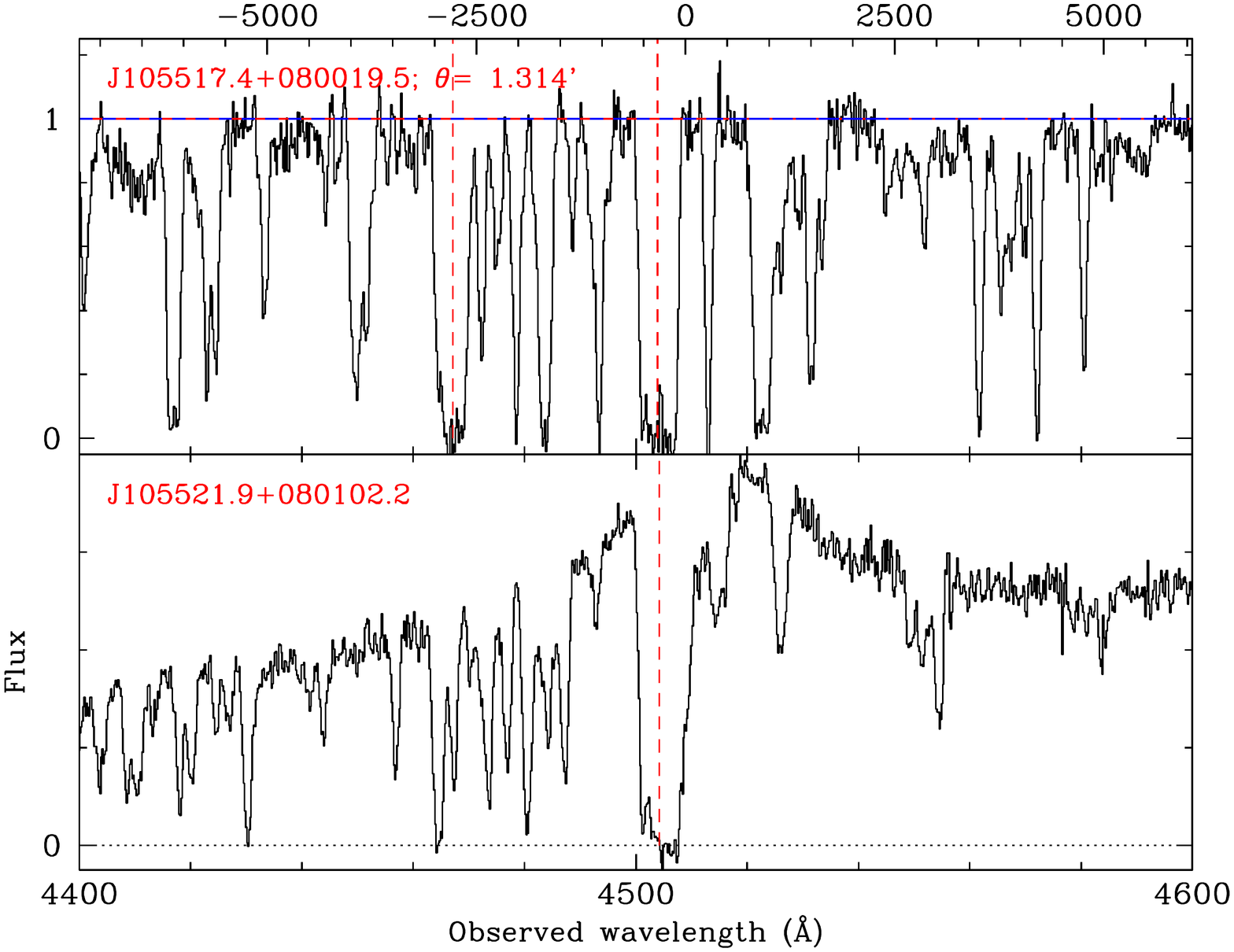}
    \caption{{\it Top:} \lya\ forest towards quasars constituting "Triplet 2" around the foreground quasar J105516.2+080216.6. DLAs are detected along all the three sightlines with redshifts close to the redshift of the foreground quasar. The fit to the DLAs are shown, the measured \HI\ column densities and the angular separation of the sightline with respect to the foreground quasars are given in each panel. Shaded region is the contamination by associated \OVI\ absorption.
    {\it Middle:} \lya\ forest towards J105517.4+080019.5 and J105521.9+080102.2 around the foreground quasar J105523.5+080326.5. 
    {\it Lower:} \lya\ forest J105517.4+080019.5 around the foreground quasar J105521.9+080102.2.
    In all cases the vertical dotted line gives the location \lya\ lines having associated the \CIV\ absorption.
    }
    \label{fig:DLA}
\end{figure}

\begin{table}
    \centering
    \caption{Quasar-\lya\ transverse correlation.}
    \setlength\tabcolsep{4.5pt}
    \begin{tabular}{cccc}
    \hline
    Quasar LOS & \lya\ LOS & Proximity & No. of  \\
     &  & Region(\ang) & absorbers \\       
    \hline
    \hline
    \multicolumn{4}{c}{\underline{\textbf{$\rm log(N_{HI})>13$} }}\\
  	J1055+0803 & J1055+0801 & $4490.2\pm 11.3$ & 7($3.7\pm 1.9$) \\
    J1055+0803 & J1055+0800 & $4490.2\pm 10.4$ & 8($4.5\pm 1.8$) \\
	J1055+0801 & J1055+0800 & $4508.9\pm 10.9$ & 1($4.8\pm 1.9$) \\
%	\\
	J1418+0700 & J1418+0657b & $3927.2\pm 4.7$ & 2($1.4\pm 1.1$) \\
%	\\
	\multicolumn{4}{c}{\underline{\textbf{$\rm log(N_{HI})>14$} }}\\
    J1055+0803 & J1055+0801 & $4490.2\pm 11.3$ & 3($1.8\pm 1.3$) \\
	J1055+0803 & J1055+0800 & $4490.2\pm 10.4$ & 3($1.1\pm 1.1$) \\
	J1055+0801 & J1055+0800 & $4508.9\pm 10.9$ & 1($1.2\pm 1.2$) \\
%	\\
	J1418+0700 & J1418+0657b & $3927.2\pm 4.7$ & 0($0.8\pm 1.0$) \\
	\hline
    \end{tabular}
    \label{tab:quasar_lya}
\end{table}

In this section, we consider the correlation between the foreground quasars and \lya\ forest along the LOS to the background quasars. In the framework of standard proximity effect one expects radiation induced voids in the transverse directions as $r_{eq}$ for most of our quasars are larger than the transverse separations probed (see Table~\ref{Tab_obs}). However, studies of quasar pairs have revealed excess \lya\ absorption in the transverse direction \citep[][]{rollinde2005,guimaraes2007,kirkman2008,Prochaska2013,lau2016, jalan2018}. Triplets allow us to probe the gas around QSOs through more than one sightlines.

When we consider spectra of all the three triplets we have (including the BOSS spectrum of J141831.7+065711.2) we cover a redshift path length $\Delta z$ of $\sim 6.9$ for detecting DLAs. From the DLA frequency distribution (i.e dN/dz) given in the table 2 of \citet{noterdaeme2012}  we expect 1.3 DLAs, while we detect 3 DLAs in Triplet 2.  Along each sightline in "Triplet 2", we cover a redshift path length of $\Delta z = 1$ where the expected number of DLAs is 0.2. Detection of a DLA along each sightline correspond to a factor 5 excess compared to the expected number of randomly distributed DLAs.
Interestingly the redshifts of the DLAs at the maximum differ by $\Delta_z = 0.088$ (or $\Delta v = 6000$ \kms; See Fig.~\ref{fig:DLA}). Given the presence of a DLA the expected number of DLAs in the other two sightline having a redshift difference within 0.088 is 3$\times10^{-4}$ (i.e $(0.088\times0.2)^2$). This implies that the chance coincidence of three DLAs with close redshift separation is less probable.

What is making this case more interesting is the fact that the fourth quasar (i.e J105516.2+080216.6 with \zem = 2.320) we identified in this field  roughly has equal angular separation with respect to the quasars in "Triplet 2". The \lya\ emission from this quasar coincides with the DLA seen along the line of sight to J1055+0803 (see Fig.~\ref{fig:DLA}). This quasar is really faint and its sphere of influence for \HI\ ionization (see last column of Table~\ref{Tab_obs}) is much smaller than separation of the other three sightlines from the quasar. Note while the presence of quasar and DLAs within a small redshift range and angular separation is interesting, the redshift space separation of $\sim 37$ pMpc may seem too large. It will be interesting to perform deep imaging in this regions to search for a possible presence of large scale density enhancements.

Based on the $r_{eq}$ provided in Table~\ref{Tab_obs}, we identify the wavelength range in the transverse direction that will be affected by isotropic  ionizing radiation from the foreground quasars. For each quasar pairs (listed in first two columns) these are listed in the third column of the Table~\ref{tab:quasar_lya}. 
The number of clouds having $N_{\rm HI}$ above a given threshold found within these wavelength range are given in last column of this table. In this column we also provide average number of clouds and associated errors found for this  similar wavelength range in our spectrum far away from quasars. 

Two background quasar sightlines [J1055+0800 (with a separation of 1.6 pMpc) and J1055+0801 (with a separation of 1.2 pMpc)] probe the proximity region of the foreground quasar J1055+0803. The \lya\ absorption along these sightlines are shown in the middle panel of Fig.~\ref{fig:DLA}. The vertical dashed lines mark the locations of \lya\ clouds with associated \CIV\ absorption. In the case of J1055+0803, we find a factor 1.8 time more absorption along J1055+0800 and J1055+0801 when we consider clouds having log$(N_{\rm HI})>13$. While based on standard proximity effect we would have expected deficit of absorption. When we consider log$(N_{\rm HI})>14$, the observed excess is 1.7 and 2.7 times the expected value towards J1055+0801 and J1055+0800 respectively. 
Given the small wavelength range probed these excess are not statistically significant. 

In the bottom panel of Fig~\ref{fig:DLA} we show the \lya\ absorption towards J1055+0800 in the proximity of the foreground quasar J1055+0801 (at a separation of 640 pkpc). It is clear from the figure that the foreground quasar has a strong associated absorption at \zabs = 2.7051. The associated \CIV\ absorption shows three components with absorption redshifts of 2.7030, 2.7043 and 2.7052. While we detect \lyb\ and \SiIV\ absorption, neither low ions like \SiII\ nor high ions like \OVI\ or $N$~{\sc v} are clearly seen in absorption at the redshifts of the \CIV\ components. 
We see \lya\ absorption and associated \CIV\ at the same redshift along the line of sight to J1055+0800. While \lya\ absorption profile matching is very good the \CIV\ absorption shows three components (at \zabs = 2.7051, 2.7064 and 2.7075) that are slightly redshifted with respect to absorption seen towards J1055+0801. Interestingly we detect \CII, \SiII, \SiIII\ and \SiIV\ absorption in addition to possible \OVI. 
These are the closest sightlines (with a physical separation of 635 pkpc) in "Triplet 2". 
From Table~\ref{tab:quasar_lya} we notice that within the proximity region considering the redshift of the quasar and not the above discussed associated \CIV\ system, we do find deficit of \lya\ absorption for log$N_{\rm HI}$$>$13 and no excess for log$N_{\rm HI}$$>$14.

\begin{figure}
    \centering
\includegraphics[viewport=30 30 600 800,height=9cm,angle=270,clip=true]{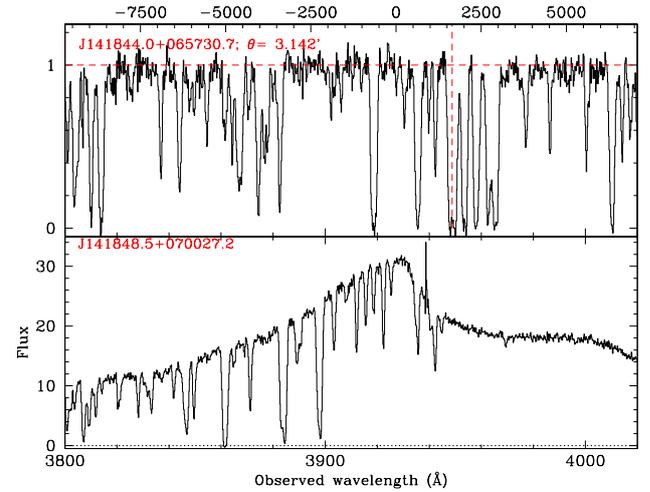}
    \caption{\lya\ forest towards quasar J141844.0+065730.7 around the foreground quasar J141848.5+070027.2.}
    \label{fig:DLA1}
\end{figure}

In Fig.~\ref{fig:DLA1}, we plot the \lya\ forest region along the line of sight to J141844.0+065730.7 around the foreground quasar J141848.5+070027.2.  The statistics of the \lya\ absorption within the expected proximity along the foreground quasar are summarised in Table~\ref{tab:quasar_lya}. While there is excess \lya\ suggested from this table, it is evident from the figure that there is a strong clustered \lya\ absorption withing the velocity range $-1000$ to $+3000$ \kms. One of this components at $z=2.2483$ (identified with red dashed line) shows \CIV\ and \MgII\ absorption. While we do not have X-Shooter spectrum of the third quasar J141831.7+065711.2 in "Field 3", the available SDSS spectrum shows strong \lya\ absorption in the same identified velocity range. Thus there appears to be a overdense region at slightly higher redshift around this quasar. 
 
In summary, in all the triplet sightlines considered here we do find signature of excess \lya\ absorption around quasars. Previous studies have shown excess absorption at the redshift of the foreground quasars in the spectrum of background quasars. Here we are able to probe the quasar environment using more than one quasar sightlines. Such studies using large number of triplets (or multiple sightlines) will allow us to probe either the geometry of the gas distribution or the nature of the ionizing radiation (isotropic or an-isotropic).  In principle one will be able to constrain the time-scale related to the quasar activities much better than what one could do with doublets.
 
\subsection{DLA-\lya\ absorption transverse correlations}

\begin{figure}
    \centering
\includegraphics[viewport=30 30 560 800,height=9cm,angle=270,clip=true]{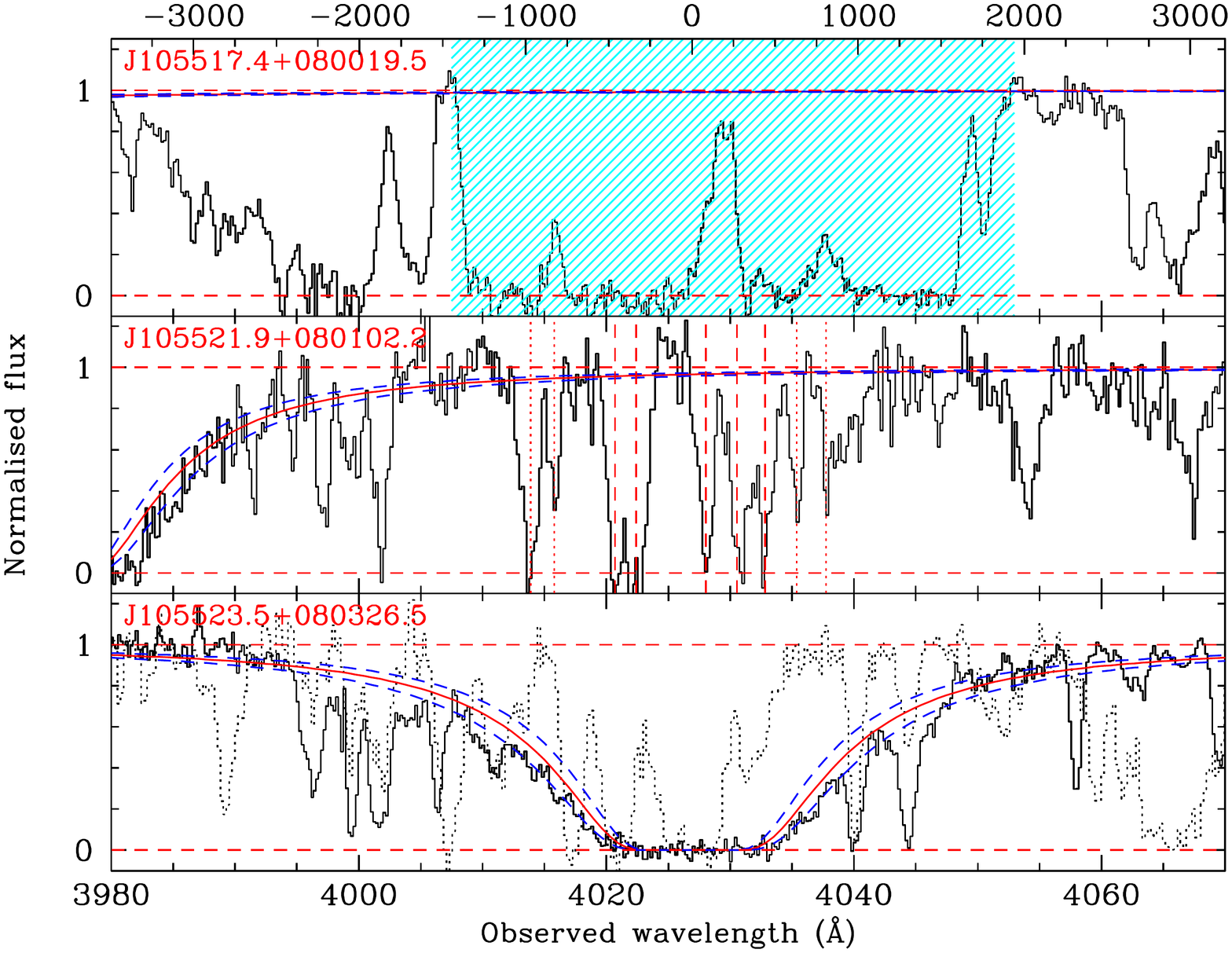}
\includegraphics[viewport=30 30 560 800,height=9cm,angle=270,clip=true]{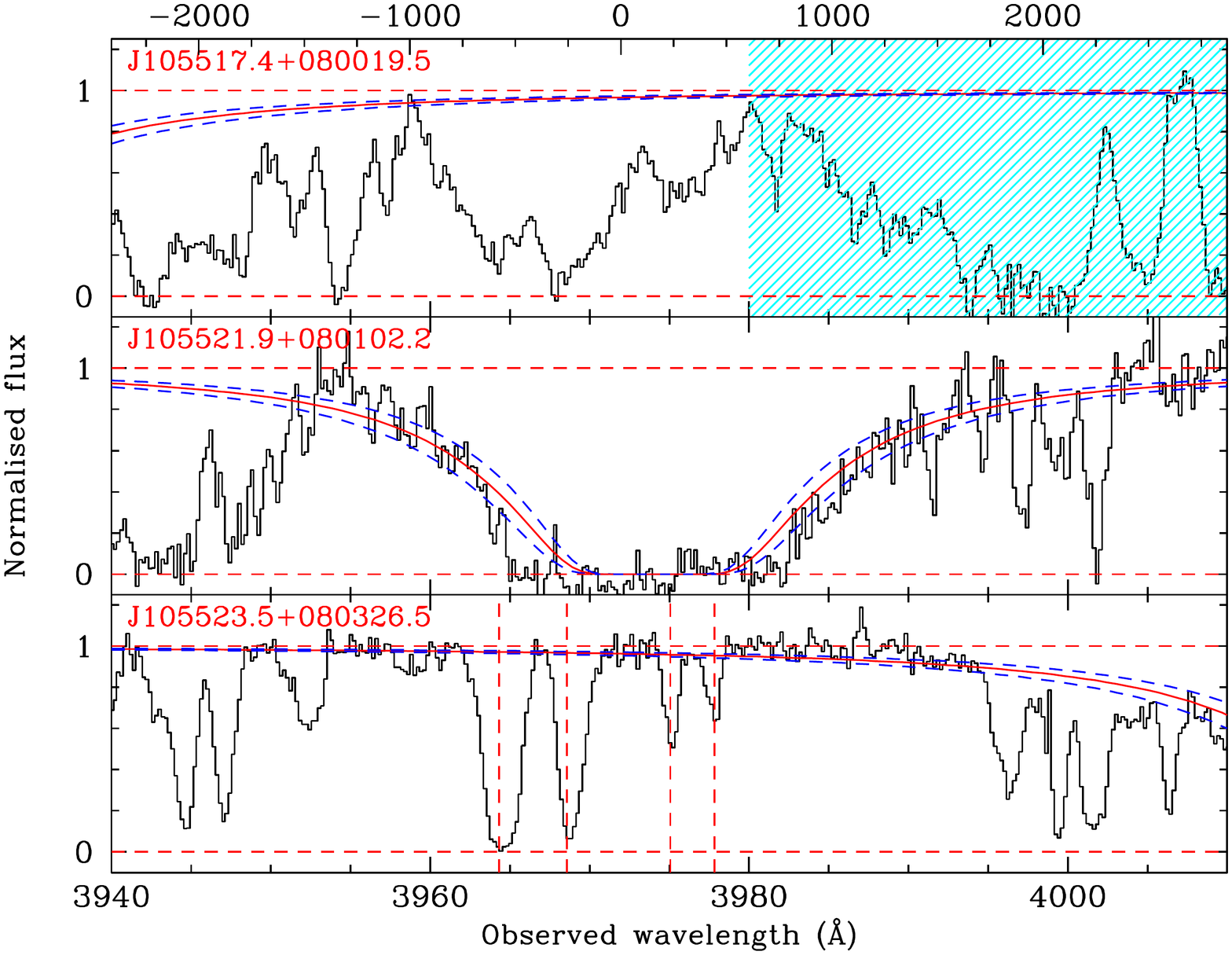}
\includegraphics[viewport=30 30 600 800,height=9cm,angle=270,clip=true]{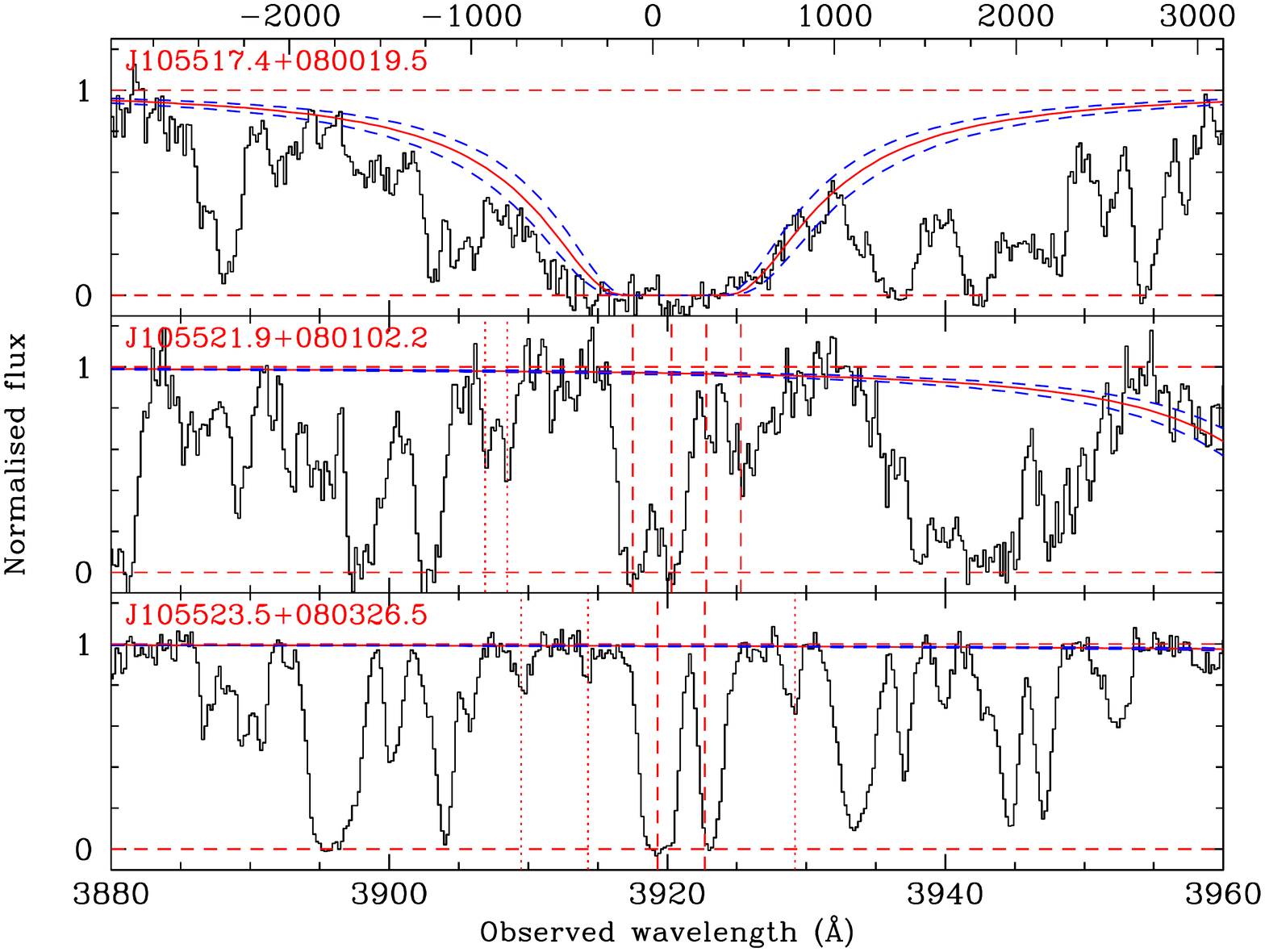}
\caption{Transverse correlations between DLAs and \lya\ absorption in "Triplet 2". Vertical dashed and dotted lines identify \lya\ absorption within a velocity separations of 600 \kms  and between 600-1000 \kms respectively. The coincident absorption from other absorption systems are masked with shaded regions.}
    \label{fig:DLA_lya}
\end{figure}

\begin{table}
    \centering
    \caption{DLA-\lya\ transverse correlation.}
    \begin{tabular}{ccccc}
    \hline
    DLA LOS & \lya\ LOS &\multicolumn{2}{c}{Number of absorbers} \\
            &           & $|v|< 600$\kms & $|v|< 1000$\kms\\
    \hline
    \hline
    \multicolumn{4}{c}{\underline{\textbf{$\rm log(N_{HI})>13$} }}\\
 	J1055+0800 & J1055+0801 &4($2.7\pm 1.6$) &7($6.0\pm 1.9$)\\
	J1055+0800 & J1055+0803 &4($3.7\pm 1.8$) &7($4.5\pm 2.2$) \\ 
    J1055+0801 & J1055+0803 &4($3.7\pm 1.8$) &6($4.5\pm 2.2$)\\
	J1055+0803 & J1055+0801 &3($2.7\pm 1.6$) &8($6.0\pm 1.9$)\\
%	\\
			\multicolumn{5}{c}{\underline{\textbf{$\rm log(N_{HI})>14$} }}\\
 	J1055+0800 & J1055+0801 & 2($1.3\pm 1.1$) & 3($2.1\pm 1.4$)\\
	J1055+0800 & J1055+0803 & 3($0.9\pm 1.1$) & 3($1.5\pm 1.3$)\\
		J1055+0801 & J1055+0803 & 2($0.9\pm 1.1$) & 3($1.5\pm 1.3$)\\
		J1055+0803 & J1055+0801 & 3($1.3\pm 1.1$) & 7($2.1\pm 1.4$)\\

	\hline
    \end{tabular}
    \label{tab:dla_lya}
\end{table}

\citet{Rubin2015} have studied the gas distribution around high-$z$ DLAs using absorption lines detected in QSO sightlines within a transverse separations of 300 kpc to 40 $z\sim 2$ DLAs. They found optically thick \HI\ absorption (i.e log $N_{\rm HI}\ge$ 17.3) up to 200~kpc with a covering fraction of $\ge 30$\%. Low ionization metals traced by Si~{\sc ii}$\lambda$ 1526 are found to have a covering fraction of 20\% within 100 kpc, while high ions traced by \CIV\ absorption seem to have higher covering factor ($\sim 57$\%) and kinematically coupled to the DLA to a larger transverse separations ($\sim$ 200 kpc) compared to the low ions.

In this section we study the DLA-\lya\ transverse correlation considering the three DLAs detected along the lines of sight towards quasars in "Triplet 2".
Note the separations we probe are at least a factor two higher than that of \citet{Rubin2015} however we have a unique opportunity to probe the gas distribution around 3 DLAs simultaneously using two sightlines.

In the bottom panel of Fig~\ref{fig:DLA_lya} we plot the \lya\ absorption towards J105521.9+080102.2 (at a projected separation of $\sim 640$ pkpc) and J105523.5+080326.5 (at a projected separation of 1.6 pMpc) at the redshift of the DLA detected along the line of sight to J105517.4+080019.5. We fit the \lya\ lines within $\pm1000$ \kms to the DLA in the redshift space with multiple component Voigt profiles.  Along J105521.9+080102.2 we find one component having 
$10^{16}\le N_{\rm HI}{\rm [cm^{-2}]\le 10^{17}}$ within $\pm 600$ \kms to the DLA. In the case of 
J105523.5+080326.5 we have three such components 
detected within $\pm 1000$ \kms.
We do not detect Si~{\sc ii} $\lambda$ 1526 or \CIV\ associated with these absorbers or any other \lya\ absorption within $\pm$ 1000 \kms to the DLA.

In Table~\ref{tab:dla_lya} we present our results for number of clouds observed and predicted by the random distribution for two different velocity intervals centered around the DLAs.  First two columns in this table indentify the DLA and \lya\ sightlines. Third and fourth columns give the number of cloud detected within $\pm 600$ \kms and $\pm 1000$ \kms respectively. The numbers in the bracket are the expected number of \lya\ absorption within the specified velocity range if we go to a random place in the observed \lya\ forest.  It is clear that the observed numbers are within 1.5 $\sigma$ to the expected number. However, detecting partial Lyman limit systems (i.e $N_{\rm HI}>10^{16}$ cm$^{-2}$) along both the sightlines is interesting.
We come back to this after looking at \lya\ absorption around other two DLAs.

In the middle panels in Fig.~\ref{fig:DLA_lya} we plot the \lya\ absorption towards J105517.4+080019.5 (transverse separation of 640 pkpc) and J105523.5+080326.5 (transverse separation of 1.2 pMpc) at the redshift of the DLA along the lines of sight to J105521.9+080102.2. The shaded region gives the wavelength affected by \lyb\ of the strong associated absorption seen along the line of sight to J105517.4+080019.5. Within $\pm 1000$ \kms we notice broad absorption with three visible components. We tentatively identify this to O~{\sc vi} $\lambda$ 1037 associated to a broad feature we identify based on \CIV\ absorption at \zabs = 2.8225. We call this tentative as profiles of \OVI\ doublets are affected by narrow intervening \lya\ absorption. While we see the two profiles do not match perfectly they do have consistent optical depths as expected. This means that we will not be able to get a handle on number of \lya\ absorptions within our velocity range of interest. However, we can conclude that there is no cloud with $N_{\rm HI}>10^{16}$ cm$^{-2}$. Along the line of sight to J105523.5+080326.5 we detect 6 \lya\ components within $\pm 1000$ \kms with $N_{\rm HI}>10^{13}$ cm$^{-2}$ (there is one component with $N_{\rm HI} = 10^{16.9}$ cm$^{-2}$ within $\pm 1000$ \kms but none with $N_{\rm HI} > 10^{16}$ cm$^{-2}$ within $\pm 600$ \kms). However as can be seen from the Table~\ref{tab:dla_lya} the observed number are consistent with the expectations.

In the top panel in Fig~\ref{fig:DLA_lya} we plot the \lya\ absorption towards J105517.4+080019.5 (transverse separation of 1.6 pMpc) and J105521.9+080102.2 (transverse separation of 1.2 pMpc) at the redshift of the DLA detected along the line of sight to J105523.5+080326.5. Unfortunately strong associated \OVI\ absorption (see the cyan shaded region) completely blanket the expected \lya\ wavelength range along the line of sight to J105517.4+080019.5. The \lyb\ range is also in the low SNR region so we will not be able to probe the DLA-\lya\ cross correlation along this sight line.  Along the 
line of sight towards J105521.9+080102.2 we find several \lya\ absorption within a velocity range of $\pm$ 1000 \kms.
In this case also we do not find any metal line associated to the \lya\ absorption. The highest H~{\sc i} column density detected is $N_{\rm HI} = 10^{16.6}$ cm$^{-2}$.
 When we consider clouds with log~$N_{\rm HI}$ $\ge$13.0 the observed number of absorption is consistent with the expected number. However, if we consider only clouds with log~$N_{\rm HI}$ $\ge$14.0 the observed number is consistent with the expected value within 1.5 $\sigma$ level for $\delta_v<600$ \kms and at 3.5$\sigma$ for $\delta_v<600$ \kms.

  %As mentioned before 
  The detection of partial LLS close to the DLAs is intriguing. We have detected 5 such absorbers within $\pm 600$ \kms and 7 absorbers within $\pm 1000$ \kms  to 3 DLAs along 5 LOS probed.
  Taken on the face value we detect at least one cloud with $N_{\rm HI}\ge 10^{16}$ cm$^{-2}$ in 60\% of DLAs within $\pm$ 600 \kms and in 80\% of DLAs within 1000 \kms  within a project separations of 0.6 to 1.6 pMpc.
  Based on the  redshift distribution of H~{\sc i} absorbers \citep[given in][]{kim2013} we expect $\sim0.1$ such absorbers in the redshift range searched within $\pm$ 600 \kms and $\sim 0.17$ within $\pm$ 1000 \kms. This together with the fact that the absorption occur close to a choosen DLA favors excess clustering of high column density 
  \lya\ absorbers around DLAs over the projected separatios 0.6 to 1.6 pMpc. Given the length scale probed these absorbers may not belong to the cirumgalatic medium of the quasars as studied by \citet{Rubin2015} but may reflect biased clustering of the IGM around DLAs\citep[see for example,][based on SDSS BOSS data]{perez2018}.

\subsection{\CIV\ correlations}

\begin{figure*}
		\includegraphics[viewport=12 24 320 270,width=7.2cm, clip=true]{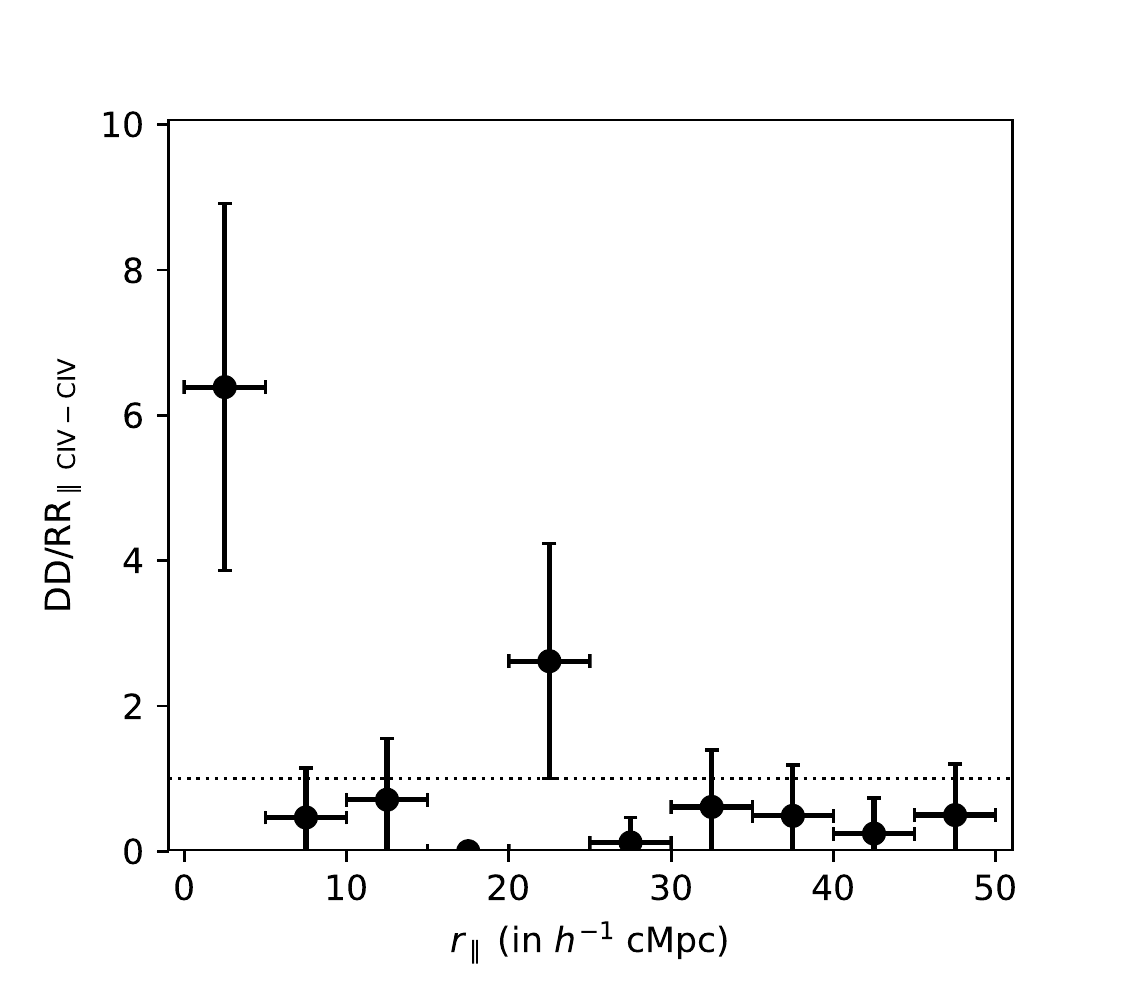}%
		\includegraphics[viewport=12 24 320 270,width=7.2cm, clip=true]{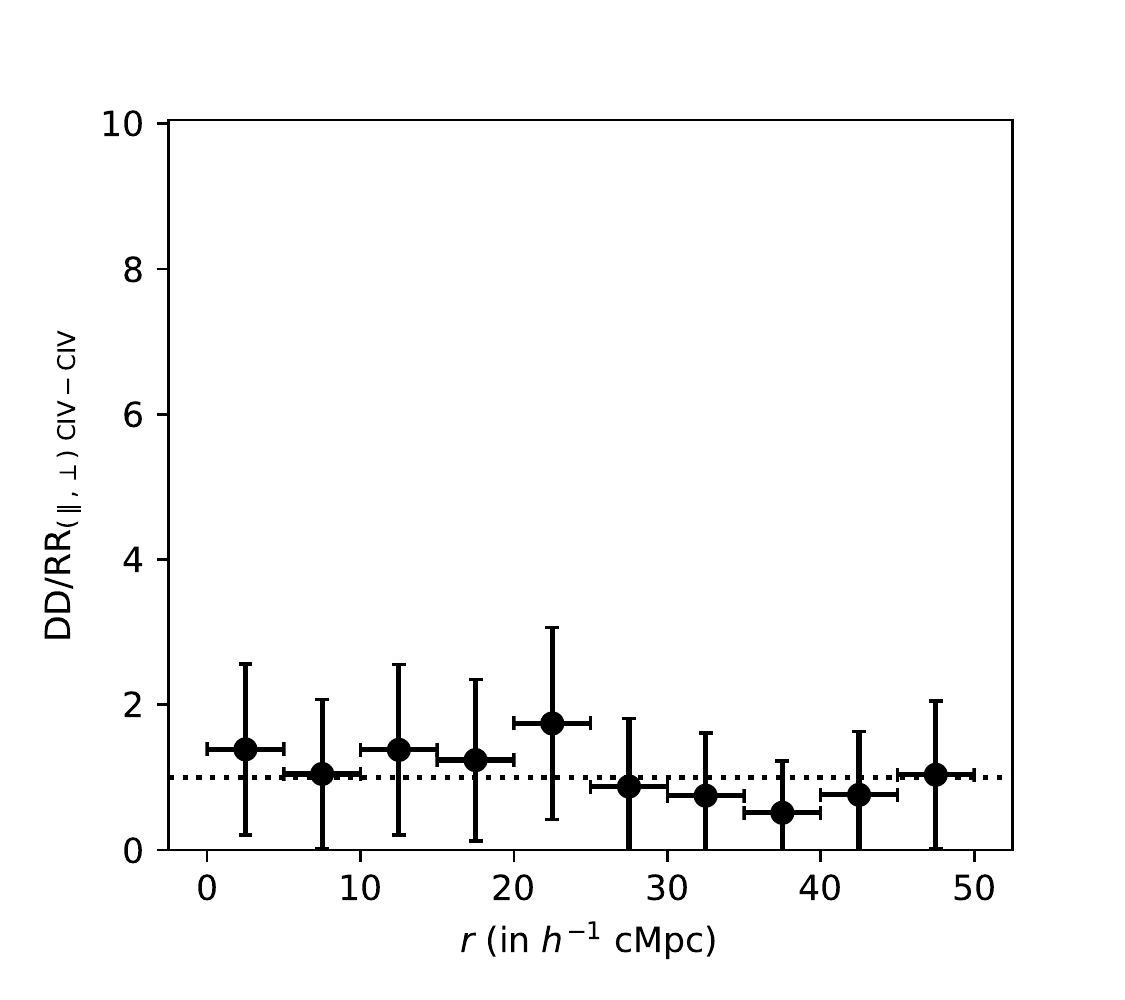}
		
		\includegraphics[viewport=12 10 320 270,width=7.2cm, clip=true]{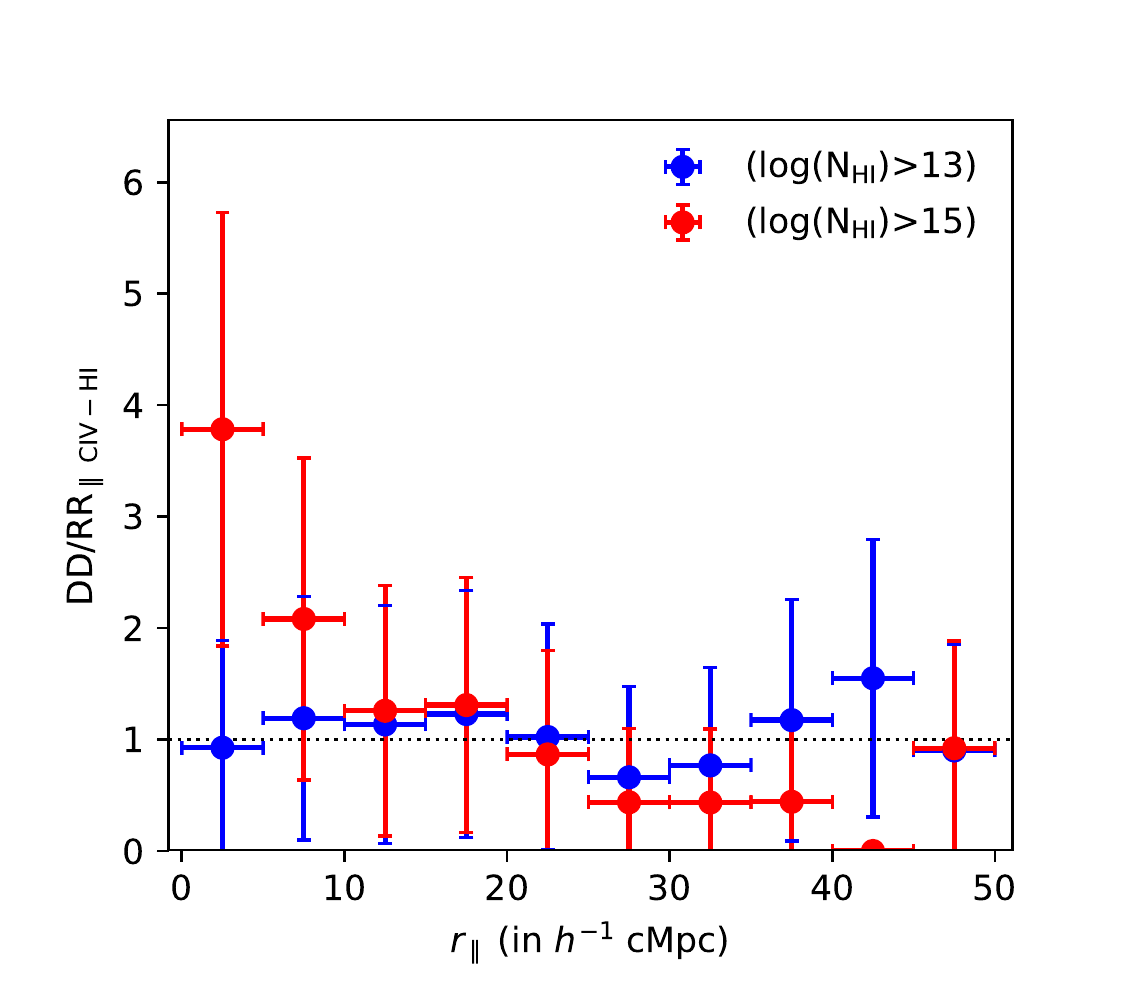}%
		\includegraphics[viewport=12 10 320 270,width=7.2cm, clip=true]{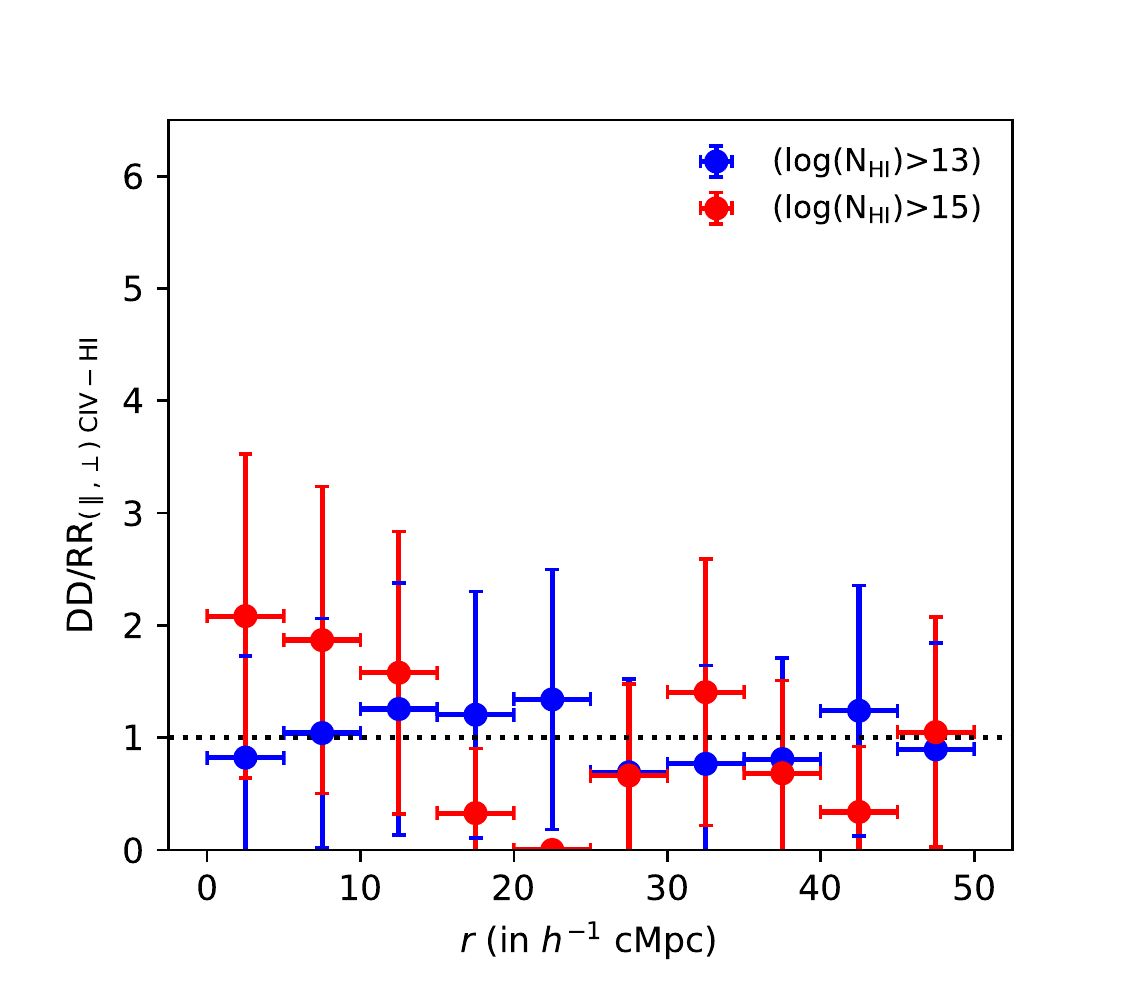}
	
		\caption{
		$Row1$ : two-point correlation using Voigt profile fits of the $\rm C_{IV}$ metals lines.
		$Row2$: two-point cross-correlation using Voigt profile fits of the $\rm C_{IV}$ metals lines with Voigt profile fits of \HI\ absorbers with different column density cut-offs. 
		The x-axis is the physical separation of the clouds in comoving units. }
	\label{fig_CIV}
\end{figure*}

In this section, we discuss the clustering properties of these \CIV\ absorbers (given in Table~\ref{tab_metal} in the Appendix) and how there clustering is associated with the underlying $N_{\rm HI}$ field. Individual \CIV\ absorbers were fitted with multiple component Voigt profiles using {\sc vpfit}. For \CIV - \CIV\ correlations, we consider \CIV\ clouds having observed wavelength greater than the \lya\ emission wavelength of that sightline and having redshift less than the \lya\ emission redshift by 5000 \kms\ to avoid the proximity regions of the quasar. To estimate the \CIV - \CIV\ correlations, we first calculate the data-data pair counts "DD". We then generate thousand random skewers to calculate the random-random pair counts "RR" keeping the number of clouds in the random skewers same as observed.
We use DD/RR as the estimator for \CIV - \CIV\ correlation with DD/RR = 1 means no correlation.

The longitudinal (transverse) correlations have been calculated by summing DD along all quasar sightlines (pairs of sightlines) and then normalizing it by the total RR. In the row 1 of Fig.~\ref{fig_CIV}, we have shown the \CIV - \CIV\ correlations both along the line of sight (left panel) as well as between two lines of sight (right panel). We have also associated a Poissonian error based on the number of the DD pair counts. For the longitudinal correlation, a sharp peak is seen in the first bin (0-5 $h^{-1}$ cMpc $\sim$ 0-500\kms). This is consistent with previous works based on \CIV-\CIV\ longitudinal correlations that have suggested a correlation length scales of about 1000 \kms \citep[][]{Rauch1996,Pichon2003,Boksenberg2003,Scannapieco2006}.
Our data also shows a peak at the 20-25 $h^{-1}$ cMpc bin (2000-2500 \kms) in the case of longitudinal correlation. Upon investigation, the longitudinal clustering at this scale, though not significant beyond 1$\sigma$ level, primarily comes from the sightlines J2117-0238 and J1055+0800.
In the case of transverse correlation of the \CIV\ absorbers, we do not see significant clustering at any scale. The correlation signal is consistent with random distribution of absorbers within $1\sigma$ of the Poissonian error. This is similar to what has been reported by \cite{coppolani2006} excluding a clustering signal at $\sim$20,000\kms\ observed due to \CIV\ groups in their quasar quartet. In passing we note that the number of detected \CIV\ absorbers are too low to attempt a three-point correlation function measurement.

Next, we probe the \CIV-\HI\ clustering using \CIV\ absorbers detected in the redshift range of the observed \lya\ forest. 
The locations of the \CIV\ absorbers are then correlated with those of the \lya\ absorbers for two different $N_{\rm HI}$ cut-offs to probe its dependence on $N_{\rm HI}$. In the second row of Fig.~\ref{fig_CIV}, the blue and red points correspond to \CIV-\HI\ cross-correlation for $\rm log(N_{HI})>$13 and 15 respectively. It is seen that there is no significant clustering of low $N_{\rm HI}$ clouds  with the \CIV\ absorbers. Rather,  higher $N_{\rm HI}$ \lya\ absorbers seem to cluster around \CIV\ absorbers up to 10$h^{-1}$ cMpc in the longitudinal direction. Similar clustering behaviour is also seen in the transverse correlations albeit with lower amplitude and significant levels. There is no observable correlation between the lower column density \HI\  and \CIV\ absorption, but for higher column densities, there is a slightly higher correlation (though with lower statistical significance) up to 10-15 $h^{-1}$ cMpc. Due to small number of sightlines involved the results presented in this sub-section can not be confirmed at high level of statistical significance. With the existing echelle spectroscopic date of high-$z$ quasars it will be possible to confirm the trend with better significant level for the longitudinal correlation. This we will do in our forthcoming paper.

\section{Summary}

We present X-Shooter observations of 8 quasars that are part of two projected triplets and a doublet (with pair separations spanning 0.5 to 1.6 pMpc). We used the absorption lines detected to study correlation properties: (a) auto-correlation of \lya\ absorption in the longitudinal and transverse directions using flux and cloud based statistics, (b) transverse cross correlation between \lya\ and foreground QSOs and DLAs and (c) auto-correlation of \CIV\  and cross-correlation of \CIV and \lya\ along longitudinal and transverse directions. To understand the results of our spatial correlation studies of \lya\ forest we use hydrodynamical simulations that are validated using observed flux probability distribution function, distribution of $N_{\rm HI}$ and observed relationship between two-point flux correlation function and radial separations. Our main finding are,
\begin{enumerate}

   \item{} Based on the transmitted flux, we derive two-point and three-point correlation functions for sightlines of our quasars.  The observed two-point correlation as a function of projected  
   separation is found to be consistent with previous studies by \citet{coppolani2006} and with our simulated results.
   
   \item{} We compute the probability of realising observed values of two-point and three-point functions and their 1$\sigma$ range in our simulations. It is found that the average clustering properties of the \lya\ forest seen along these two triplets are reproduced with the probability ranging from 20 to 75\%. Therefore the observed triplets do not seem to probe any abnormal regions of the IGM.
   
   \item{} The conclusions we derived based on the transmitted flux statistics are also confirmed using cloud based statistics obtained using Voigt profile decomposition of \lya\ forest. Even though this approach requires more resources it has distinct advantages: (a) as the measured correlation strengths are directly related to real correlations, this allows straight forward interpretations of $\xi$, $\zeta$ and Q and (b) it also allows us to study the dependence of clustering on $N_{\rm HI}$. Cloud based correlation function studies are now possible thanks to the availability of automatic Voigt profile fitting codes like {\sc viper} and high performance computing.

    \item{} Our simulations show that two-point and three-point correlation functions and the reduced three-point correlation (i.e Q) depend strongly of $N_{\rm HI}$ for a given spatial separations probed. {We observe this trend in the observation too, though the actual dependence of $N_{\rm HI}$ seems weaker than what we find in our simulations in case of the two-point correlations.}

    \item{} We searched for large void regions using 
     concurrent gaps found among the three sightlines. The largest concurrent flux gap of 17\AA\ is observed in "Triplet 1" which corresponds to a length scale of  $14.2h^{-1}$cMpc. This is bigger than such gaps reported in the literature\citep{rollinde2003,cappetta2010}. A search for similar gap in the simulations have yields 14.2\% sighlines showing a concurrent gap similar to or larger than what we have observed.  
     We also probe the gap statistics using cloud distributions. The above mentioned void is also picked by this statistics.
    
    \item{} We study the clustering of \lya\ absorption around quasars using "Triplet 2" and "Doublet" sightlines.  As all three quasars are at very similar redshift in the case of "Triplet 1" we did not use them for this study. Based on the inferred Lyman continuum luminosity of the quasars and UV-background intensity we expect all the background quasar sightlines  to be affected transverse proximity effect. However, consistent with previous studies of quasar pairs, we do find excess H~{\sc i} absorption at the redshifts of the foreground quasars within the expected proximity regions. 
    
    \item {} In the case of "Triplet 2"  we detect DLAs along all three sightlines within a redshift interval of $2.22\le z\le 2.32$.  The small probability associated with such a coincidence and the presence of a fourth quasar with a projected separation of $\sim$ 1 Mpc to all the three sightlines suggests a possible presence of large overdense region. It will be interesting to confirm this by deep imaging (or integral field spectroscopic) observations of this region.
    
    \item{} For the DLAs detected in the spectra of quasars in "Triplet 2" we study the transverse clustering of \lya\ absorption along other two sight lines. While we do not find any significant excess number of \lya\ absorption within $\pm 1000$ \kms to the DLA we do find the presence of significant excess \lya\ absorption with $N_{\rm HI}\sim$ few times $10^{16}$ cm$^{-2}$. Given the length scale probed these absorbers may not belong to the cirumgalatic medium of the quasars as studied by \citet{Rubin2015} but may reflect biased clustering of the IGM around DLAs\citep[see for example][]{perez2018}. As there are indications that these DLAs may be in the over-dense regions it will be intereting to check whether occurrence of partial LLS at $\sim$ 1 pMpc is special to this region or generic to high-$z$ DLAs. 
    
    \item {} Consistent with previous studies we detect correlation signal within 5 h$^{-1}$ cMpc  (or velocity scale of $\sim$500 \kms) for longitudinal correlation between \CIV\ absorbers. We do not find any excess clustering in the transverse direction between the \CIV\ absorbers over the projected separations probed in this study.
    \item{} We have studied the \CIV-\lya\ cross-correlation  along longitudinal and transverse directions using pairs of sightlines. We find this correlation to be higher (up to 10 h$^{-1}$cMpc) when we consider higher column density (i.e $N_{\HI} > 10^{15}$ cm$^{-2}$) \lya\ absorbers. It will be important to confirm these findings as through this one can get independent constraints on the objects hosting \CIV\ absorbers.
\end{enumerate}

Using our simulations, we estimate the total redshift path length required to detect $\zeta$ at 5$\sigma$ level for two triplets considered here.

We found that for flux based statistics, we need a redshift path length $\geq$ 16 (for $\Delta r_{\parallel}=0h^{-1}$cMpc) and 24 (for $\Delta r_{\parallel}=2h^{-1}$cMpc) for resolving the three-point correlation with $5\sigma$ detectibility for "Triplet 1". The corresponding values are $\geq$ 55 (for $\Delta r_{\parallel}=0h^{-1}$cMpc) and 65 (for $\Delta r_{\parallel}=2h^{-1}$cMpc) for "Triplet 2".
For the cloud based statistics with $N_{\rm HI}>13$, the required redshift path length is 44 for "Triplet 1". In case of "Triplet 2" configuration, the cloud based statistics essentially gives 0 three-point correlation upto 2 decimal places, thus making it difficult to resolve with $5\sigma$ detectibility.

\section*{Acknowledgement}

RS and PPJ gratefully acknowledge the support of the Indo-French Centre for the Promotion of Advanced Research (Centre Franco-Indien pour la Promotion de la Recherche Avance\'e) under contract no. 5504-2. We acknowledge the use of High performance computing facility PERSEUS at IUCAA, Pune and thank Kandaswamy Subramanian and Aseem Paranjape for useful discussions.


\bibliographystyle{mnras}
\bibliography{Lya_correlation} % if your bibtex file is called example.bib
%    \end{document}	

%%%%%%%%%%%%%%%%%%%% Appendix %%%%%%%%%%%%%%%%%%
%\Appendix

%%%%%%%%%%%%%%%%% APPENDICES %%%%%%%%%%%%%%%%%%%%%

\appendix

\section{Quasar associated absorption }

\begin{figure*}
	
	\center
	\includegraphics[viewport=10 0 345 290,width=5.8cm, clip=true]{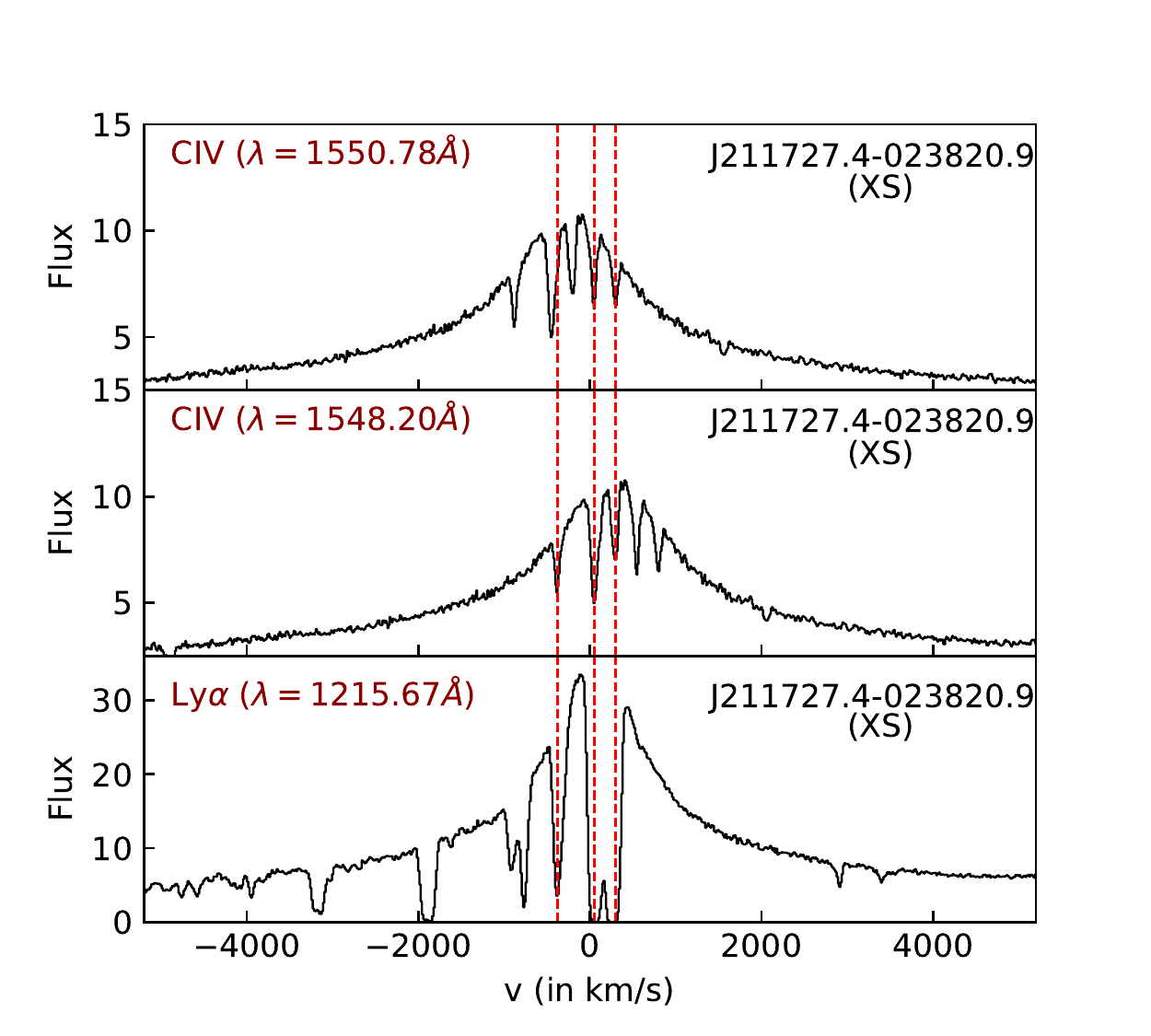}%
	\includegraphics[viewport=10 0 345 290,width=5.8cm, clip=true]{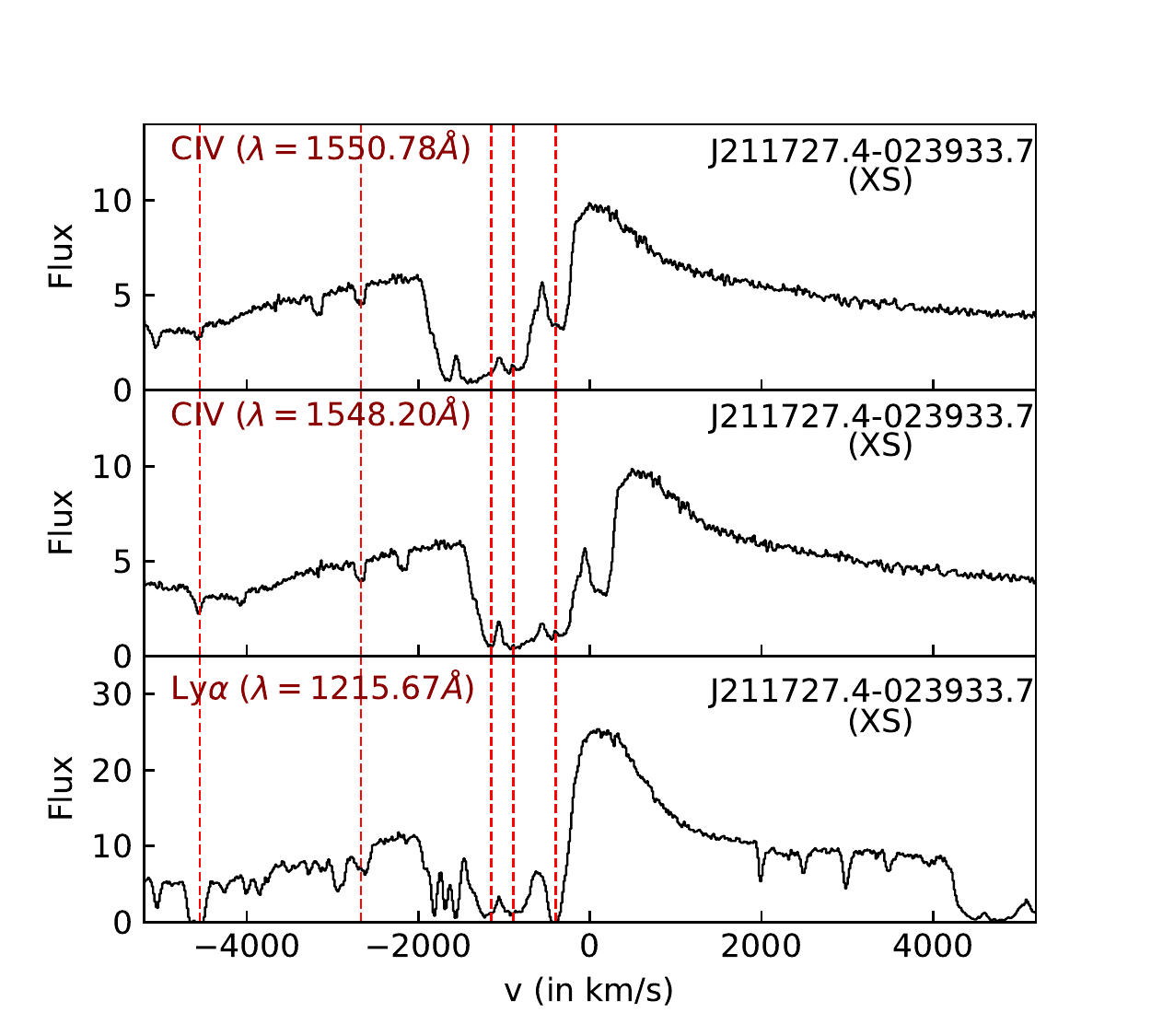}%
	\includegraphics[viewport=10 0 345 290,width=5.8cm, clip=true]{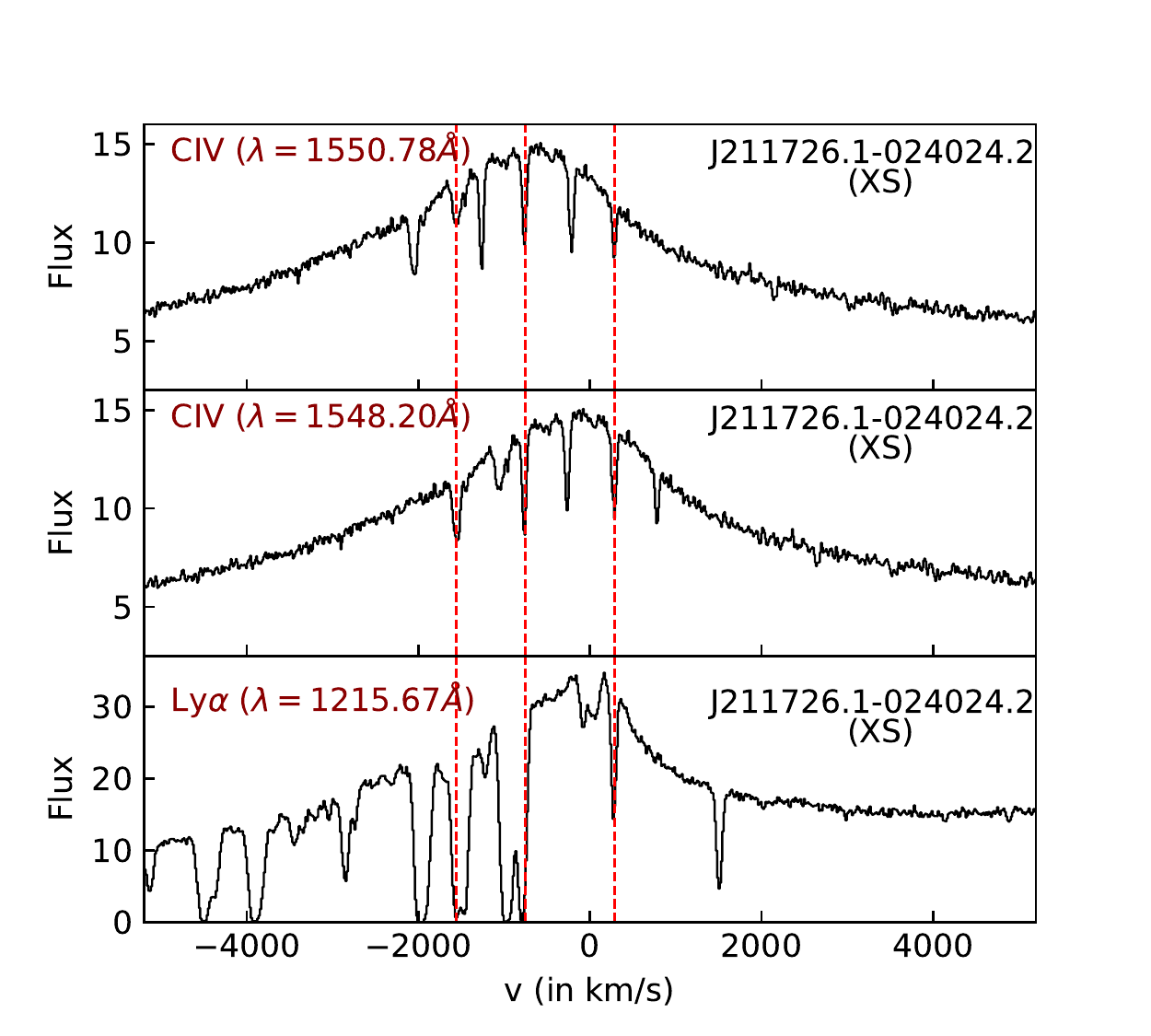}
	
	\includegraphics[viewport=10 0 345 290,width=5.8cm, clip=true]{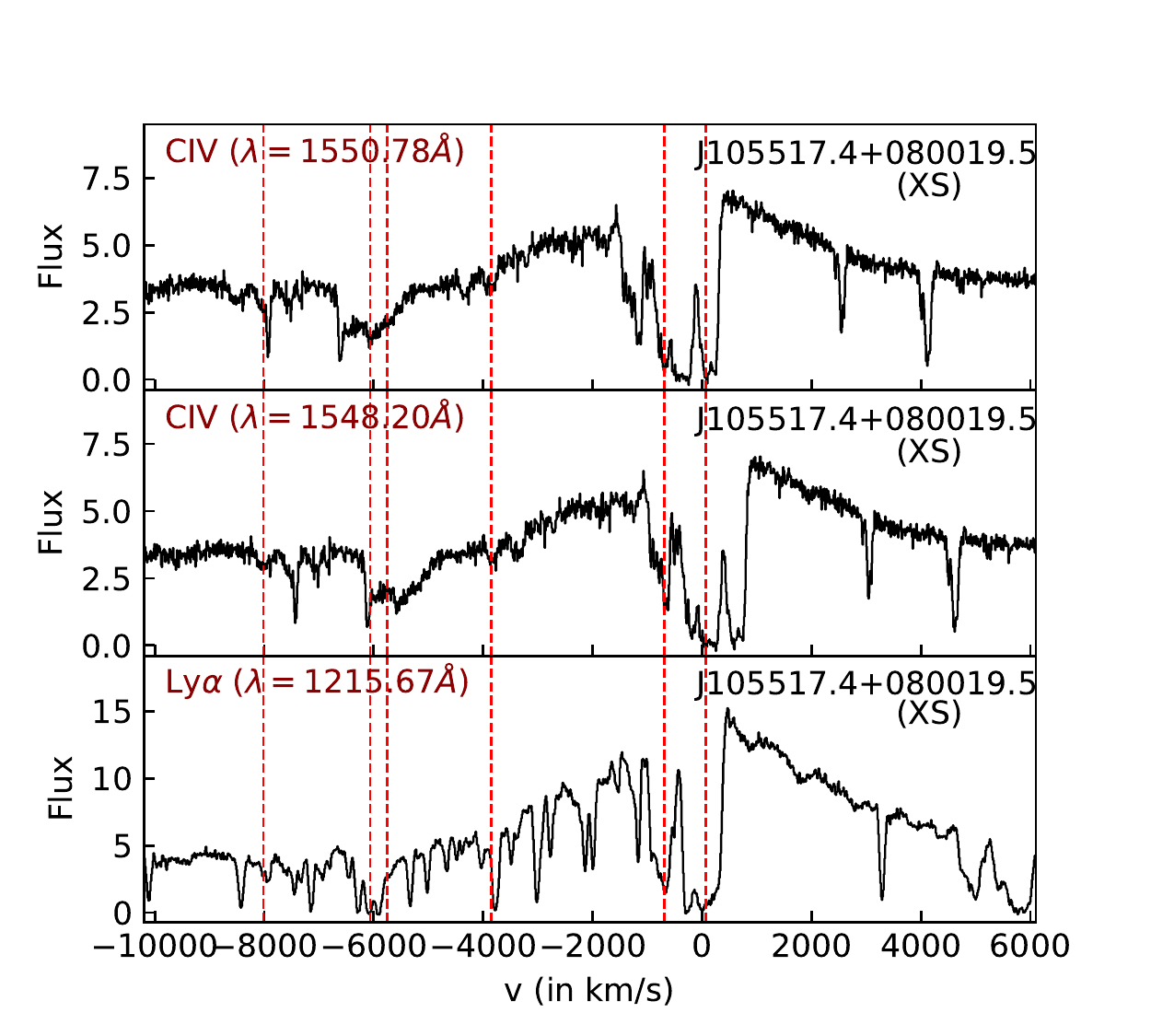}%
	\includegraphics[viewport=10 0 345 290,width=5.8cm, clip=true]{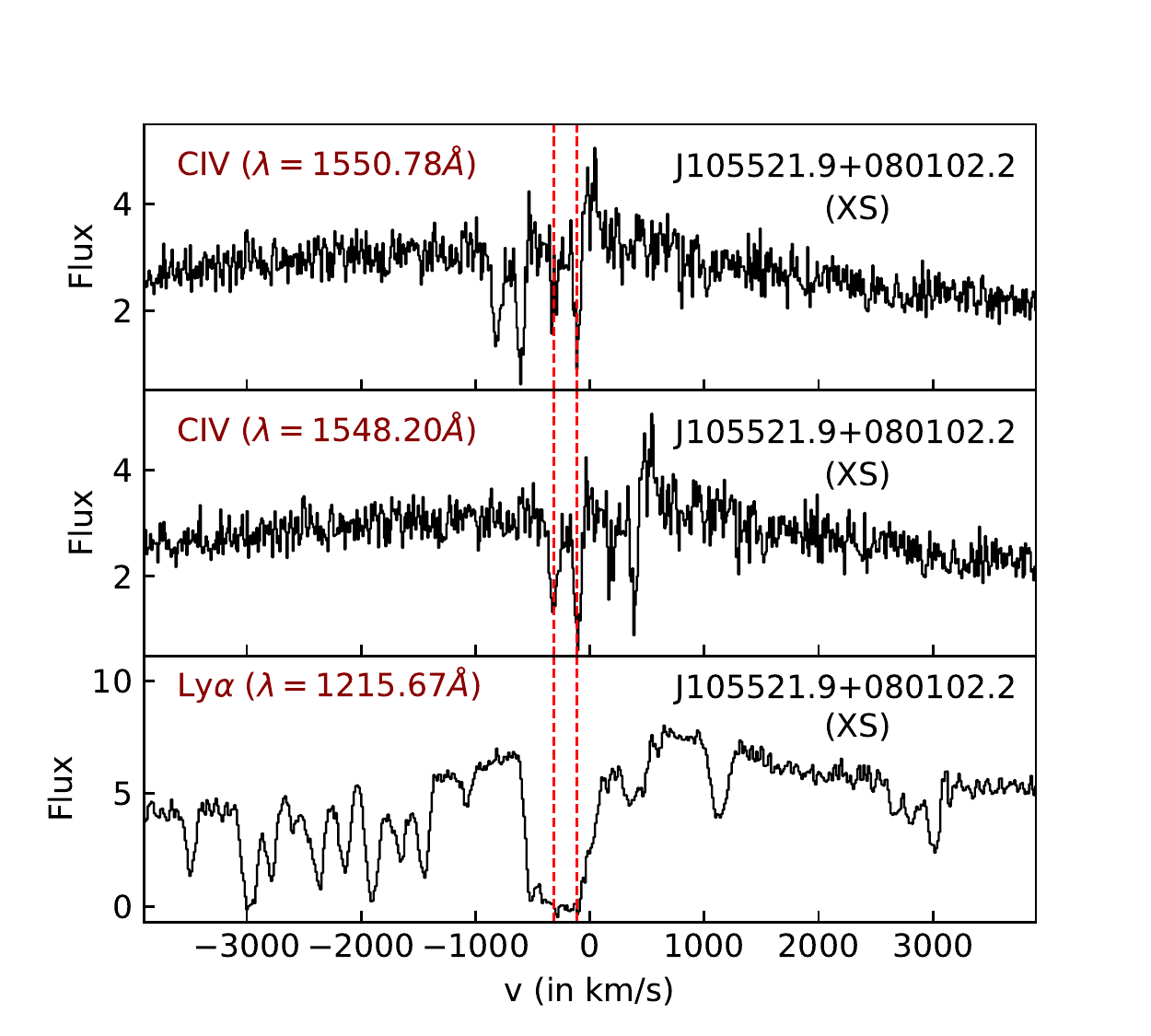}%
	\includegraphics[viewport=10 0 345 290,width=5.8cm, clip=true]{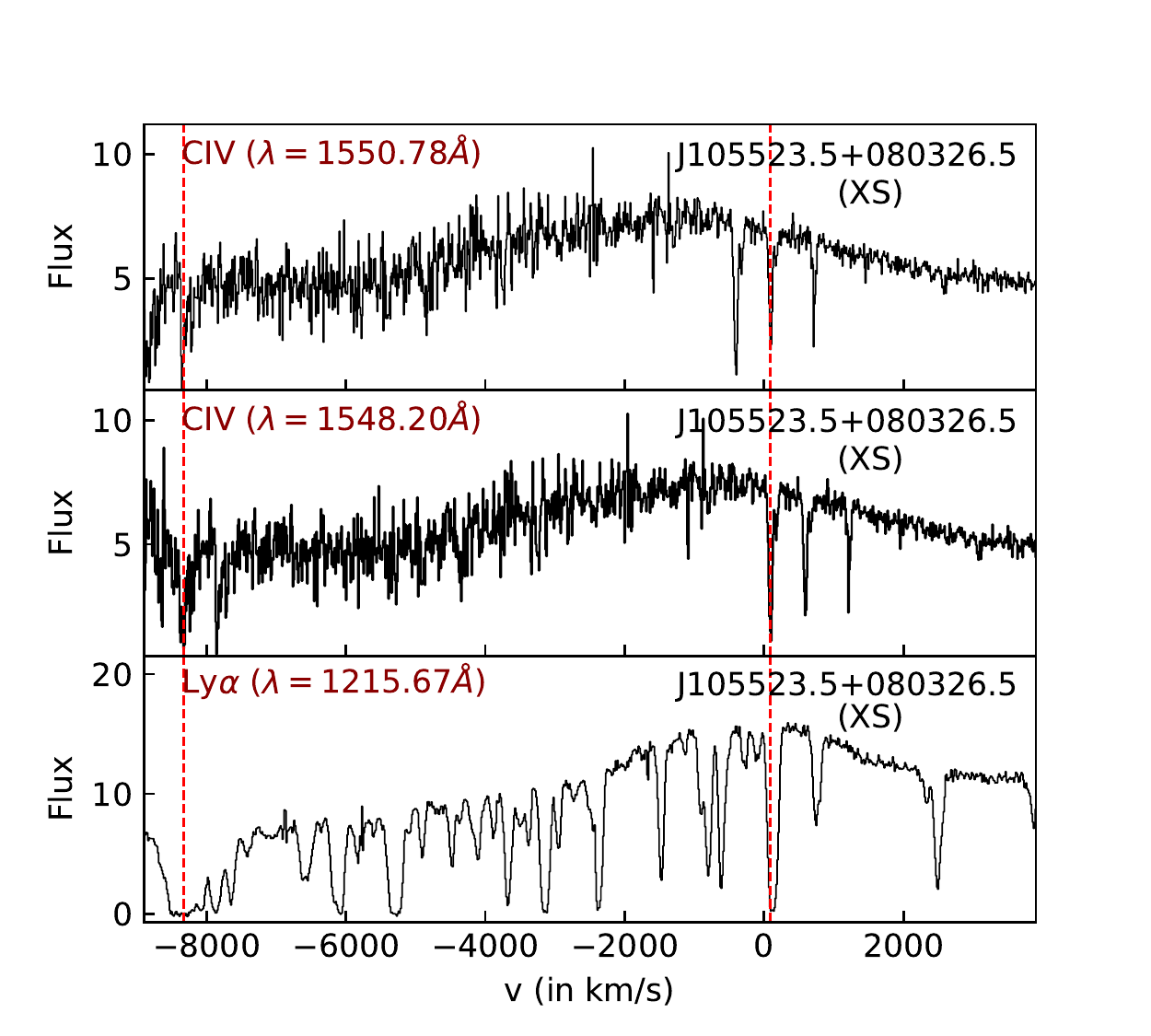}
	
	\includegraphics[viewport=10 0 345 290,width=5.8cm, clip=true]{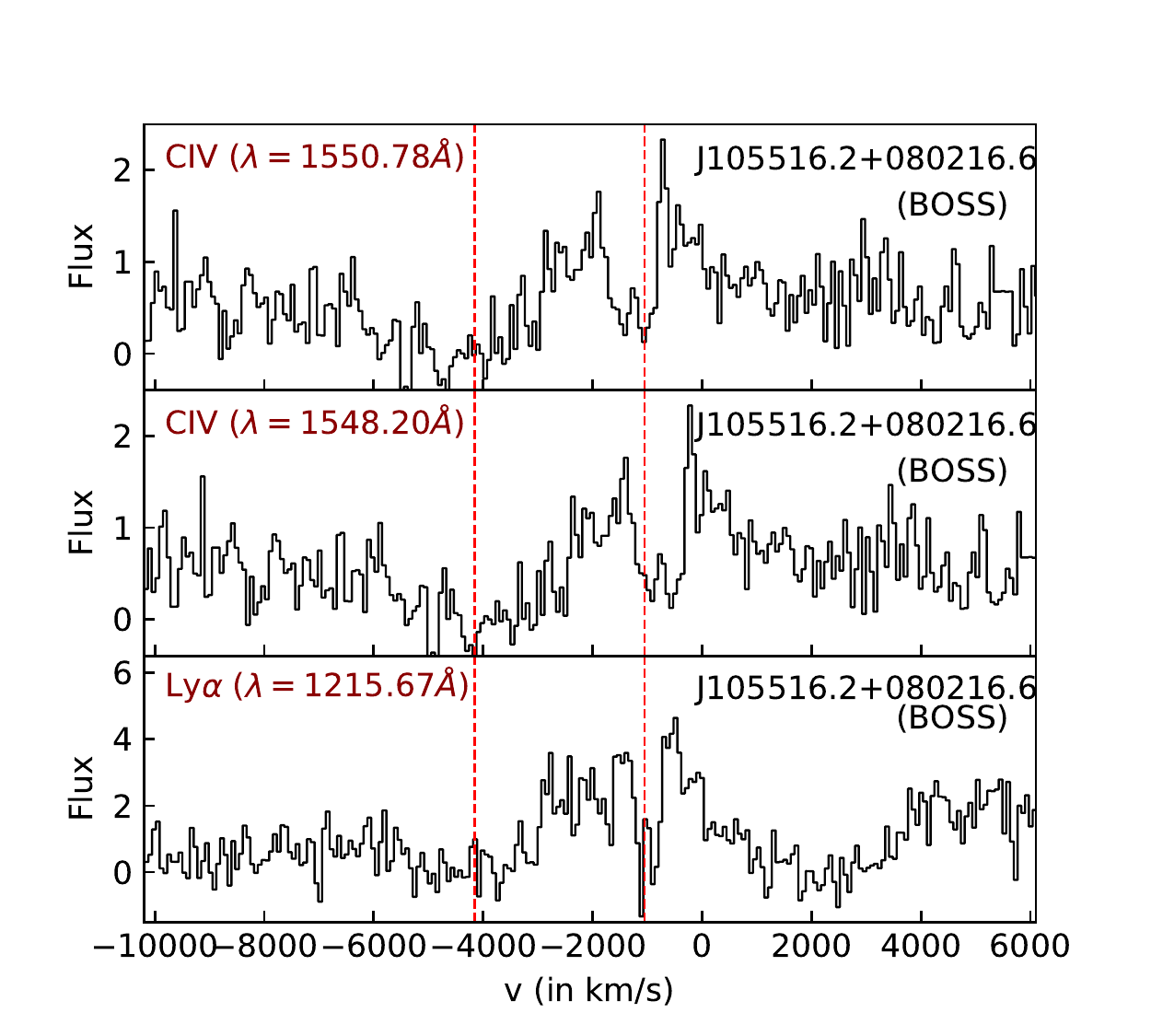}

	\caption{\CIV \  and \HI\ \lya\ absorptions associated with the observed quasars.}
	\label{fig_CIV_a}
\end{figure*}

In Figure~\ref{fig_CIV_a}, we show the associated absorptions with the quasar redshift for "Field 1" and "Field 2" quasars. The top panel is for "Triplet 1" quasars, the middle panel is for "Triplet 2" quasars and the bottom panel is for the 4th faint quasar (with lower SNR from SDSS) that we see in the "Field 2". In each of the plots, we show the spectrum as a function of velocity considered in rest wavelength of \HI\ \lya \ (bottom) and \CIV \ doublets (top and middle). We take the emission redshift of the quasars given in Table~\ref{Tab_obs} as zero velocity in the plots. The red vertical lines denote the positions of \HI\ \lya\ absorptions having associated \CIV\ absorptions near the redshift of the quasar.

\section{Metal lines in the observed sightlines}
\begin{table}
	\caption{List of metal line and DLA systems}
	\centering
	\begin{tabular}{lcl}
		\hline
		\multicolumn{1}{c}{QSO} & Species & \multicolumn{1}{c}{ z$_{\rm abs}$} \\
		\hline
		\hline
		J211727.4-023820.9 &\CIV & 1.6347, 1.7162, 1.7362, 1.7855, 1.8424,\\
		&     & 1.9325, 2.0475, 2.0565, 2.0584, 2.1978,\\ 
		&     & 2.2366, 2.2380, 2.2637, 2.3188, 2.3236, \\
		&     & 2.3263 \\
		&\MgII& 0.9325 , 1.3109\\ 
		\\
		J211727.4-023933.7  &\CIV & 1.3552, 1.3611, 1.6157, 1.6243, 2.1185,\\
		&     & 2.2592, 2.2797, 2.2963, 2.2992, 2.3046\\
		&\MgII& 0.6962, 0.8967\\
		\\
		J211726.1-024024.2  & \CIV& 1.5089, 1.5095, 1.5135, 1.7321, 1.7328,\\
		&     & 1.8307, 2.0406, 2.0963, 2.2919, 2.3007,\\ 
		&     & 2.3122\\
		& \MgII& 0.8979, 1.5135\\
		\\
		J105517.4+080019.5  &\CIV & 1.7445, 1.7575, 2.1375, 2.1732, 2.2228,  \\
		&     & 2.2382, 2.3292, 2.5956, 2.6746, 2.7043,  \\
		&     & 2.7942, 2.8190, 2.8230, 2.8473, 2.8880, \\
		&     & 2.8979 \\
		& \MgII& 1.3565, 1.6610, 1.6630, 1.7446, 1.7571,\\
		&     &2.1375, 2.1735, 2.2237 \\
		%                    & DLA & 2.2235
		& DLA & 2.2235\\
		\\
		J105521.9+080102.2&\CIV & 2.1539, 2.2674, 2.5232, 2.7051, 2.7075\\
		&\MgII& 2.2673\\
		& DLA & 2.2673\\
		\\
		J105523.5+080326.5&\CIV &2.2040, 2.3133, 2.5277, 2.6281\\
		&\MgII&1.2852,2.3118, 2.3126, 2.5274\\
		&DLA  & 2.3125\\
		\\
		J105516.2+080216.6&\CIV &1.9534, 2.2743, 2.3084\\
		\\
		J141844.0+065730.7&\CIV & 1.6332, 2.2483 \\
		&\MgII& 1.0149, 1.2569, 2.2486\\
		\\
		J141848.5+070027.2&\CIV & 1.5422, 1.6174, 1.9203, 2.0051, 2.0224\\
		&\MgII& 0.3519, 0.5779, 0.7847, 0.8262, 1.6173,\\
		&     & 2.0223 \\
		\\
		J141831.7+065711.2&\CIV & 1.8210, 2.1945, 2.3919\\
		&\MgII& 0.8423\\
		\hline
	\end{tabular}
	\label{tab_metal}
\end{table}	

Table~\ref{tab_metal} shows the list of metal line systems in all the quasar spectra observed in "Field 1", "Field 2" and "Field 3". All of them are X-Shooter spectra except J105516.2+080216.6 and J141831.7+065711.2, where the spectra has been obtained from SDSS.The table shows the positions of \CIV and \MgII lines observed in the spectra in redshift. For  "Triplet 2", we have also tabulated the positions of the observed DLAs in the spectra.

\section{Observed spectra}
\begin{figure*}
	\center
	\includegraphics[viewport=7 2 570 430,width=14.5cm, clip=true]{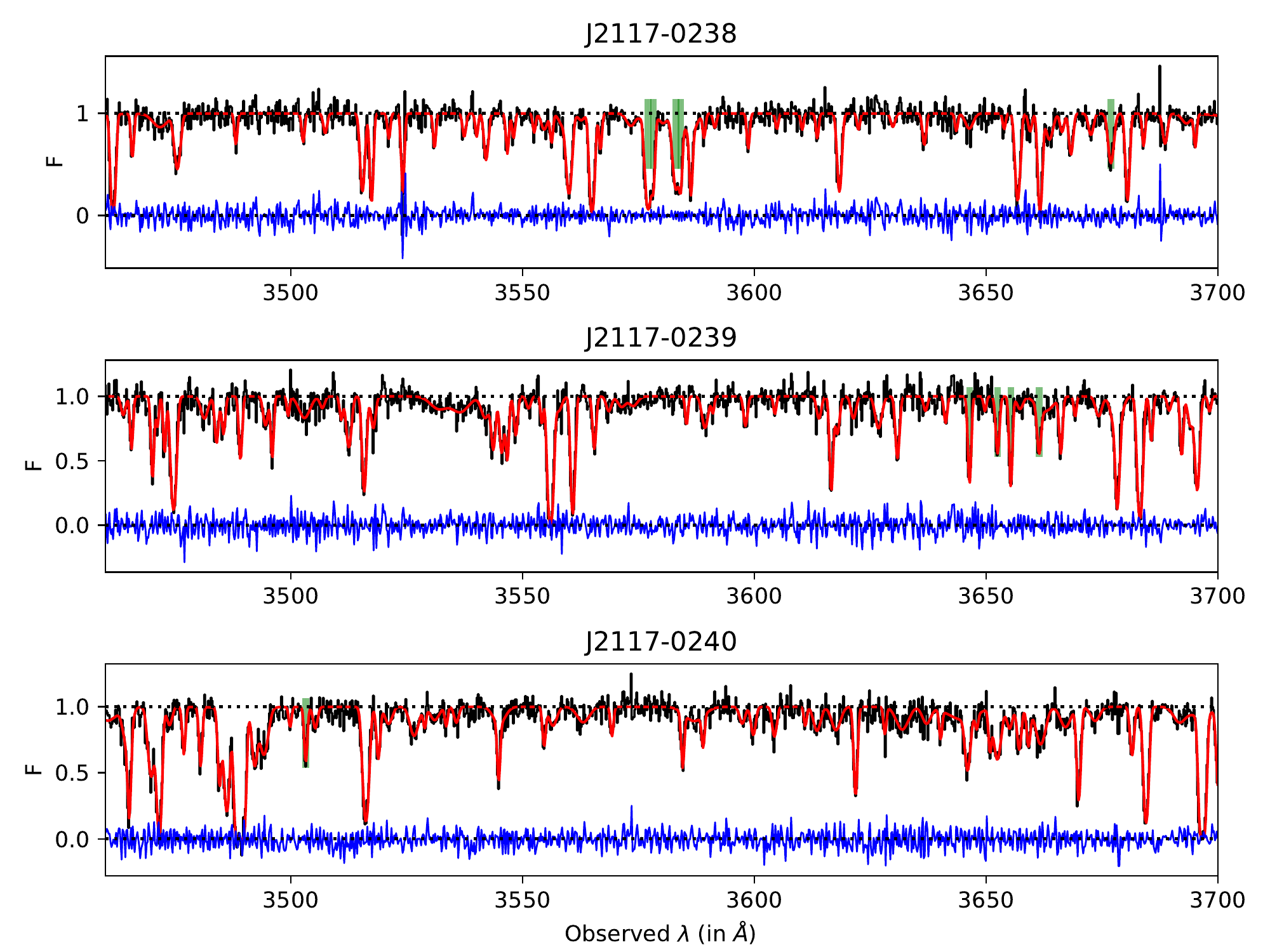}
	\vspace{0.7cm}
	\includegraphics[viewport=7 2 570 430,width=14.3cm, clip=true]{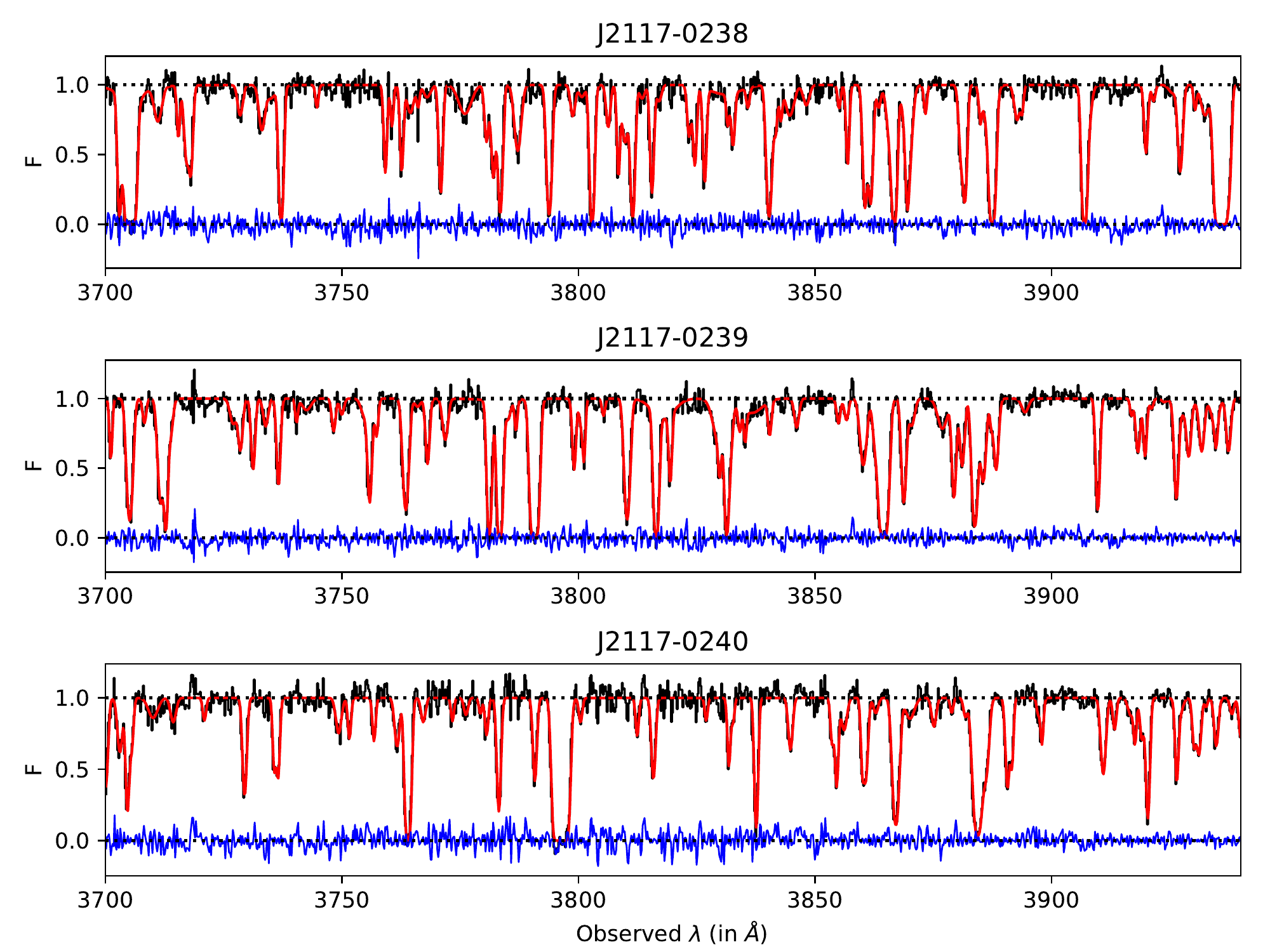}
	\caption{Observed \lya\ forest spectra of "Triplet 1". The observed transmitted flux is in black and their Voigt profile fitting is in red. The corresponding error for the Voigt profile fit has been shown in blue. The positions of the metal lines identified in the \lya\ forest using green vertical ticks.}
	\label{fig_J2117_spectra}
	
\end{figure*}

\begin{figure*}
	\center
	\includegraphics[viewport=30 5 680 440,width=15cm, clip=true]{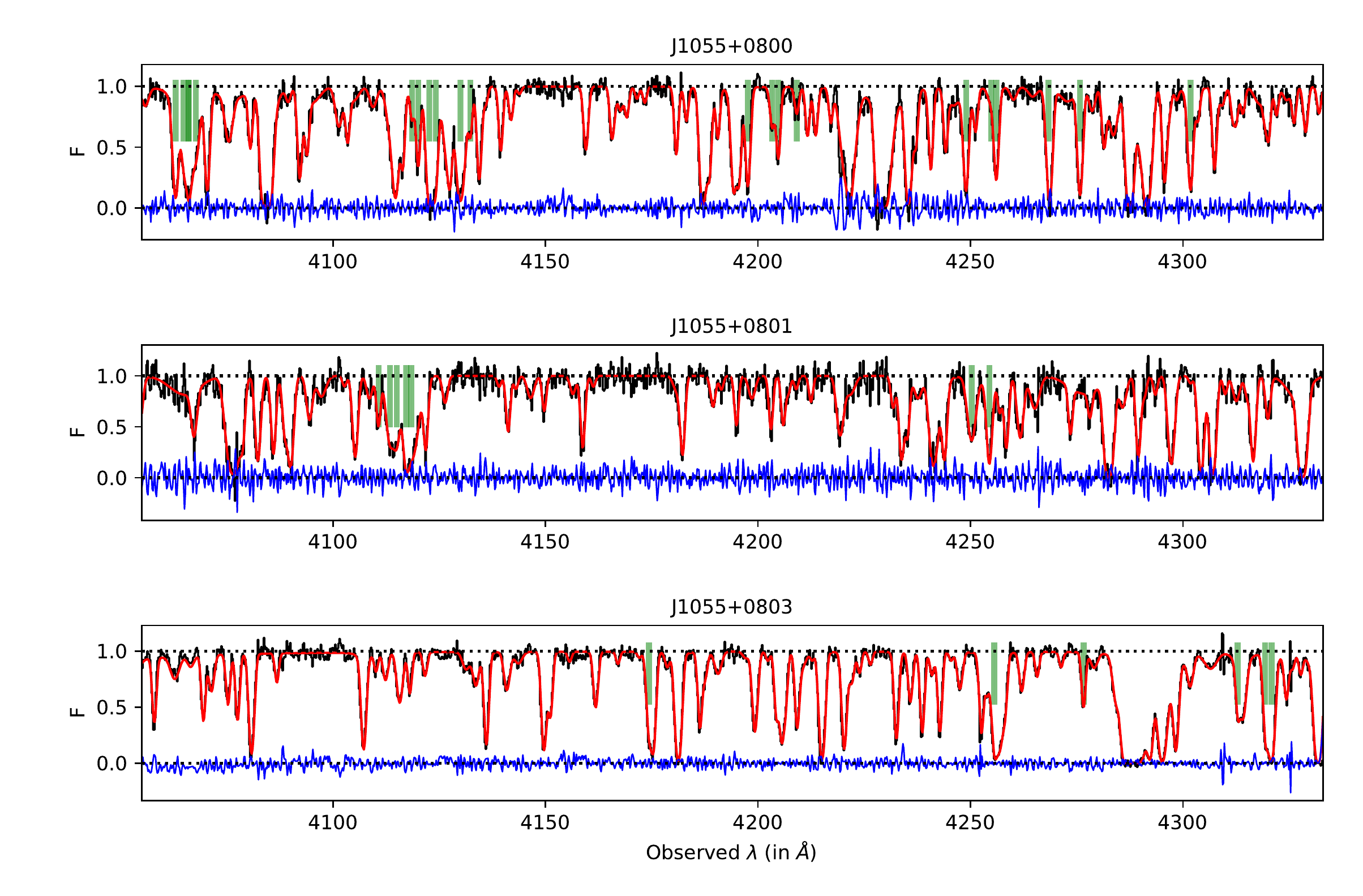}
	
	\caption{Observed \lya\ forest spectra of "Triplet 2". The plotting convention used is same as in Fig.~\ref{fig_J2117_spectra}}
	\label{fig_J1055_spectra}
\end{figure*}%
\begin{figure*}
	\center
	\includegraphics[viewport=70 5 670 280,width=15cm, clip=true]{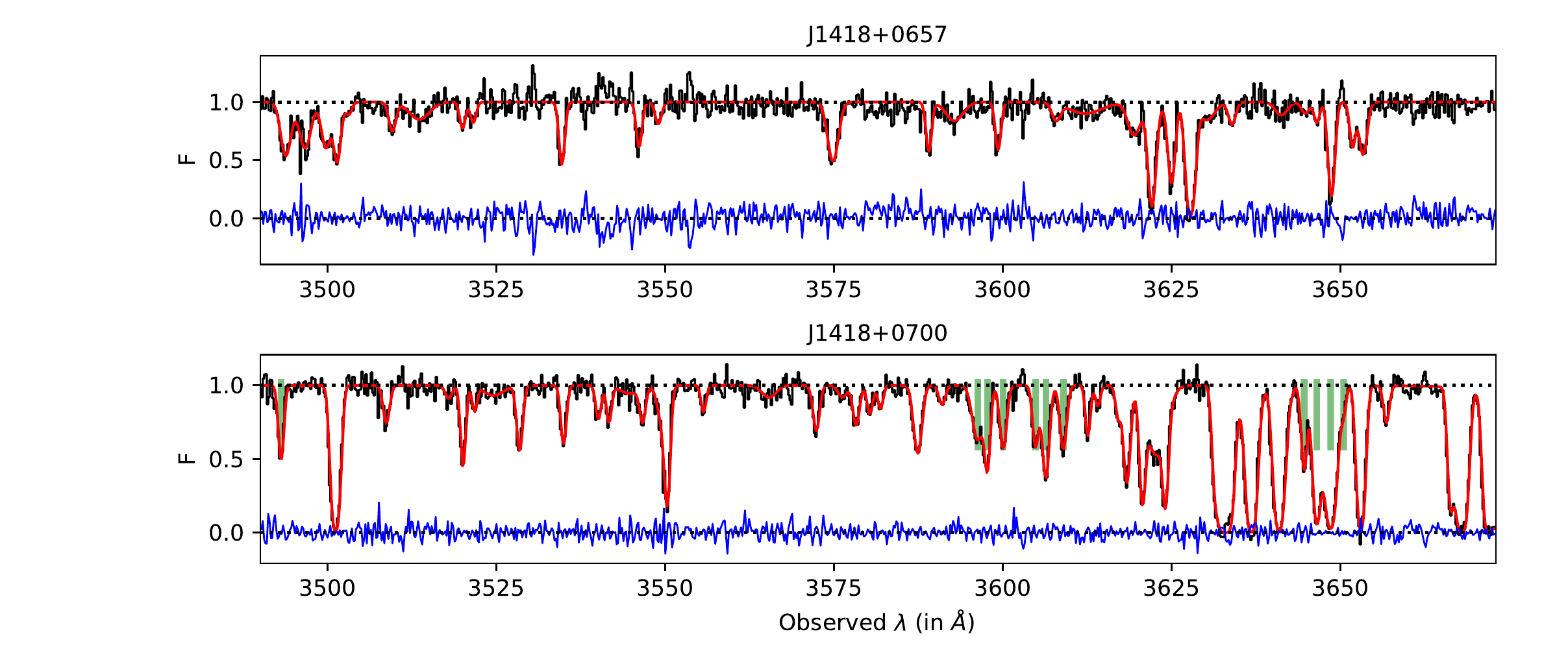}
	\includegraphics[viewport=60 5 670 280,width=15.5cm, clip=true]{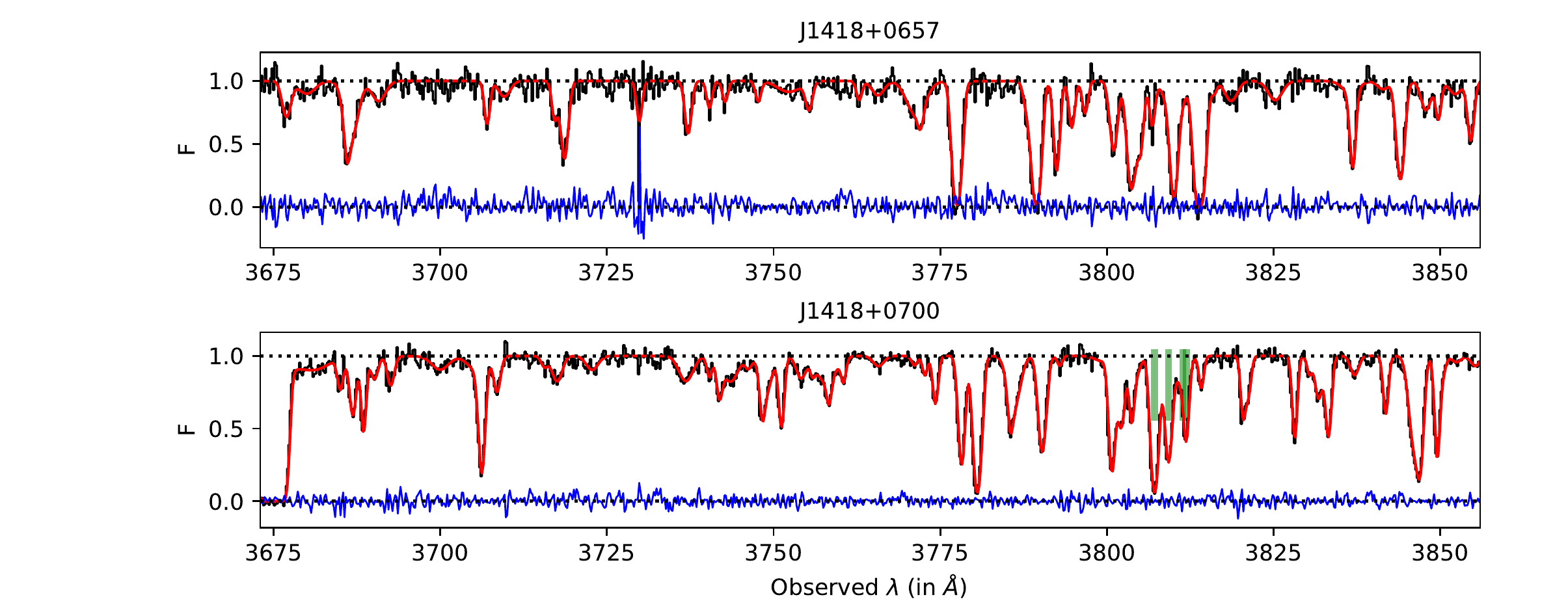}
	\caption{Observed \lya\ forest spectra of "Doublet". The plotting convention used is same as in Fig.~\ref{fig_J2117_spectra}.}
	\label{fig_J1418_spectra}

\end{figure*}

Figure~\ref{fig_J2117_spectra},~\ref{fig_J1055_spectra} and ~\ref{fig_J1418_spectra} shows the \lya\ forest normalized flux from the X-Shooter spectra for "Triplet 1", "Triplet 2" and "Doublet", respectively, that has been used for our correlation analysis. The ranges for these spectra have been give in Table~\ref{Tab_obs}. The observed transmitted flux has been shown in black and their Voigt profile fits using {\sc VIPER} has been shown using red. The corresponding error for the Voigt profile fit has been shown in blue. In the figure, we also show the positions of the metal lines identified in the \lya\ forest using green vertical ticks. It is to be noted that the transmitted flux plotted in the figure is before the removal of these metal lines.

\section{Dependence of transverse cloud correlation on redshift space binning}	
\begin{figure*}
	\begin{minipage}{0.3\textwidth}
		\includegraphics[viewport=5 10 340 270,width=6cm, clip=true]{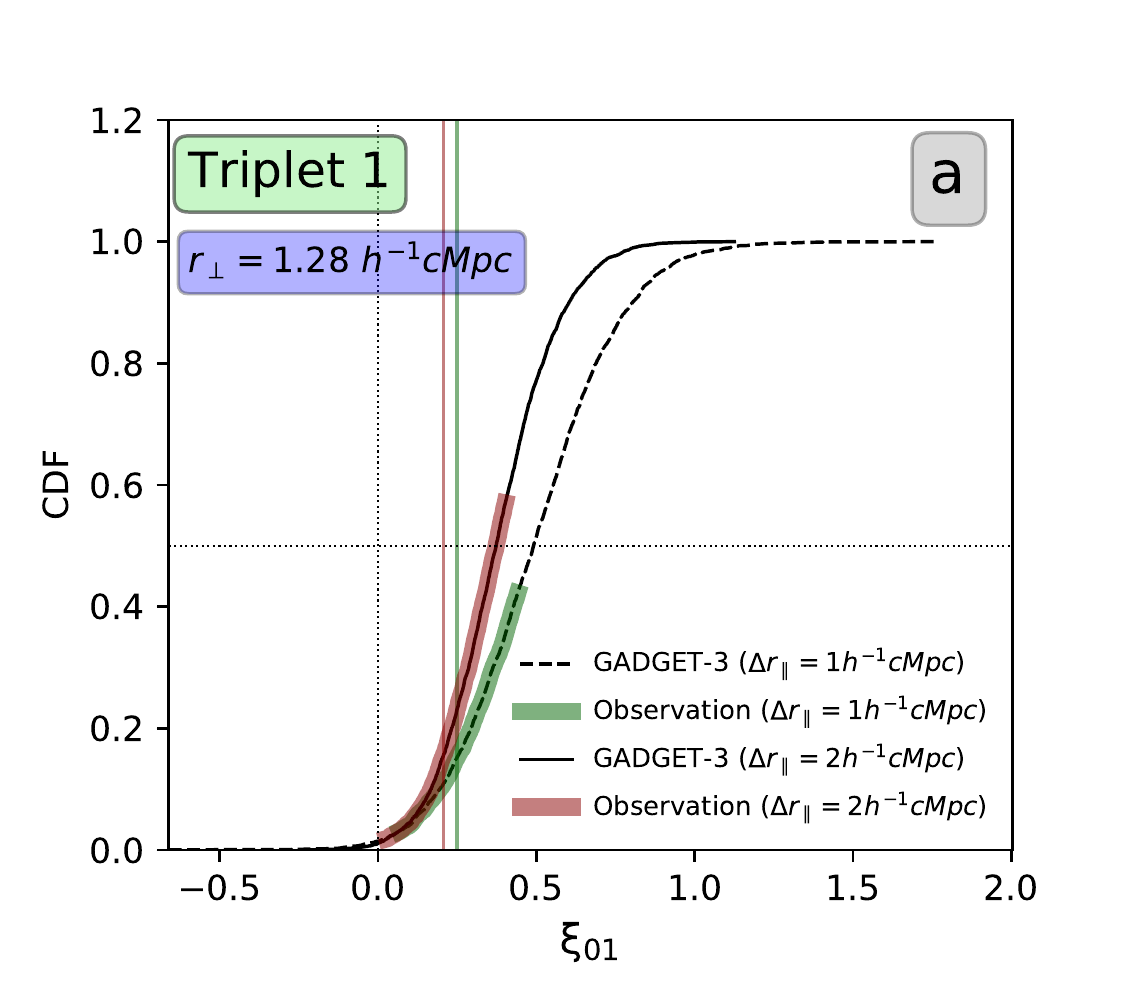}
	\end{minipage}%
	\begin{minipage}{0.3\textwidth}
		\includegraphics[viewport=5 10 340 270,width=6cm, clip=true]{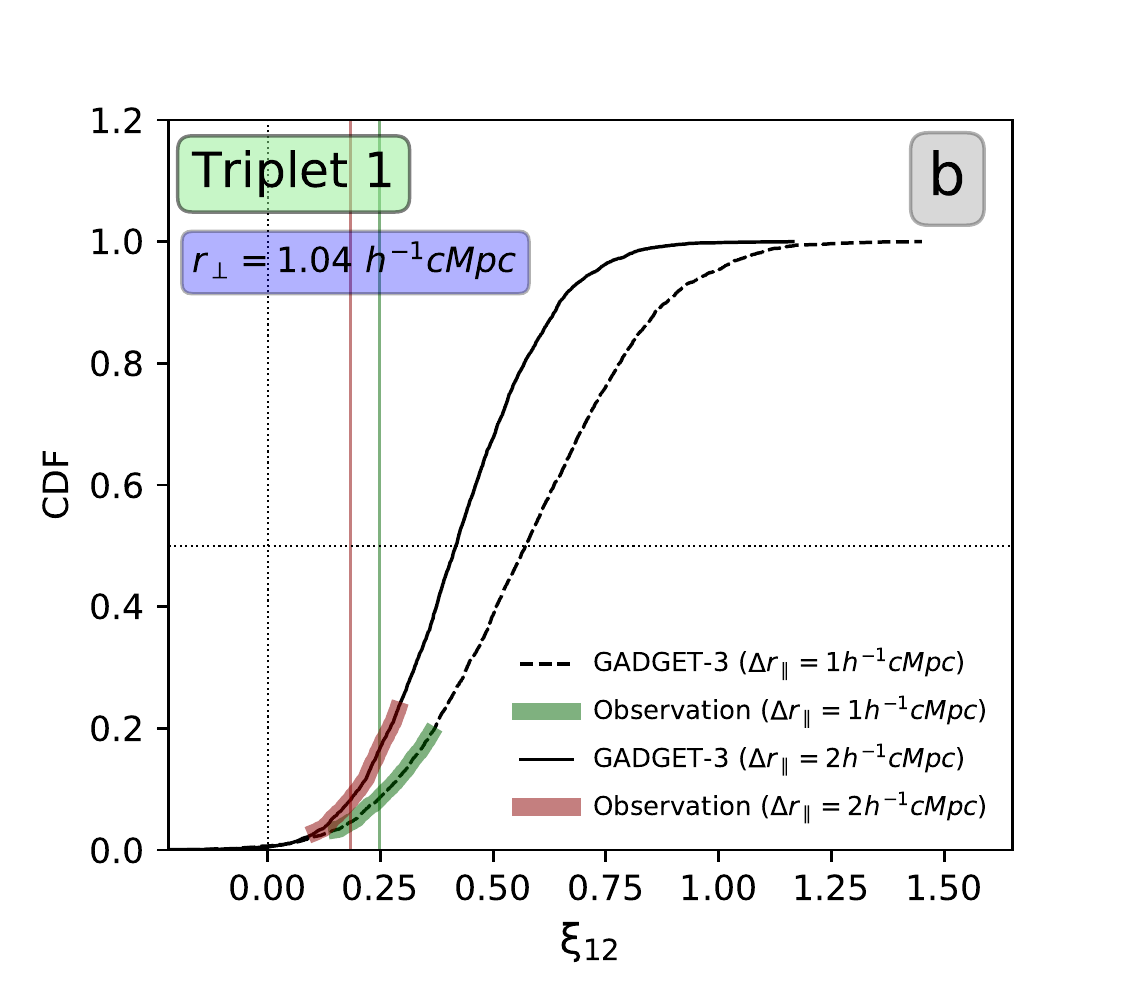}
	\end{minipage}
	\begin{minipage}{0.3\textwidth}
		\includegraphics[viewport=5 10 340 270,width=6cm, clip=true]{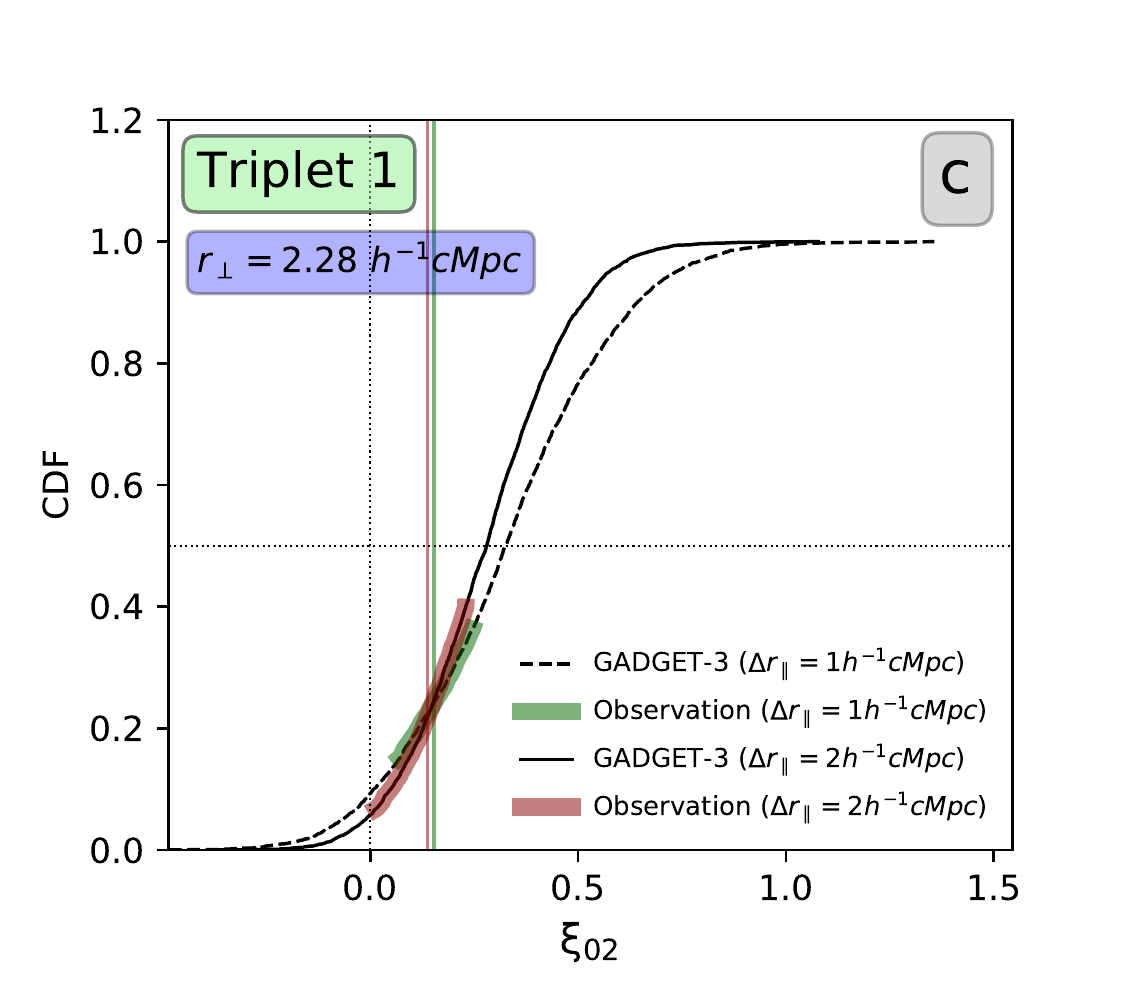}
	\end{minipage}%
	
	\begin{minipage}{0.3\textwidth}
		\includegraphics[viewport=5 10 340 270,width=6cm, clip=true]{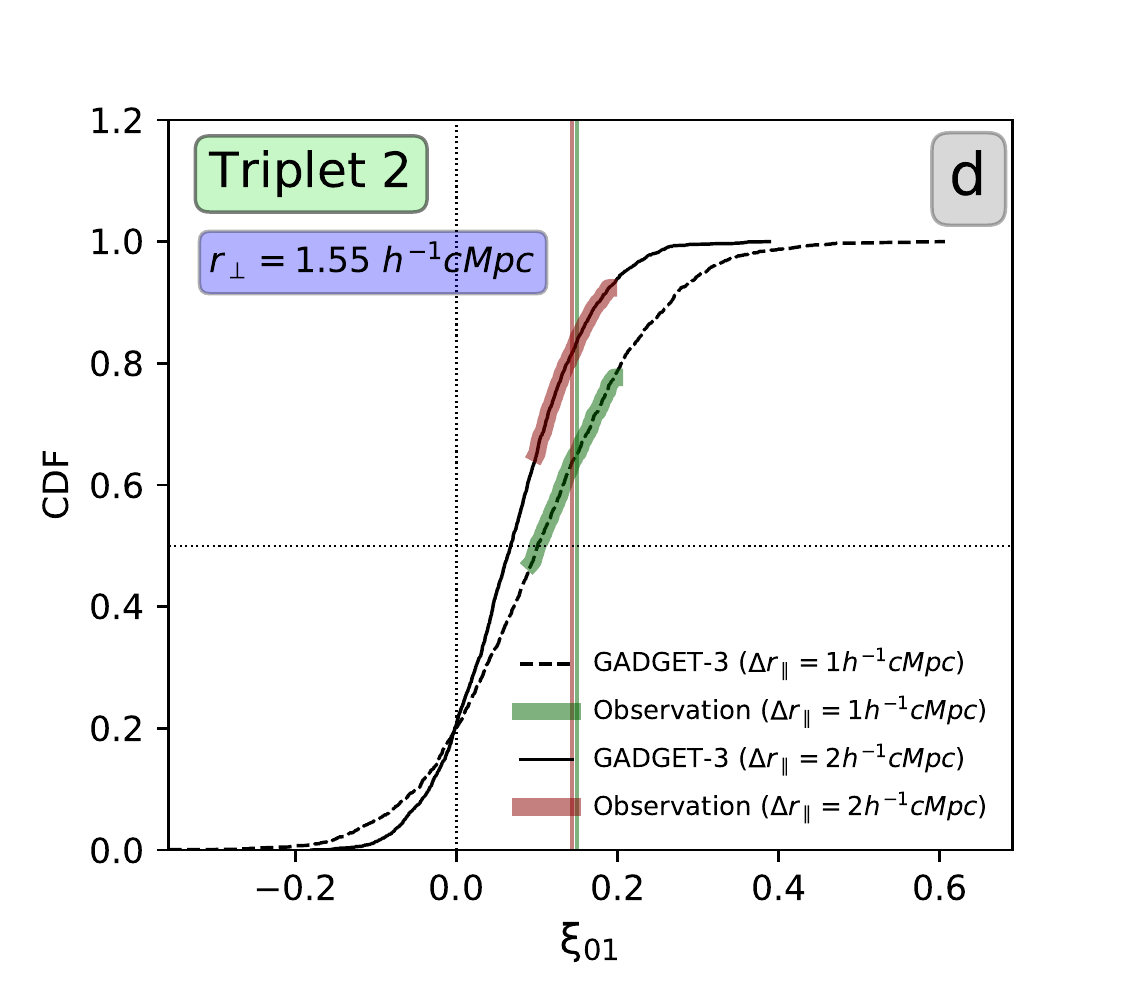}
	\end{minipage}%
	\begin{minipage}{0.3\textwidth}
		\includegraphics[viewport=5 10 340 270,width=6cm, clip=true]{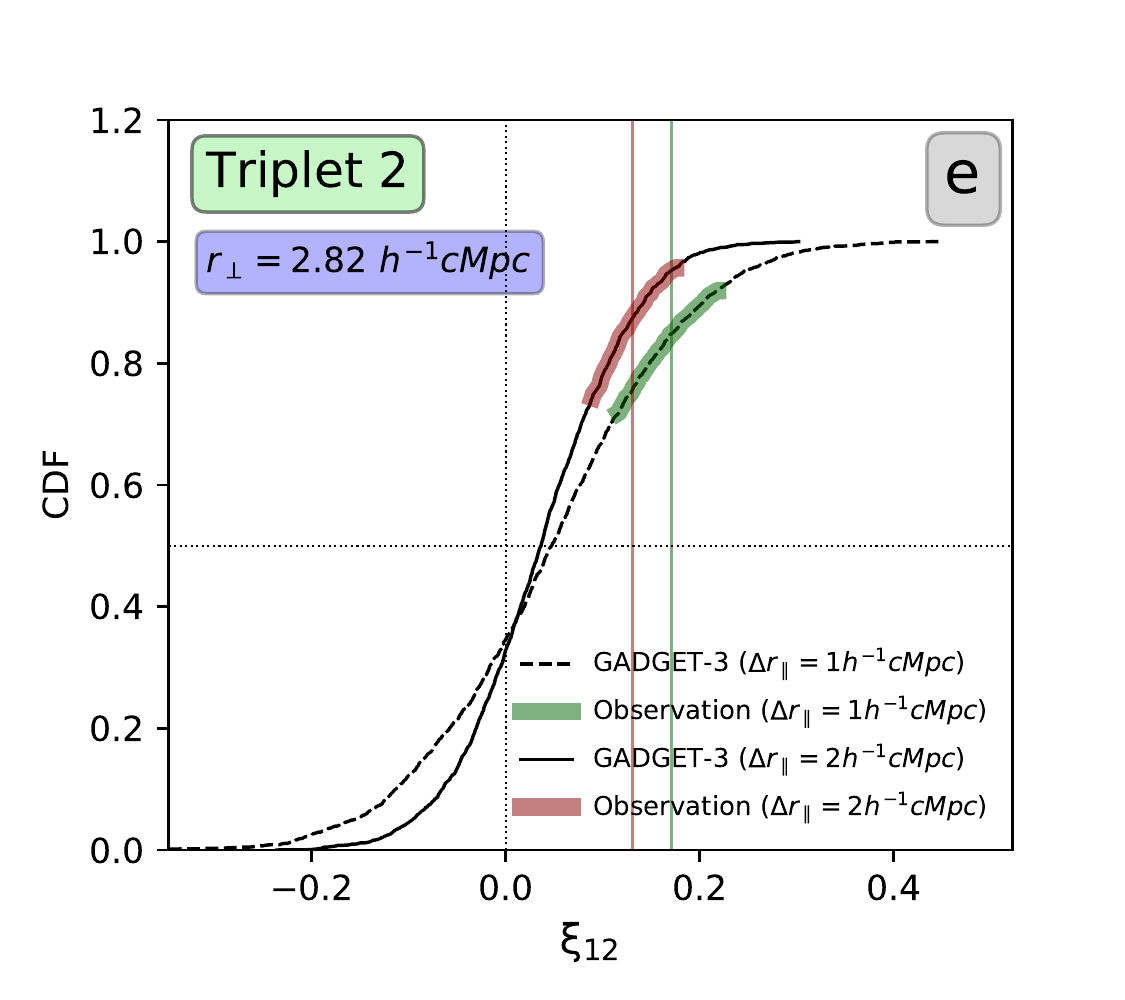}
	\end{minipage}
	\begin{minipage}{0.3\textwidth}
		\includegraphics[viewport=5 10 340 270,width=6cm, clip=true]{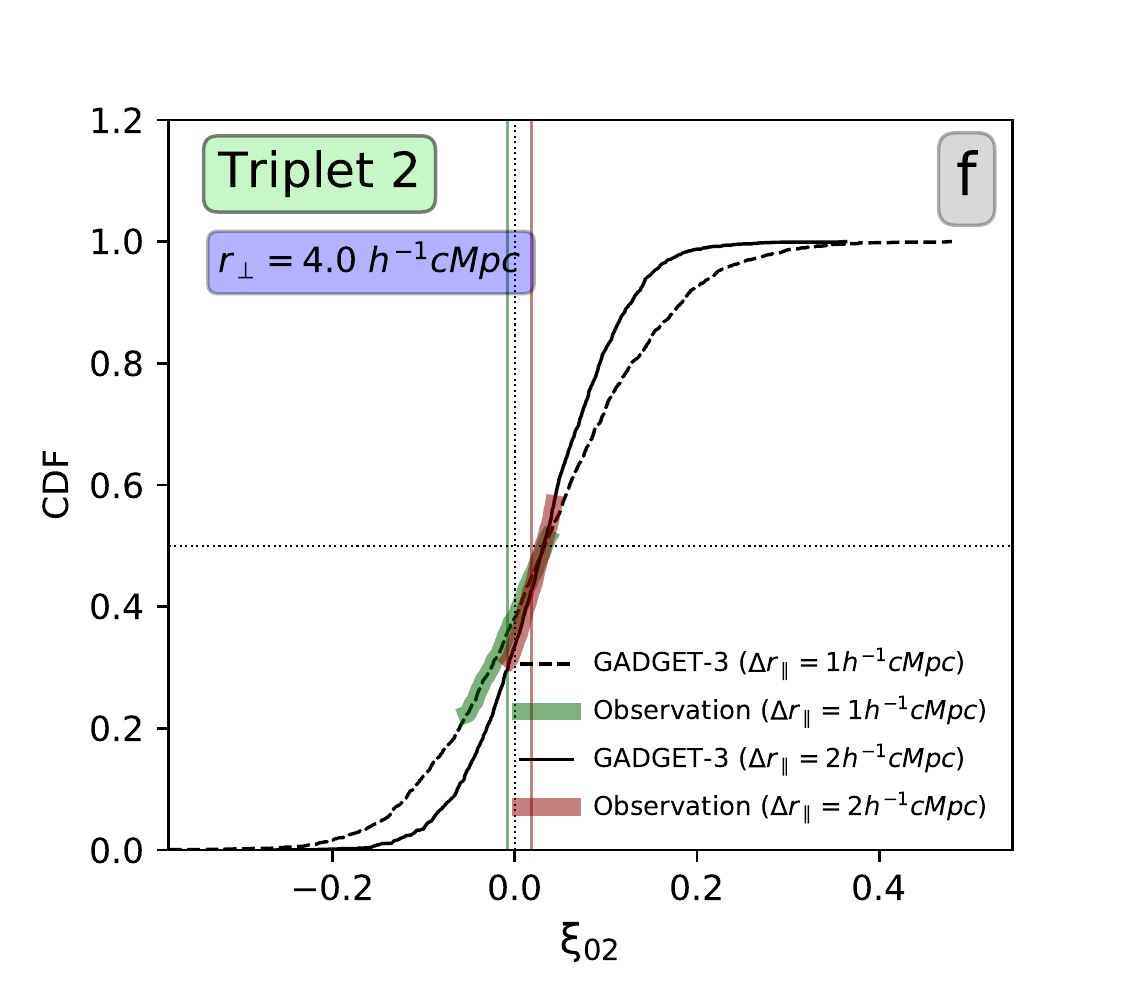}
	\end{minipage}%
	
	\begin{minipage}{0.3\textwidth}
		\includegraphics[viewport=5 10 340 270,width=6cm, clip=true]{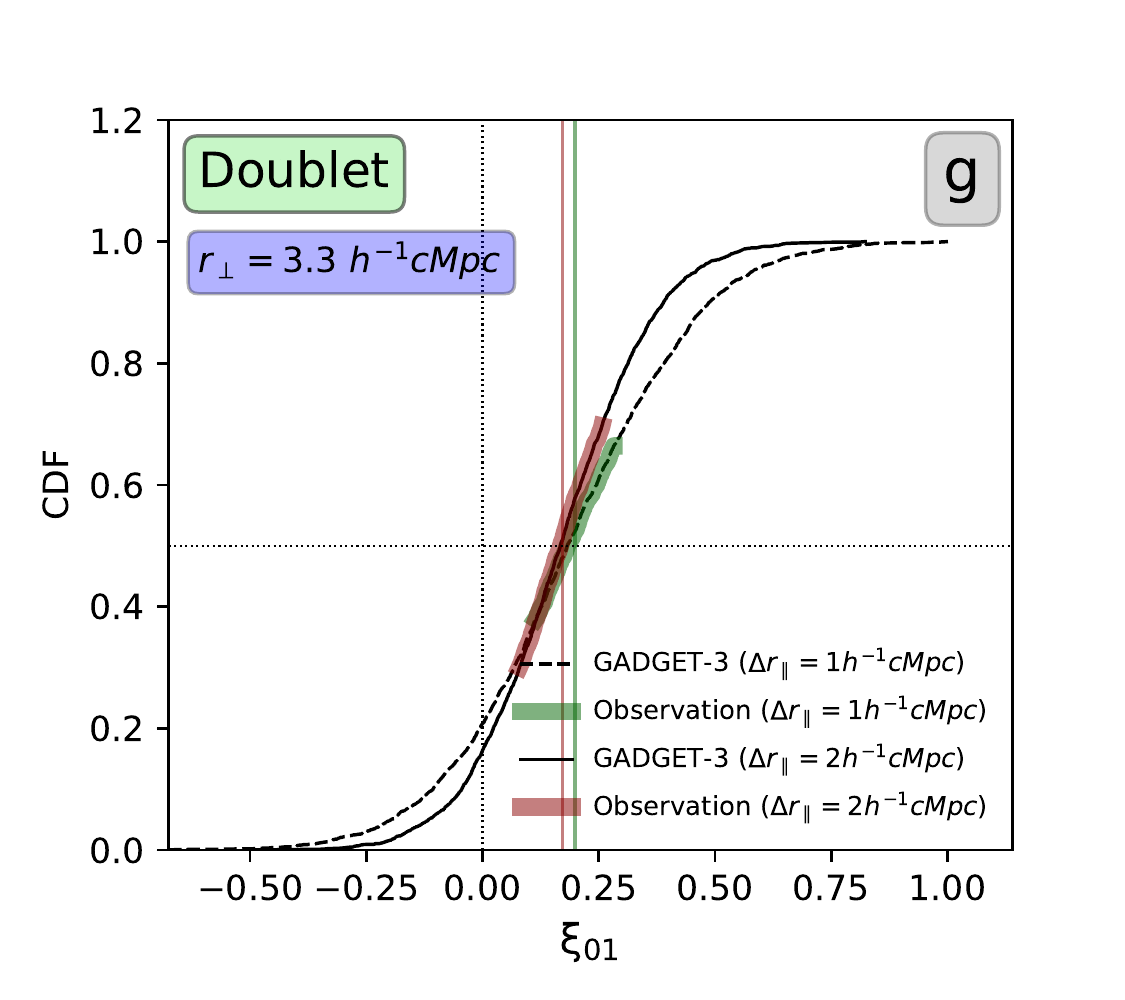}
	\end{minipage}%
	\begin{minipage}{0.3\textwidth}
		\includegraphics[viewport=5 10 340 270,width=6cm, clip=true]{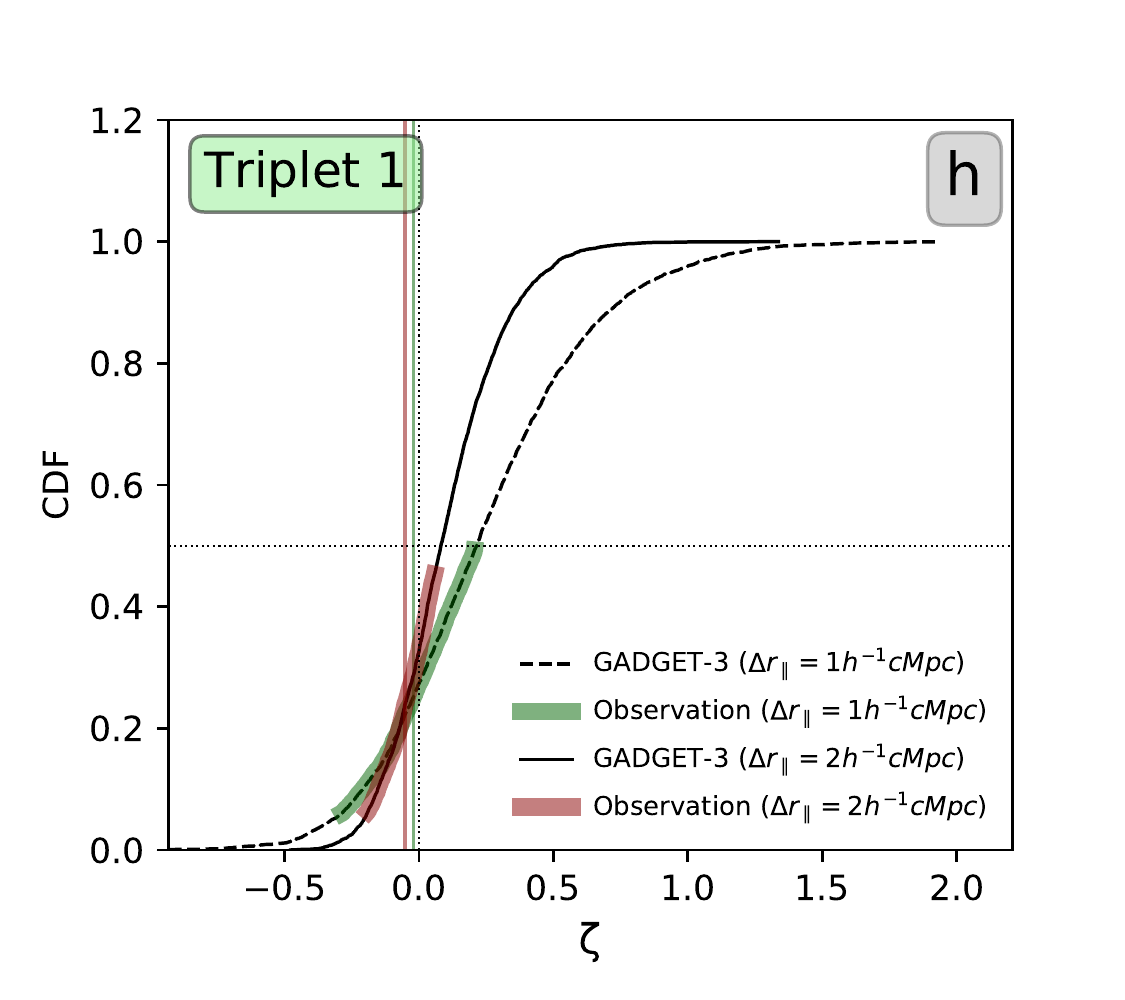}
	\end{minipage}
	\begin{minipage}{0.3\textwidth}
		\includegraphics[viewport=5 10 340 270,width=6cm, clip=true]{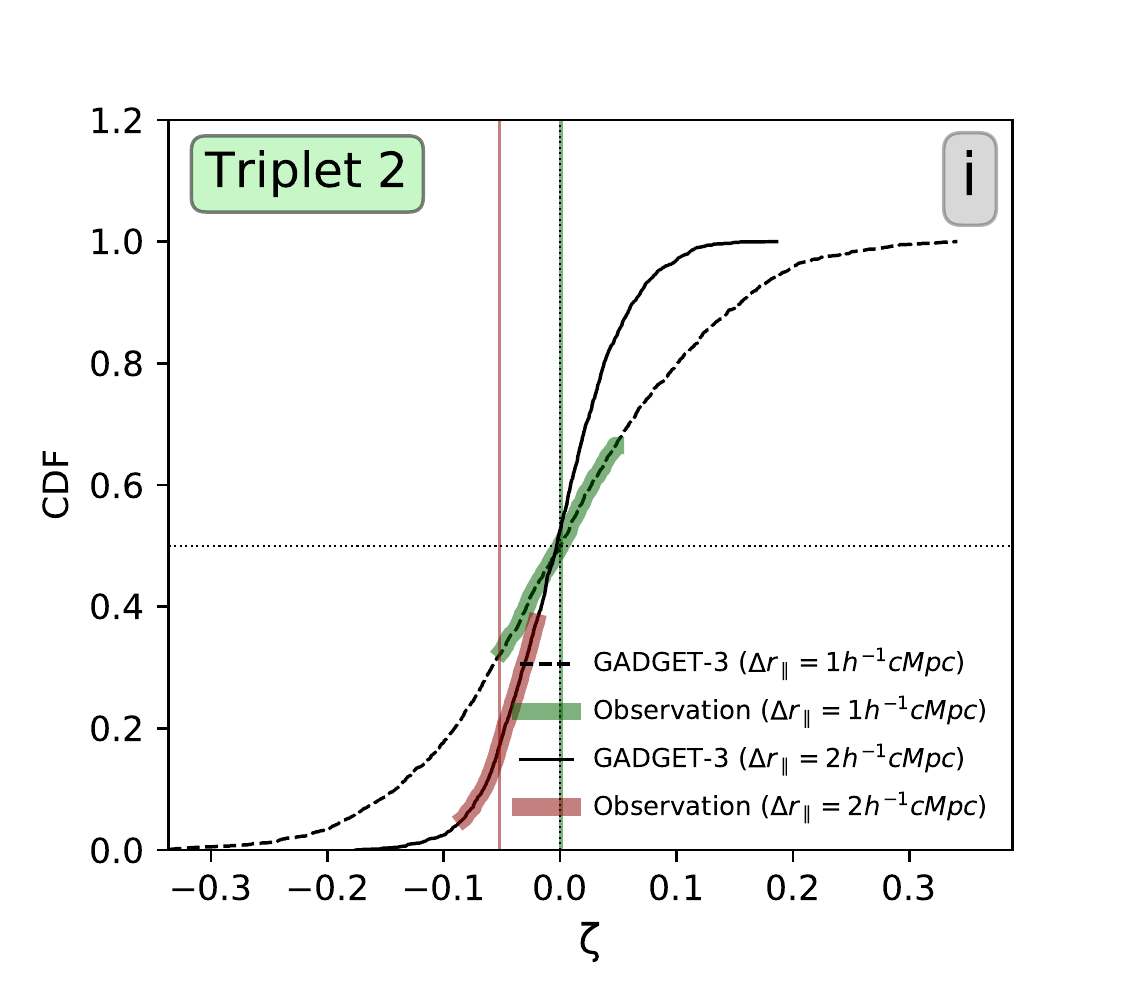}
	\end{minipage}%
	
	\caption{Cloud based transverse two-point and three-point correlation cumulative distribution for the two triplets. The dashed and solid curves are the cumulative distribution functions for $1 h^{-1}$cMpc and $2 h^{-1}$cMpc gridding respectively in the longitudinal direction. The red and green vertical lines denote the mean of the sub-sampled observed two-point and three-point correlation.  The red and green regions overlaying the simulated curve represents the $1\sigma$ confidence interval obtained from sub-sampling of the observed sightlines for the two different longitudinal grids. 
	}
	\label{fig_cloud_cdf_ap}
	
\end{figure*}
In Fig~\ref{fig_cloud_cdf_ap}, we have plotted the observed and simulated cumulative distributions for the cloud based transverse two-point (from panel "a" to "g") and three-point correlation (panel "h" and "i") for the X-Shooter triplets and doublets. The correlations for the observed spectra are obtained by subsampling the spectra with lengths similar to the simulation box size. 1000 random subsamples are generated and the mean and 1$\sigma$ confidence intervals are assigned based on these subsamples. The colored vertical lines denote the mean of the sub-sampled observed two-point and three-point correlation. The colored regions overlaying the simulated curve represents the 1$sigma$ confidence interval obtained from sub-sampling of the observed triplet sightlines. The figure emphasizes the effect of redshift space binning on the computed transverse two-point and three-point correlation. The plots have been done with $\Delta r_{\parallel}=\ \pm 1h^{-1}$cMpc and $\pm 2h^{-1}$cMpc. In case of two-point correlation, the general trend is that the correlation decreases as one increases the redshift space binning, both in observations and simulations. In case of three-point correlation, the magnitude of the correlation decreases (i.e, negative will become less negative) with increase in redshift space binning.

	% Don't change these lines
	\bsp	% typesetting comment	\label{lastpage}
\end{document}